\pdfoutput=1
\documentclass[sigconf, screen]{acmart}
\AtBeginDocument{%
  \providecommand\BibTeX{{%
    \normalfont B\kern-0.5em{\scshape i\kern-0.25em b}\kern-0.8em\TeX}}}

\copyrightyear{2023}
\acmYear{2023}
\setcopyright{cc}
\setcctype[4.0]{by}
\acmConference[CHI '23]{Proceedings of the 2023 CHI Conference on Human Factors in Computing Systems}{April 23--28, 2023}{Hamburg, Germany}
\acmBooktitle{Proceedings of the 2023 CHI Conference on Human Factors in Computing Systems (CHI '23), April 23--28, 2023, Hamburg, Germany}\acmDOI{10.1145/3544548.3580790}
\acmISBN{978-1-4503-9421-5/23/04}

\acmSubmissionID{9553}

\usepackage{acmart-taps}

\usepackage{xspace}
\usepackage{amsmath}

\usepackage{wrapfig}
\usepackage{enumitem}
\usepackage{url}
\usepackage[utf8]{inputenc}

\usepackage{rotating}
\usepackage{amsmath}
\usepackage{microtype}
\usepackage{bm}
\usepackage{soul}
\usepackage{mathtools}
\usepackage{balance}
\usepackage{CJKutf8}
\usepackage{stfloats} %

\definecolor{orange}{RGB}{250,130,49}
\definecolor{red}{RGB}{234,59,90}
\definecolor{agreen}{RGB}{74, 198, 148}
\definecolor{purple}{RGB}{158, 62, 177}
\definecolor{darkpurple}{RGB}{170, 70, 210}
\definecolor{aqua}{RGB}{87, 180, 181}
\definecolor{lightblue}{RGB}{72, 123, 232}
\definecolor{hotpink}{RGB}{255, 83, 115}
\definecolor{teal}{RGB}{90, 200, 250}
\definecolor{linkColor}{RGB}{0, 128, 229}
\definecolor{lightgreen}{RGB}{33, 222, 128}
\definecolor{gray}{RGB}{75, 101, 132}

\definecolor{germanred}{RGB}{234, 59, 90}
\definecolor{germanorange}{RGB}{250, 130, 49}
\definecolor{germanyellow}{RGB}{254, 211, 48}
\definecolor{germangreen}{RGB}{32, 191, 107}
\definecolor{germanblue}{RGB}{0, 128, 229}
\definecolor{germanviolet}{RGB}{56, 103, 214}
\definecolor{germanpurple}{RGB}{136, 84, 208}
\definecolor{anglergreen}{HTML}{00897B}
\definecolor{anglerorange}{HTML}{FB8C00}
\definecolor{equationgreen}{HTML}{009688}
\definecolor{equationblue}{HTML}{42A5F5}

\definecolor{myred}{RGB}{224, 49, 119}
\definecolor{myorange}{RGB}{250, 130, 49}
\definecolor{myyellow}{RGB}{254, 211, 48}
\definecolor{mygreen}{RGB}{14, 152, 136}
\definecolor{myblue}{RGB}{0, 128, 229}
\definecolor{myviolet}{RGB}{56, 103, 214}
\definecolor{mypurple}{RGB}{136, 84, 208}
\definecolor{mybrown}{RGB}{132, 99, 88}
\definecolor{mygray}{RGB}{220, 220, 220}

\newcommand{\tool}{\textsc{\textsf{Angler}}}
\newcommand{\tableview}{\textit{Table View}}
\newcommand{\preview}{\textit{Set Preview}}
\newcommand{\previews}{\textit{Set Previews}}
\newcommand{\detailview}{\textit{Detail View}}

\newcommand{\timeline}{\textit{Timeline}}
\newcommand{\spotlight}{\textit{Spotlight}}
\newcommand{\thumbnail}{\textit{Thumbnail}}
\newcommand{\thumbnails}{\textit{Thumbnails}}
\newcommand{\filter}{\textit{Filter Bar}}
\newcommand{\samplelist}{\textit{Sentence List}}

\newcommand{\figpart}[1]{\textcolor{ACMPurple}{#1}}

\newcommand*{\vcenteredhbox}[1]{\begingroup\setbox0=\hbox{#1}\parbox{\wd0}{\box0}\endgroup}

\newcommand{\link}[1]{{\href{#1}{\color{linkColor}\textbf{\texttt{#1}}}}}

\newcommand{\aci}[1]{\todo[linecolor=myorange,backgroundcolor=myorange!25,bordercolor=myorange]{AC1: ``#1''}}
\newcommand{\acii}[1]{\todo[linecolor=myblue,backgroundcolor=myblue!25,bordercolor=myblue]{AC2: ``#1''}}
\newcommand{\revieweri}[1]{\todo[linecolor=mygreen,backgroundcolor=mygreen!25,bordercolor=mygreen]{R2: ``#1''}}
\newcommand{\reviewerii}[1]{\todo[linecolor=mypurple,backgroundcolor=mypurple!25,bordercolor=mypurple]{R3: ``#1''}}

\newcommand{\reviewerclean}{
  \renewcommand{\aci}[1]{}
  \renewcommand{\acii}[1]{}
  \renewcommand{\revieweri}[1]{}
  \renewcommand{\reviewerii}[1]{}
}
\reviewerclean{}

\usepackage{xspace,xpunctuate}
\newcommand{\ie}{{i.e.,}\xspace}
\newcommand{\eg}{{e.g.,}\xspace}

\newcommand{\numParticipants}{{13}\xspace}
\newcommand{\numParticipantsObs}{{7}\xspace}

\newcommand{\userOne}{{UX1}\xspace}
\newcommand{\userTwo}{{E1}\xspace}
\newcommand{\userThree}{{E2}\xspace}
\newcommand{\userFour}{{UX2}\xspace}
\newcommand{\userFive}{{UX3}\xspace}
\newcommand{\userSix}{{E3}\xspace}
\newcommand{\userSeven}{{UX4}\xspace} 
\begin{document}
\begin{CJK*}{UTF8}{gkai}

\title{\tool{}: Helping Machine Translation Practitioners Prioritize Model Improvements}

\author{Samantha Robertson}
\authornote{Work done at Apple.}
\authornote{Both authors contributed equally to this research.}
\affiliation{%
  \institution{University of California, Berkeley}
  \city{Berkeley}
  \state{CA}
  \country{USA}
}
\email{samantha\_robertson@berkeley.edu}

\author{Zijie J. Wang}
\authornotemark[1]
\authornotemark[2]
\affiliation{%
  \institution{Georgia Institute of Technology}
  \city{Atlanta}
  \state{GA}
  \country{USA}
}
\email{jayw@gatech.edu}

\author{Dominik Moritz}
\author{Mary Beth Kery}
\authornote{Both authors contributed equally to this research.}
\author{Fred Hohman}
\authornotemark[3]
\affiliation{%
  \institution{Apple}
  \city{Seattle}
  \state{WA}
  \country{USA}
}
\email{{domoritz,mkery,fredhohman}@apple.com}

\renewcommand{\shortauthors}{Robertson, Wang, et al.}

\begin{abstract}
Machine learning (ML) models can fail in unexpected ways in the real world, but not all model failures are equal.
With finite time and resources, ML practitioners are forced to prioritize their model debugging and improvement efforts.
Through interviews with \numParticipants ML practitioners at Apple, we found that practitioners construct small targeted test sets to estimate an error's nature, scope, and impact on users.
We built on this insight in a case study with machine translation models, and developed \tool{}, an interactive visual analytics tool to help practitioners prioritize model improvements.
In a user study with \numParticipantsObs machine translation experts, we used \tool{} to understand prioritization practices when the input space is infinite, and obtaining reliable signals of model quality is expensive.
Our study revealed that participants could form more interesting and user-focused hypotheses for prioritization by analyzing quantitative summary statistics and qualitatively assessing data by reading sentences.
\end{abstract}

\ccsdesc[500]{Human-centered computing~Visualization systems and tools}
\ccsdesc[500]{Human-centered computing~Interactive systems and tools}
\ccsdesc[300]{Computing methodologies~Machine learning}
\ccsdesc[300]{Computing methodologies~Artificial intelligence}

\keywords{Model evaluation, machine translation, visual analytics}

\begin{teaserfigure}
  \centering
  \includegraphics[width=0.8\textwidth]{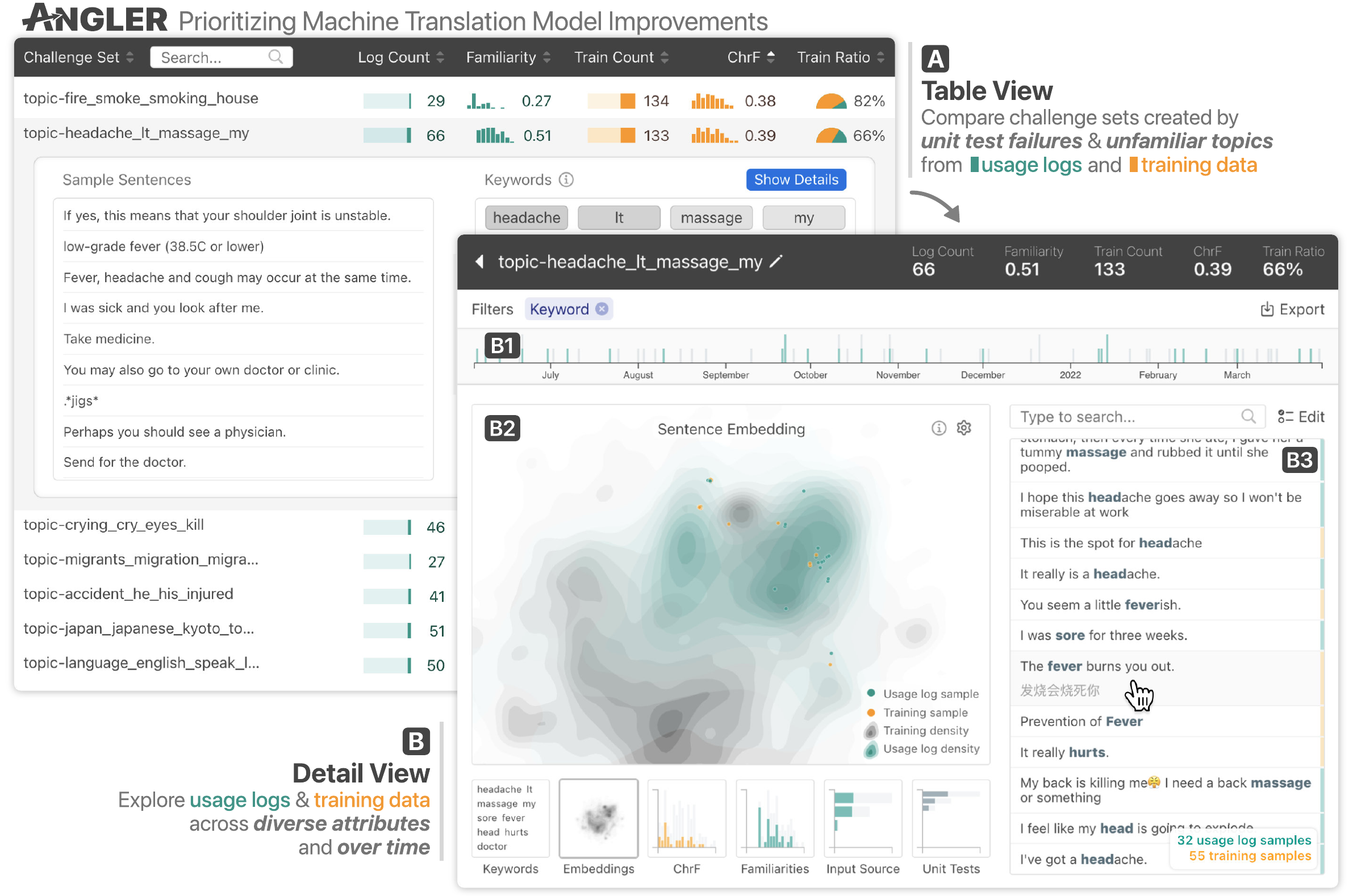}
  \caption[]{
    \tool{} enables ML developers to easily explore and curate challenge sets for machine translation.
    \textbf{(A) The \tableview{}} lists all challenge sets, allowing users to compare them by metrics such as sample count, model performance, and familiarity score.
    After selecting a set, \textbf{(B) the \detailview{}} allows users to further explore samples in this set across various dimensions.
    \textbf{(B1) The \timeline{}} enables users to query data samples by time.
    \textbf{(B2) The \spotlight{}} presents visualizations with linking and brushing to help users characterize the set from different angles.
    \textbf{(B3) The \samplelist{}} shows all selected data samples and allows users to further fine-tune before exporting this challenge set for downstream tasks.
  }
  \Description{
    Two components: (A) a screenshot of Angler's Table View, and (B) a screenshot of Angler's Detail View.
    The Table View aims to compare multiple challenge sets created by unit test failures and unfamiliar topics.
    The Table View contains a sortable table with column headers: Challenge Set, Log Count, Familiarity, Train Count, ChrF, and Train Ratio.
    Each row in the table represents a challenge set.
    A challenge set named ``topic-headache_it_massage_my'' is extended, where sample sentences and representative keywords are shown below that row.
    The Detail View explores training and traffic data across diverse attributes and over time.
    At the top of the Detail View, there is a header describing the Name, Log count, Familiarity, Train Count, ChrF, and Train Ratio of the challenge set ``topic-headache_it_massage_my''.
    Below the header, there is a filter bar with a filter tag ``Keyword''.
    Below the filter bar, there is a timeline histogram from June to April.
    There are two components below the timeline.
    On the left, there is a large embedding view visualizing the embedding spaces of the training data and usage logs.
    Below the embedding view, there are six thumbnails labeled as ``Keywords'', ``Embeddings'', ``ChrF'', ``Familiarities'', ``Input Source'', and ``Unit Tests''.
    On the right, there is a list of sentences with a search bar and an edit button.
    Each sentence is an English input.
    A clicked sentence has a Chinese translation below the input.
  }
  \label{fig:teaser}
\end{teaserfigure}

\maketitle

\section{Introduction}
\label{sec:introduction}

In machine learning (ML), out-of-sample evaluation metrics are used to approximate how well a model will perform in the real world. However, numerous high-profile failures have demonstrated that aggregate performance metrics only estimate how a model will perform \textit{most of the time} and obscure harmful failure modes~\cite[e.g.][]{koeneckeRacialDisparitiesAutomated2020, devriesDoesObjectRecognition2019, bolukbasiManComputerProgrammer2016, obermeyerDissectingRacialBias2019}. In response, researchers have explored how to anticipate model failures before they impact end users. For example, disaggregated error analysis has helped identify errors that impact people with marginalized identities. Prominent examples include facial recognition models failing to recognize women with dark skin tones~\cite{buolamwiniGenderShadesIntersectional2018} or translation models perpetuating gender stereotypes~\cite{stanovskyEvaluatingGenderBias2019}. However, subgroups where a model fails can be highly contextual, specific, and may not match any social category (\ie ``men wearing thin framed glasses''~\cite{cabreraDiscoveringValidatingAI2021} or ``busy/cluttered workspace''~\cite{deonSpotlightGeneralMethod2022}).
It remains an open challenge for ML practitioners to detect \textit{which} specific use case scenarios are likely to fail out of a possibly infinite space of model inputs ---and prioritize \textit{which} failures have the greatest potential for harm~\cite{holsteinImprovingFairnessMachine2019b, barocasDesigningDisaggregatedEvaluations2021}.\looseness=-1

With finite time and resources, where should machine learning practitioners spend their model improvement efforts? \textbf{In this work, we aim to help practitioners detect and prioritize underperforming subgroups where failures are most likely to impact users}.  Towards this goal, we contribute the following research:

\begin{itemize}[topsep=5pt, itemsep=1mm, parsep=1mm, leftmargin=10pt]

    \item \textbf{A formative interview study with 13 ML practitioners at Apple} to understand their process for prioritizing and diagnosing potentially under-performing subgroups~(\autoref{sec:interview}). Practitioners rely on the model type, usage context, and their own values and experiences to judge error importance. To test \textit{suspected} issues, practitioners collect similar data to form \textbf{challenge sets}. Using a challenge set, practitioners rely on a combination of signals from model performance and usage patterns to gauge the prevalence and severity of a failure case. The most common fix for an under-performing subgroup is dataset augmentation to increase the model's \textbf{coverage} for that subgroup.

    \item \textbf{\tool{}~(\autoref{fig:teaser}), an open-source}\footnote{\tool{} code: \link{https://github.com/apple/ml-translate-vis}.} \textbf{interactive visualization tool for supporting error prioritization for machine translation} (MT)~(\autoref{sec:ui}). Since our research centers on the issue of \textit{prioritization} (rather than specific error identification) we chose an ML domain where practitioners cannot directly observe model errors. MT developers do not speak all the languages that their translation models support. They rely on proxy metrics like BLEU~\cite{papineniBleuMethodAutomatic2002} to estimate model performance but ultimately depend on human expert translators to obtain any ground-truth. Since gathering feedback from human translators is expensive and time-consuming, careful allocation of annotation resources is crucial. To help MT practitioners prioritize suspected error cases that most align with user needs, we worked with an industry MT product team to develop \tool{}.
    By adapting familiar visualization techniques such as \textit{overview + detail}, \textit{brushing + linking}, and \textit{animations}, \tool{} allows MT practitioners to explore and prioritize potential model performance issues by combining multiple model metrics and usage signals.

    \item \textbf{A user study of \numParticipantsObs MT practitioners using \tool{}} to assess the relative importance of potentially under-performing subgroups~(\autoref{sec:study}). MT practitioners completed a realistic exercise to allocate a hypothetical budget for human translators. Observing MT practitioners using \tool{} revealed how they use their intuition, values, and expertise to prioritize model improvements. Direct inspection of data showed the potential to encourage more efficient allocation of annotation resources than would have been possible by solely relying on quantitative metrics. While rule-based error analysis allowed participants to more successfully find specific model failure patterns, exploring data grouped by topic encouraged practitioners to think about how to improve support for specific use cases. The study also prompted discussion for future data collection and helped practitioners imagine new features for translation user experiences.
\end{itemize}

\noindent No model is perfect, and large production models have a daunting space of potential error cases. Prioritization of subgroup analysis is a practical challenge that impacts model end users. By exploring prioritization in the context of MT, where there are no reliable quality signals for previously unseen model inputs, we highlight the value of flexible visual analytics systems for guiding choices and trade-offs. Our findings support the potential for mixed-initiative approaches: where automatic visualizations \& challenge sets help reveal areas of model uncertainty, and human ML practitioners use their judgment to decide where to spend time and money on deeper investigation.\looseness=-1

\section{Related Work}
This research builds on substantial prior work across general ML evaluation practices, visualization tooling for ML error analysis, and a broad body of work from our target domain, machine translation.

\subsection{ML Evaluation and Error Analysis}
\label{sec:related:auditing}

First, we review standard evaluation practices in ML, and discuss how visualization tools can support ML error discovery.

\subsubsection{How do Practitioners Evaluate ML Models?}
Standard practice in machine learning is to evaluate models by computing aggregate performance metrics on held-out test sets before using them in the real world (offline evaluation)~\cite{amershiSoftwareEngineeringMachine2019,mReviewEvaluationMetrics2015}.
The goal of using held-out test sets, \ie data that was not used during model development, is to estimate how well the model will generalize to real world use cases.
However, offline evaluations are limited.
For example, held-out datasets can be very different from real usage data~\cite{patelInvestigatingStatisticalMachine2008, rechtImageNetClassifiersGeneralize2019}, as data in the wild is often noisy~\cite{kielaDynabenchRethinkingBenchmarking2021} and the real world is ever-changing~\cite{kohWILDSBenchmarkIntheWild2021}.
Held-out datasets tend to contain the same biases as the training data so they cannot detect potentially harmful behaviors of the model~\cite{rajpurkarKnowWhatYou2018, gevaAreWeModeling2019}.
While summarizing a model's performance in aggregate metrics is undeniably useful, it is insufficient for ensuring model quality.

To overcome these limitations, researchers have proposed additional approaches to help discover model weaknesses~\cite[e.g.,][]{burlot2017evaluating, gardner2020evaluating, hookerCharacterisingBiasCompressed2020}.
For example, practitioners can apply subgroup analysis to discover fairness issues~\cite{dudikFairlearnToolkitAssessing2020}, use perturbed adversarial examples to evaluate a model's robustness to noise~\cite{bhatt2021case, wuPolyjuiceGeneratingCounterfactuals2021, rajpurkarKnowWhatYou2018}, create rule-based unit tests to detect errors~\cite{ribeiroAccuracyBehavioralTesting2020, rottgerHateCheckFunctionalTests2021}, and conduct interactive error analysis to expand known failure cases~\cite{naik2018stress, wuErruditeScalableReproducible2019, ribeiroAdaptiveTestingDebugging2022}. ML practitioners also continuously monitor a deployed model's performance and distribution shifts over time~\cite{barocasDesigningDisaggregatedEvaluations2021}.

We build on this work by focusing on the question of \textit{prioritization}: how ML practitioners judge where to spend their time and resources among many possible model failure cases. This understanding can help inform the design of future tooling and techniques for surfacing model issues that are more attuned to urgency or severity.\looseness=-1

\subsubsection{Visualization Tools for Supporting Error Discovery}
\label{sec:related:vis-error}

Interactive visualization is a powerful method for helping ML developers explore and interpret their models~\cite{hohmanVisualAnalyticsDeep2019,beauxis-aussaletRoleInteractiveVisualization2021}.
While many visualizations have been built to help practitioners evaluate models over time, one area of recent work has focused on designing and developing analytic tools for ML error discovery~\cite[e.g.][]{wexlerWhatIfToolInteractive2019,liuVisualDiagnosisTree2018,liUnifiedUnderstandingDeep2022,onoPipelineProfilerVisualAnalytics2021, chung2020automated, zhang2022sliceteller}.
For example, \textsc{FairVis}~\cite{cabreraFAIRVISVisualAnalytics2019} uses visualizations to help ML developers discover model bias by investigating known subgroups and exploring similar groups in the tabular data.
Similarly, \textsc{Visual Auditor}~\cite{munechikaVisualAuditorInteractive2022} automatically surfaces underperforming subgroups and leverages graph visualizations to help practitioners interpret the relationships between subgroups and discover new errors.
For image data, \textsc{explAIner}~\cite{spinnerExplAInerVisualAnalytics2019} combines interactive visualization and post-hoc ML explanation techniques~\cite[e.g.,][]{ribeiroWhyShouldTrust2016, lundbergUnifiedApproachInterpreting2017} to help practitioners diagnose problems with image classifiers.
For text data, \textsc{Seq2seq-Vis}~\cite{strobeltEq2sEqv2018} helps practitioners debug sequence-to-sequence models by visualizing the model's internal mechanisms throughout each of its inference stages.

The success of these recent visual ML diagnosis systems highlights the outstanding potential of applying visualization techniques to help ML developers detect errors.
Instead of visualizing a model's internals ~\cite[e.g.,][]{strobeltEq2sEqv2018,hooverExBERTVisualAnalysis2020}, we treat ML models as black-box systems and focus on probing their behaviors on different data subsets.
While prior systems have focused on ML applications like image captioning where errors are directly observable \cite{cabreraFAIRVISVisualAnalytics2019, cabreraDiscoveringValidatingAI2021}, we designed \tool{} for the more challenging modeling domain where practitioners cannot always spot-check errors and must rely on proxy metrics to estimate the likelihood of an error.

\subsection{Evaluating Machine Translation Models}
\label{sec:related:mt-evaluation}

Evaluating translation quality is extremely nuanced and difficult~\cite{hanMachineTranslationEvaluation2018, koehnEuroMatrixMachineTranslation2007, vilarErrorAnalysisStatistical2006, saiSurveyEvaluationMetrics2022}. Language can mean different things and be written in different ways by different people at different times.
There are also often multiple ``correct'' translations of the same input~\cite{kingIssuesNaturalLanguage1998}.\looseness=-1

The gold standard for machine translation evaluation is to have professional translators directly judge the quality of model outputs, for instance, by rating translation fluency and adequacy~\cite{callison-burch-etal-2007-meta}. There are also automatic metrics for machine translation---such as BLEU~\cite{papineniBleuMethodAutomatic2002}, ChrF~\cite{popovicChrFCharacterNgram2015} and METEOR~\cite{banerjeeMETEORAutomaticMetric2005}---which measure the similarity between a candidate text translation (model output) and one or more reference translations.
Intuitively, these metrics apply different heuristics to measure token overlap between two sentences. While these metrics are less reliable and nuanced than human judgment ~\cite{postCallClarityReporting2018, callison-burchReevaluatingRoleBleu2006, linORANGEMethodEvaluating2004, reiterStructuredReviewValidity2018, mathurTangledBLEUReevaluating2020, stentEvaluatingEvaluationMethods2005, saiSurveyEvaluationMetrics2022}, they are intended to correlate as much as possible with human judgments and are widely used for comparing the aggregate performance of different MT models.

An overarching challenge in MT evaluation is that it is especially resource intensive. Both human and automatic evaluation depends on the expertise of human translators, to either directly judge translation quality, or generate reference translations. Since translators have high levels of expertise and are often difficult to find for rare language pairs~\cite{freitagExpertsErrorsContext2021, moreMachineTranslationnessMachinelikeness2014}, it is expensive to evaluate translation quality if the input data does not already have a reference translation (\eg users' requests to a model). In addition, it is difficult to maintain consistent quality within and across human evaluators~\cite{popovicAgreeDisagreeAnalysis2021, saldiasfuentesMoreEffectiveHuman2022, ipeirotisQualityManagementAmazon2010}. Since evaluating the quality of translations from real world use cases requires human annotation, online monitoring and debugging MT models presents a resource allocation problem. In this work, we explore how interactive visualization of online model usage might help MT practitioners prioritize data for human evaluation.\looseness=-1

\subsubsection{Subgroups in Machine Translation}
\label{sec:related:challenge-set}

More recently, researchers have explored how to systematically identify specific kinds of errors in MT models~\cite{stanovskyEvaluatingGenderBias2019, isabelleChallengeSetApproach2017, popovicChallengeTestSets2019}. Many of these are language-dependent challenge sets to probe the syntactic competence of MT models~\cite{burchardtLinguisticEvaluationRulebased2017, avramidisLinguisticEvaluationGermanEnglish2019, macketanzFinegrainedEvaluationGermanEnglish2018}.
For example,~\citeauthor{isabelleChallengeSetApproach2017} introduces a dataset of difficult sentences designed to test linguistic divergence phenomena between English and French.
\citeauthor{stanovskyEvaluatingGenderBias2019} analyzed sentences with known stereotypical and non-stereotypical gender-role assignments in MT, which falls in a broader body of work on detecting gender bias in MT~\cite{zhaoGenderBiasCoreference2018, rudingerGenderBiasCoreference2018, vanmassenhoveGENderITAnnotatedEnglishItalian2021, trolesExtendingChallengeSets2021}.

While these approaches deepen our understanding of specific model failure modes, it is unclear how different errors impact end users of MT models. As a recent survey suggests~\cite{popovicChallengeTestSets2019}, most of these challenge test sets are created either manually by MT experts or automatically with linguistic rule-based heuristics~\cite[e.g.,][]{choshenAutomaticallyExtractingChallenge2019, raganatoMuCoWTestSuite2019, burlotWMT18Morpheval2018}.
An alternative approach has been to examine the performance of MT models in specific domains like law~\cite{kitComparativeEvaluationOnline2008, yatesScalingTowerBabel2006, prabhuDidTheyDirect2021} or healthcare~\cite{khoongAssessingUseGoogle2019, dasDangersMachineTranslation2019, tairaPragmaticAssessmentGoogle2021}.
These domain-specific challenge sets are deeply informed by knowledge of a particular use case, but are limited in scope. It is difficult for researchers to develop broader challenge sets guided by real users' needs because we lack a clear understanding of how people use MT models, and how they can be impacted by errors.
In this work we strive to narrow this gap by working with an industry MT team to understand how practitioners might prioritize model improvements based on their users' needs.

\subsubsection{Visualization Tools for Evaluation in Machine Translation}
\label{sec:related:visualization}

There is a growing body of research focused on designing visual analytics tools for MT researchers and engineers~\cite[e.g.,][]{leeInteractiveVisualizationManipulation2017, parkVANTVisualAnalytics2022}.
For example, with the \textsc{Chinese Room} system MT practitioners can interactively decode and correct translations by visualizing the source and translation tokens side by side~\cite{albrechtChineseRoomVisualization2009}.
Similarly, \textsc{NMTVis}~\cite{munzVisualizationbasedImprovementNeural2022}, \textsc{SoftAlignments}~\cite{riktersVisualizingNeuralMachine2017}, and \textsc{NeuralMonkey}~\cite{helclNeuralMonkeyOpensource2017} use interactive visualization techniques such as parallel coordinate plots, node-link diagrams, and heatmaps to help MT practitioners analyze attention weights and verify translation results.
MT researchers also use visual analytics tools~\cite[e.g.,][]{klejchMTComparEvalGraphicalEvaluation2015,neubigComparemtToolHolistic2019, wangVizSeqVisualAnalysis2019} to better understand MT evaluation metrics such as BLEU, ChrF, and METEOR scores.
For example, \textsc{iBLEU}~\cite{madnaniIBLEUInteractivelyDebugging2011} allows researchers to visually compare BLEU scores between two MT models at both corpus and sentence levels.
\textsc{VEMV}~\cite{steeleVisevalMetricViewer2018} uses an interactive table view and bar charts to help researchers compare MT models across various metric scores.

In contrast, our work focuses on evaluation based on \textit{usage} of a deployed model. We use interactive visualization as a probe to understand how practitioners prioritize model evaluation when reliable quality signals are expensive to obtain. Our findings provide insight into what kind of information practitioners need to assess potential model failures with respect to their impact on users.
\section{Interview Study: Prioritizing Model Improvements}
\label{sec:interview}

This work began in close collaboration with an industry machine translation team, with the goal of helping them prioritize model debugging and improvement resources on problems that had the greatest potential to impact end users. From initial conversations with team members, we learned that their existing process for identifying and addressing problems was largely driven by specific errors (e.g., bug reports, or biases surfaced in academic research), or based on random sampling of online requests.
Further, this process was limited to team members with the technical expertise to conduct one-off data analyses. To gain insight into a broader range of existing approaches to prioritization, we turned to practitioners across other ML domains.

We conducted a semi-structured interview study with 13 ML practitioners at Apple. In this section, we describe how practitioners identify and solve specific issues with ML models that impact the user experience. First, we discuss how practitioners navigate a large space of possible failure cases~(\autoref{sec:interview-sourcing}). Next, we describe how they build challenge sets to assess the cause, scope and severity of an issue, which then informs which issues they address and how they fix them~(\autoref{sec:interview-challenge-set}). At each stage, we highlight how practitioners bring a range of approaches and perspectives to the task of prioritization. We synthesize these findings into four design implications for tooling to support prioritization in model debugging~(\autoref{sec:design-implication}).

\subsection{Data Collection and Analysis}\label{sec:interview-methods}

We recruited practitioners from both internal mailing lists related to ML and snowball sampling at Apple.
Each interview lasted between 30 to 45 minutes.
We recorded the interviews when participants gave permission, and otherwise took detailed notes. The study was approved by our internal IRB.
We recruited practitioners who have worked on developing and/or evaluating models that are embedded in user-facing tools and products.
Incorrect, offensive, or misleading model predictions are detrimental to users' experiences with these models.
Therefore, engineers and data scientists that are working on evaluating and improving user-facing ML models are more likely to consider how their models shape users' experiences than other kinds of ML practitioners.
Indeed, we found in our interviews that participants often considered how different kinds of model failures may impact end users.
An overview of the participants' primary ML application is shown in \autoref{tbl:interview-participants}.

\begin{table}
\caption[]{Primary type of machine learning application that each interview participant works on.}
\label{tbl:interview-participants}
\begin{tabular}{ cll}
\textbf{Participant} & \textbf{ML Application} & \textbf{Role}\\
\midrule
P1 & Business forecasting & Data Scientist\\
P2 & Multiple NLP tasks & Data Scientist\\
P3 & Image segmentation & ML Engineer\\
P4 & ML Tooling & ML Engineer\\
P5 & Image classification & Data Scientist\\
P6 & Image classification & Research Scientist\\
P7 & Various CV tasks & ML Manager\\
P8 & Image classification & ML Engineer\\
P9 & Resource use forecasting & ML Engineer\\
P10 & Image segmentation & Research Scientist\\
P11 & Recommender systems & ML Manager\\
P12 & Image captioning & Robustness Analyst\\
P13 & Gesture recognition & Research Scientist \\
\end{tabular}
\Description{
    A table that links each interview participant with their primary machine learning application area.
    P1 works on business forecasting.
    P2 works on multiple NLP tasks.
    P3 works on image segmentation.
    P4 works on ML tooling.
    P5 works on image classification.
    P6 works on image classification.
    P7 works on various CV tasks.
    P8 works on image classification.
    P9 works on resource use forecasting.
    P10 works on image segmentation and classification.
    P11 works on recommender systems.
    P12 works on image captioning.
    P13 works on gesture recognition.
}
\end{table}

Two authors conducted inductive qualitative analysis of the interview data.
One author conducted three rounds of open coding, synthesizing and combining codes each round \cite{qualresearch}.
Next, a second author took the code book in development and independently coded two interviews, adding new codes where relevant and noting disagreements.
These two authors then discussed these transcripts and codes to ensure mutual understanding and shared interpretation of the codes, and converged on a final code book.
Lastly, they used this code book to code half of the transcripts each.

\subsection{Sourcing Potential Issues}
\label{sec:interview-sourcing}

Out of the many ways an ML model could fail, we found that practitioners want to prioritize those that are most consequential for end users. Participants discussed three approaches to find such issues: (1) analyzing errors reported by users; (2) brainstorming potential errors in collaboration with domain experts; and (3) comparing usage patterns against model training data to find areas where the distributions of these two data sources differ.

\subsubsection{User Testing and User-Reported Errors}

Six participants discussed identifying potential issues through direct feedback from users or from user testing [P3, P4, P5, P7, P10, P13]. Even \textit{ad hoc} testing with a small sample of people can reveal issues that are not surfaced in standard offline evaluations. For example, P5 once \textit{``just showed the [model-driven] app to other people''} and found a \textit{``weird edge case''} where the model always (and often erroneously) classified images containing hands as a certain output class. This was an error that was not surfaced in offline model evaluations because the test data was drawn from the same distribution as the training data, and both contained a spurious correlation between images with hands and this particular output class. \textit{``That's why you do your user testing''} (P5).

User feedback outside of user testing can be difficult to source. In some settings, failures are detectable from signals in usage data, \eg whether a user accepts or rejects a suggested word from a predictive text model [P5]. More often, real users need to take additional steps to report errors, which they are unlikely to do for every error they encounter: \textit{``I think it takes a lot of \textup{[effort]} and willingness to go and file these things''} [P4]. In the most difficult case, users are not \textit{able} to assess prediction quality themselves (\eg if someone is using MT to translate to or from a language they do not know). In such contexts, direct user feedback is particularly rare.

\subsubsection{Brainstorming with Domain Experts}
\label{sec:interview-sourcing:brainstorming}

Another approach is to brainstorm potential failure modes with people who hold specialized knowledge of what may be both \textit{important} to users and \textit{challenging} for the model [P1, P4, P9, P12, P13].\looseness=-1

\begin{quote}
    \textit{``Sometimes we involve partners, like other teams or providers who are specialized in the area, to attack the model make the model fail.''} --- P4
\end{quote}

P13 works on gesture recognition models and followed this approach of brainstorming potential errors. P13 had a deep understanding of how their model worked, and thus what kinds of inputs might be difficult to classify accurately.
They then collaborated with designers and accessibility experts, who have a deep understanding of users' needs, to identify how the model's weaknesses lined up with \textit{realistic} and important use cases.

Sometimes ML practitioners have built this kind of expertise themselves over years of working with a similar model type [P1], or through precedent with prior reported model failures [P7]. P4 and P12 pointed to academic research (\eg work published at venues focused on fairness and ethics in AI), the press, and social media as additional helpful sources of potential failure cases. These sources, while not necessarily directly related to any specific model they are working on, can help practitioners understand patterns in ML failures more systematically, and anticipate high-stakes failures.

Brainstorming is particularly useful for identifying \textit{types} of failures that could impact users \cite{ribeiroAccuracyBehavioralTesting2020, ribeiroAdaptiveTestingDebugging2022}. However, it is difficult to translate these types into actual failure cases and keep them up to date \cite{ribeiroAdaptiveTestingDebugging2022, bhatt2021case}.

\begin{quote}
    \textit{``We can go with things like what's known as the OVS list---offensive, vulgar, slur. Those are quite obvious, but things can be more subtly offensive... Frankly, there are ways to be offensive that we just simply probably haven't anticipated and language evolves and slang becomes apparent, and even the global situation changes and things that weren't offensive before could become offensive.''} --- P7
\end{quote}

\subsubsection{Identifying Usage Patterns with Low Training Data Coverage}
\label{sec:interview:coverage}

A third approach is to identify suspected areas of weakness for the model by looking for differences between how users are interacting with a model and the data with which that model was trained. We use the term \textbf{coverage} to describe how well a model's training data ``covers'' the space of inputs that the model receives after deployment (i.e., use cases). The coverage of a particular use case is a measure of how much the model ``knows'' about those kinds of inputs, and can be used as a proxy for model performance when other quality signals are not available.

\begin{quote}
    \textit{``We know that the \textup{[training]} datasets that we have, however large they are, they don't cover the entire space. So wherever we don't have coverage we don't expect the \textup{[model]} performance to be that great.''} --- P3
\end{quote}

To detect coverage issues, practitioners monitor online data to see how people are using a model, and compare that against their training data:

\begin{quote}
    \textit{``If it is a classification task, you were expecting to have a very balanced dataset, but online [you see] that almost 90\% of the traffic is coming for 1 class. That means your offline [data] was not representative of what is going to happen in an online setting. So, by monitoring and looking at all data distributions, you will get a sense of those discrepancies.''} --- P2
\end{quote}

Considering coverage allows practitioners to move beyond the kinds of failures that they already know of or suspect based on past experience and identify new failure modes that they were not previously aware of [P7].

\subsection{Creating Challenge Sets to Validate and Evaluate Issues}
\label{sec:interview-challenge-set}

When practitioners identify reported or suspected failures, they still need to determine whether this is a systematic problem with the model and, if so, assess the scope and severity of the error. Participants first wanted to understand if a potential failure is a one-off error (as can be expected given ML is probabilistic) or a more systematic problem [P4, P7, P12]. We found that practitioners shared a common general approach of:

\begin{quote}
    \textit{``Collecting more similar data and testing the model behavior and seeing if it's systemically failing.''} --- P7
\end{quote}

Practitioners in our sample referred to these curated data subsets as \textit{aggressor sets}, \textit{golden test sets}, and \textit{stress tests}. In the remainder of this paper, we refer to these kinds of datasets as \textbf{challenge sets}.

Challenge sets differ from standard test sets in ML because they are designed to target a specific failure case, and are thus often more reflective of how people really use a model in practice. As P6 described, \textit{``that kind of test while we call it stress test is probably closer to what happens in reality than when you do random sampling for testing.''}
Creating these sets can be challenging. In particular, it might not be immediately clear what kind of ``similar'' data will replicate a failure mode, and the axis of similarity that matters might not be annotated explicitly in the data.

\begin{quote}
    \textit{``The length of the beard seemed to play a role [in a failure mode]. It [the dataset] was just annotated as has beard or not, and not so much the length.''} --- P6
\end{quote}

Once practitioners have built a challenge set and determined that a suspected failure case is indeed a systematic problem with a model, they can then conduct quantitative and qualitative analyses of the challenge set to deeply understand the cause of the issue and how it might impact users. This understanding is critical to prioritizing issues to solve and informs the choice of solution.

\subsubsection{Assessing the Cause of the Problem}

Practitioners look for patterns in challenge sets to understand the potential cause of a problem. A first step is usually to compare the challenge set to the model's training data to identify coverage issues or other data problems, \eg spurious correlations. This analysis could be a simple process of \textit{``manually going through \textup{[the challenge set]} and looking for any general trends''} [P5], although models with larger output spaces or high dimensional data may require more sophisticated techniques like embedding space visualization and dimensionality reduction techniques [P7].

\subsubsection{Assessing the Impact of the Problem on Users}

Practitioners also want to assess the impact of the problem on users to judge its urgency. Model failures might be prioritized if they impact many users, happen frequently [P1], or if they produce a negative user experience [P7, P11, P12]. In this way, prioritization is \textit{``not just a pure data science question''} [P1], but involves considering different and possibly conflicting perspectives and values.

For example, in P7's work, \textit{``certain mispredictions could be more offensive than others,''} so when gathering feedback from quality annotators, they ask annotators to \textit{``exercise some judgment,''} and specifically flag anything they feel are \textit{``potentially offensive or egregious mistakes.''}

Practitioners might also prioritize improvements to ensure no subpopulation of users is experiencing particularly poor performance compared to others:

\begin{quote}
    \textit{``What we want to do is, reduce the length of the tail end of users that have poor experience and talk more about, how can we bring these people up and what is it about \textup{[their use context]} that causes the models to perform poorly.''} --- P11
\end{quote}

\subsubsection{Assessing Potential Solutions}

The choice of appropriate solution depends on understanding the scope and nature of the problem, and discussing these with reference to how the issues impact end users. Often, problems are the result of poor coverage and can be addressed by increasing training data in a specific area and retraining the model [P9]. For some participants, this was the default approach: \textit{``the answer is pretty much always going to be more representative data across all classes.''} [P5]. However, our findings highlight a wider range of approaches that practitioners can take when they deeply understand the nature and stakes of a problem.

For example, three participants talked about strategies to augment the model's output space. This could mean adding or removing classes from a classification taxonomy [P4], preventing specific outputs using a block list or an additional classification model, or hard-coding outputs for certain inputs using a lookup table [P4, P7, P12]. Other approaches included improving annotation quality [P7, P10], removing problematic data from the training set [P5], changing the user interaction with the model to control the input environment in production [P8], adjusting the model architecture or loss function [P12], or adding additional data pre-processing steps [P5].\looseness=-1

These approaches differ in complexity, cost, and effectiveness. The choice of solution is not solely based on technical and resource constraints, but could involve negotiating trade-offs, considering conflicting values, and accounting for the urgency of the error. For example, practitioners might select a fix that is faster to implement if an error impacts many users or is particularly offensive. Such decisions require input from stakeholders with a broad range of expertise. Therefore, ML practitioners must be able to discuss problems with reference to business metrics and user experience and in terms that are accessible to stakeholders without ML expertise [P1].\looseness=-1

\subsection{Design Implications}
\label{sec:design-implication}

Our findings demonstrate how challenge sets allow practitioners to develop a deep understanding of a problem's cause, scope, and impact on users. This understanding is necessary to effectively prioritize resources on the most egregious and urgent model failures. Existing tooling for model evaluation and debugging have largely focused on \textit{identifying} model weaknesses rather than \textit{prioritizing resources} on weaknesses with the greatest potential impact on users. Based on the practices uncovered through our interview study, we developed four design implications for tooling to support prioritization:

\aptLtoX[graphic=no,type=env]{
\begin{enumerate}%
    \item[\textbf{D1.}] \label{item:explore-suspected}
    Compare usage patterns to training data to support exploration of suspected model weaknesses in addition to known errors.
    \item[\textbf{D2.}]\label{item:similar-data}
    Build collections of similar data (challenge sets) to assess and prioritize problems, and allow users to compare challenge sets.
    \item[\textbf{D3.}] \label{item:different-info}
    Provide information about model performance and usage patterns to surface issues that matter most to users.
    \item[\textbf{D4.}] \label{item:no-right-answer}
    Since prioritization is not solely a technical question and does not have a singular solution, account for prioritization subjectivity, and make the tools easy to use for stakeholders with diverse backgrounds.
\end{enumerate}
}{
\begin{enumerate}[topsep=5pt, itemsep=0mm, parsep=1mm, leftmargin=18pt, label=\textbf{D\arabic*.}, ref=D\arabic*]
    \item\label{item:explore-suspected}
    Compare usage patterns to training data to support exploration of suspected model weaknesses in addition to known errors.
    \item\label{item:similar-data}
    Build collections of similar data (challenge sets) to assess and prioritize problems, and allow users to compare challenge sets.
    \item \label{item:different-info}
    Provide information about model performance and usage patterns to surface issues that matter most to users.
    \item \label{item:no-right-answer}
    Since prioritization is not solely a technical question and does not have a singular solution, account for prioritization subjectivity, and make the tools easy to use for stakeholders with diverse backgrounds.
\end{enumerate}
}

Interactive visualizations have successfully helped ML practitioners discover ML errors~(\autoref{sec:related:vis-error}) and understand model behaviors~(\autoref{sec:related:visualization}).
Interactive visualization techniques are especially useful for exploring data to support hypothesis generation and serendipitous discoveries~(\ref{item:explore-suspected}), comparing and contrasting slices of data~(\ref{item:similar-data}), analyzing data from multiple perspectives~(\ref{item:different-info}), and supporting collaborative interpretation of data among stakeholders with diverse skill sets~(\ref{item:no-right-answer}). For these reasons, visual analytics is a promising choice for supporting prioritization.

The remainder of this paper focuses on \tool{}~(\autoref{fig:teaser}), a visual analytics tool to support prioritization in the context of machine translation (MT). While our interview study revealed common practices across ML domains, prioritization depends on measures of prediction quality and insight into usage patterns, both of which are extremely specific to a particular model. Therefore, to understand these practices more deeply and begin to explore what tooling support for prioritization might look like, it is useful to choose a specific ML task as a case study. We chose MT because it poses unique challenges that make prioritization both especially important and especially difficult: it is difficult and expensive to attain reliable measures of prediction quality \cite{callison-burch-2009-fast}; MT models accept open-ended input from users, opening a vast space of possible failures; and we know relatively little about how people use MT models in the real world \cite{liebling2020unmet}.\looseness=-1 %
\section{Designing \tool{}: Exploring Machine Translation Usage with Challenge Sets}
\label{sec:ui}

Given the design implications~(\ref{item:explore-suspected}--\ref{item:no-right-answer}) described in \autoref{sec:design-implication}, we present \tool{}~(\autoref{fig:teaser}), an interactive visualization tool that helps MT developers prioritize model improvements by exploring and curating challenge sets.
\tool{} leverages both usage logs and training data to help users discover model weaknesses~(\ref{item:explore-suspected}, \autoref{sec:ui:creation}).
\tool{} introduces two novel techniques to automatically surface challenge sets and expand challenge sets with similar data~(\ref{item:similar-data}, \autoref{sec:ui:creation}).
\tool{} uses the \textit{overview + detail} design pattern~\cite{cockburnReviewOverviewDetail2009} to tightly integrate two major components: the \tableview{} that summarizes challenge sets as table rows~(\autoref{sec:ui:table}) and the \detailview{} that enables users to explore one challenge set in depth with different attributes over time~(\ref{item:different-info}, \autoref{sec:ui:detail}).
Finally, to lower the barrier for different stakeholders to easily prioritize model improvements~(\ref{item:no-right-answer}), Angler allows users to conduct quantitative and qualitative analyses without needing to write custom code and manipulate complex data and model pipelines.  We develop \tool{} with modern web technologies so that anyone can access it without installation overhead, and we open-source our implementation~(\autoref{sec:ui:implementation}).

We designed and developed \tool{} in conversation with an industry MT product team. To contextualize our design in the team's practices and gain iterative feedback, we used one of the team's MT models (English $\rightarrow$ Spanish), with a sample of their training data and usage logs. Usage logs are only available from users who have opted-in. For privacy and security reasons, members of our research team required special permissions and security protocols to access this data. Therefore, we cannot show \tool{} with the original model or dataset from our design process.
Moreover, to demonstrate how \tool{} can support many different MT models and language pairs, we instead describe the \tool{} interface in this section using a public MT model (English $\rightarrow$ Simplified Chinese) and public datasets~(\autoref{sec:ui:implementation}).

\begin{figure*}[tb]
  \includegraphics[width=\linewidth]{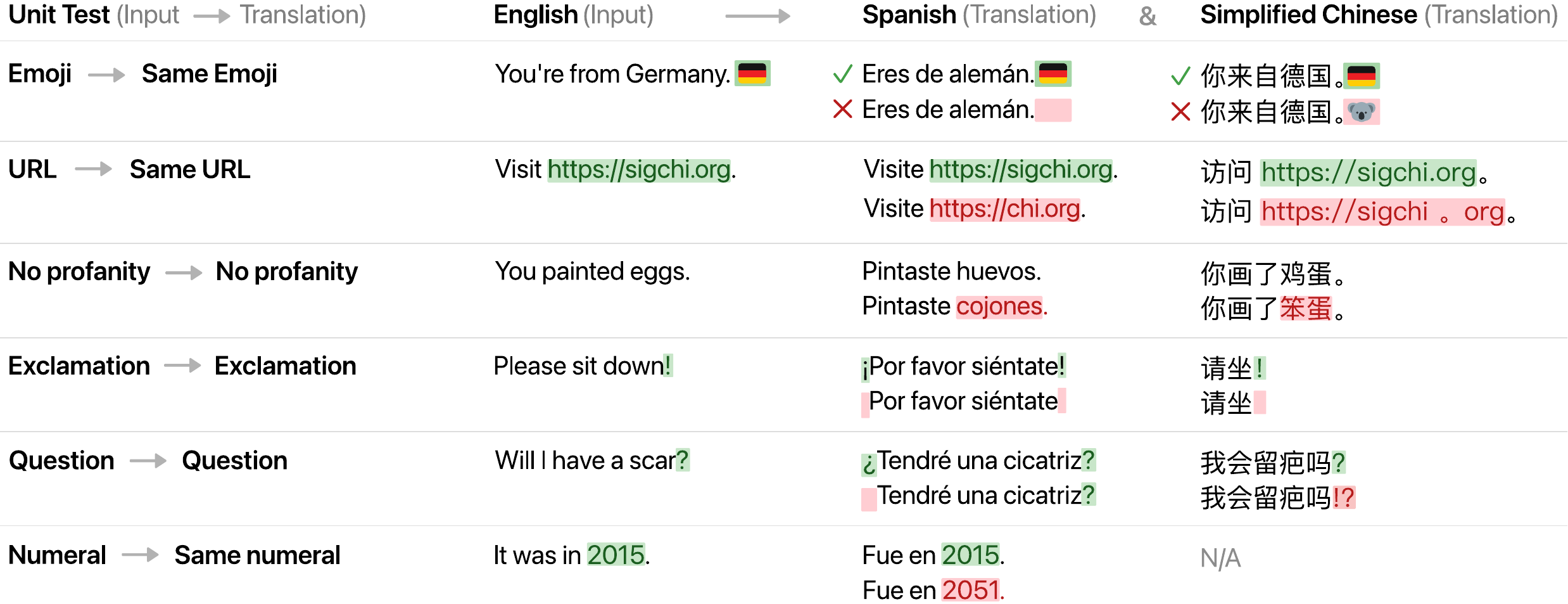}
  \caption[]{
  We create a suite of regex-based unit tests to detect translation errors without the need for ground-truth translation.
  For example, some tests check if the source and translation contain the same special words (\eg Emoji, URLs, and roman numerals).
  Some tests check punctuation (\eg question marks and exclamation points).
  Another test validates that a translation does not contain profane words if its source does not contain any profane words, by matching language-specific offensive, vulgar, slur word lists.
  In the translation columns, each row lists two examples that pass (top) and fail (bottom) the unit test.
  }
  \label{fig:unit-test}
\end{figure*}

\subsection{Subgroup Analysis through Challenge Sets}
\label{sec:ui:creation}

In \tool{}, we introduce two novel techniques to surface interesting subsets from the usage logs and training data.
We automatically extract \textit{challenge sets} by sampling data that either fails our model performance unit tests~(\autoref{sec:ui:unit-test}) or involves topics the model is less familiar with~(\autoref{sec:ui:topic}).

\subsubsection{Unit Test Failures}
\label{sec:ui:unit-test}
The current state-of-the-art approach to building challenge sets for machine translation is to build rule-based unit tests (\ref{sec:related:challenge-set}).
In line with this practice, the first type of challenge sets that we include in \tool{} extends the team's existing suite of unit tests to identify unexpected model behavior~(\autoref{fig:unit-test}).
These unit tests use regex search to find patterns in a source-translation pair and verify that each match meets some pre-defined rules.
For example, when a source includes an emoji, we expect the translation to have the same emoji.
Similarly, when a source does not contain offensive, vulgar, or slur (OVS) words, we expect the translation not to include OVS words either.
Some unit tests are language-specific: consider translating English to Spanish; if a source is a question, we expect the translation output to have both \texttt{``¿''} and \texttt{``?''} characters.
For simplified Chinese, however, we would expect the translation output to end with the ``？'' unicode instead.
For our English $\rightarrow$ Chinese demo in this section, we apply 5 unit tests to both usage logs and training data~(\ref{item:explore-suspected}), and collect data samples that fail any unit test into challenge sets~(\autoref{fig:unit-test}).
Each unit test corresponds to one challenge set.\looseness=-1

\subsubsection{Unfamiliar Topics}
\label{sec:ui:topic}

While unit tests can reveal some straightforward errors, they do not offer insight into issues of \textbf{coverage}, which our interview participants highlighted as critical to identifying failure modes that are highly consequential for end users~(\autoref{sec:interview:coverage}).
In the context of MT, coverage refers to how much a translation model ``knows'' from its training data about a particular topic or way of speaking. For example, coverage is a major concern with \textit{domain-specific language}: \eg doctors use domain-specific phrases to talk about medicine, while video game players use language specific to their game. Coverage can be improved by collecting more training data to give the model more exposure to that particular language pattern.

Extending existing techniques for building challenge sets in MT, we sought to help MT practitioners prioritize \textit{which} domains may need better coverage based on what their users are requesting. To identify topics that are not well represented in the training data, we first use a sentence transformer~\cite{reimersSentenceBERTSentenceEmbeddings2019} to extract high-dimensional latent representations of sentences in the training data.
This latent representation is trained to cluster sentences with similar meaning close together in high-dimensional space.
We then apply UMAP~\cite{mcinnesUMAPUniformManifold2018}, a dimensionality reduction method, to project the latent representation into 2D.
We choose to use UMAP instead of other dimensionality reduction techniques, such as PCA~\cite{pearsonLinesPlanesClosest1901} and t-SNE~\cite{vandermaatenVisualizingDataUsing2008}, because UMAP is faster and has better preservation of the data's global structure~\cite{mcinnesUMAPUniformManifold2018}.
We use the cosine similarity to measure the distance between two samples in the high-dimensional space, as previous works have shown that cosine distance provides better and more consistent results than the Euclidean distance~\cite{vermeulenApplicationUniformManifold2021}.
Following the suggested practice in applying UMAP~\cite{coenenUnderstandingUMAP2019}, we fine-tune UMAP's hyperparameters \texttt{n\_neighbors} and \texttt{min\_dist} through a grid search of about 200 parameter combinations; we choose hyperparameters that spread out the training samples in the 2D space while maintaining local clusters.
With about 47,000 training sentences and a latent representation size of 768, it takes about 50 seconds on average to fit a UMAP model with one parameter combination on a MacBook Pro laptop.

We use Kernel Density Estimation (KDE)~\cite{scottMultivariateDensityEstimation2015} to estimate the \textcolor[HTML]{FA8231}{training data's distribution}. For the KDE, we choose a standard \textcolor{equationblue}{multivariate Gaussian kernel with a Silverman bandwidth $\mathbf{H}$}~\cite{silvermanDensityEstimationStatistics2018}.
It only takes about 1 second to fit a KDE model on \textcolor[HTML]{FA8231}{47,000 training sentences' 2D representations}.
Then, we use this trained KDE model to compute the \textit{familiarity score} (FA)~\cite{hopkins2023designing} for \textcolor[HTML]{009688}{each sentence from the usage logs}.
We define the familiarity score~(\autoref{eq:kde}) of a sentence from the usage log as the log-likelihood of observing \textcolor{equationgreen}{that sentence's UMAP 2D coordinate $\left( x, y \right)$} under the \textcolor{germanorange}{training data's UMAP distribution $\left[ \left(x_1, y_1 \right), \left(x_2, y_2 \right), \dots, \left(x_i, y_i \right) \right]$ }.
This concept of familiarity can be generalized to other data types and ML domains, and has shown to be a powerful tool for debugging data~\cite{hopkins2023designing}.\looseness=-1

\begin{equation}
  \label{eq:kde}
  \text{FA}\textcolor[HTML]{009688}{\left({x, y}\right)} = \log \left( \textcolor[HTML]{FA8231}{\frac{1}{n} \sum_{i = 1}^{n}} \textcolor[HTML]{42A5F5}{\frac{\exp \left( -\frac{1}{2} \begin{bmatrix}  \textcolor[HTML]{009688}{x} - \textcolor[HTML]{FA8231}{x_i} & \textcolor[HTML]{009688}{y} - \textcolor[HTML]{FA8231}{y_i} \end{bmatrix} \mathbf{H}^{-1} \begin{bmatrix} \textcolor[HTML]{009688}{x} - \textcolor[HTML]{FA8231}{x_i} \\ \textcolor[HTML]{009688}{y} - \textcolor[HTML]{FA8231}{y_i} \end{bmatrix}  \right)}{2\pi \sqrt{\mid \mathbf{H} \mid}}}  \right)
\end{equation}

Computing FA is slow when the \textcolor[HTML]{FA8231}{training data} is large (\ie \textcolor{orange}{$n$} is large in \autoref{eq:kde}), because the algorithm needs to iterate through all \textcolor[HTML]{FA8231}{$n$ points in the training data} for each \textcolor[HTML]{009688}{sentence from the usage logs}.
Therefore, to accelerate FA computation, we apply a 2D binning approximation approach.
We first pre-compute the log-likelihoods over a 2D grid of \textcolor[HTML]{FA8231}{training data's} UMAP 2D space $\mathbf{F}\in \mathbb{R}^{200 \times 200}$, constrained by the range of the \textcolor[HTML]{FA8231}{training data's} UMAP coordinates.
Then, to approximate the FA of \textcolor[HTML]{009688}{a sentence}, we only need to (1) locate the cell $\mathbf{F}_{i, j}$ in the grid that \textcolor[HTML]{009688}{the sentence} falls into, and (2) look up the pre-computed log-likelihood associated with that cell $\mathbf{F}_{i, j}$.
If \textcolor[HTML]{009688}{a sentence} falls out of the 2D grid, we extrapolate its FA by using the log-likelihood associated with the closest grid cell.
Note that one can choose a different grid density other than $200\times200$; we tune the grid density ($d = 200$) to balance the computation time and the approximation accuracy.
Our binning approximation is scalable to large \textcolor[HTML]{009688}{usage logs ($m$)} and \textcolor[HTML]{FA8231}{training data ($n$)}, as it decreases the FA computation's time complexity from a quadratic time $\mathcal{O}(k \textcolor[HTML]{009688}{m} \textcolor[HTML]{FA8231}{n})$ to a linear time $\mathcal{O}(\textcolor[HTML]{009688}{m} + d^2 k \textcolor[HTML]{FA8231}{n})$, where $k$ is the dimension of the UMAP space ($k = 2$ in our case), and $d$ is the grid density ($d = 200$).
In addition, we use the KDE implementation from \textit{Scikit-Learn}~\cite{pedregosaScikitlearnMachineLearning2011}, which leverages KD Tree~\cite{bentleyMultidimensionalBinarySearch1975} for more efficient distance computation.
With a tree-based KDE, our FA computation method has a logarithmic time complexity $\mathcal{O}(\textcolor[HTML]{009688}{m} + d^2k \textcolor[HTML]{FA8231}{\log{\left(n\right)}})$ on average and a linear time complexity $\mathcal{O}(\textcolor[HTML]{009688}{m} + d^2k \textcolor[HTML]{FA8231}{n})$ in the worst case.

After estimating the FAs for all sentences in the \textcolor[HTML]{009688}{usage logs}, we use BERTopic~\cite{grootendorstBERTopicNeuralTopic2022} to build a topic model on a sample of 50,000 sentences from the \textcolor[HTML]{009688}{usage logs} with the lowest FA and select the 100 largest topics from this model.
To estimate the model's performance on these topics, we need labeled \textcolor[HTML]{FA8231}{training data}.
Therefore, we extend each extracted topic set with a sample of training sentences that are close to the topic set in the high-dimensional space~(\ref{item:similar-data}).
To reduce the computational cost of this search, we randomly sample 15 \textcolor[HTML]{009688}{``seed sentences''} from each topic and add any sentences from the \textcolor[HTML]{FA8231}{training data} that are close to at least one of the \textcolor[HTML]{009688}{``seed sentences''} in the high-dimensional space (threshold $\ell_2 < 0.6$ selected through manual inspection).
We have tuned the number of \textcolor[HTML]{009688}{``seed sentences''} to balance the computational cost and the number of \textcolor[HTML]{FA8231}{close training sentences} that we can find.
Finally, we have controlled random seeds for random sampling, UMAP computation, and BERTopic, so that our topic results are reproducible.

\subsubsection{Limitations}

Identifying new model failure modes and collecting examples to replicate the failure is extremely challenging. Developing automatic, expert-driven, and crowd-sourcing methods for identifying failures is an active area of research in machine learning and human-computer interaction  \cite{zhang2022sliceteller, chung2020automated, devos2022toward, shen2021everyday, wexlerWhatIfToolInteractive2019, cabreraFAIRVISVisualAnalytics2019, munechikaVisualAuditorInteractive2022}. Compared to prior research, it is especially difficult to automatically identify MT model failures because there are no explicit, interpretable features or metadata on which to slice data into subgroups, and automatic evaluation metrics are very noisy.
Further, prior work largely focuses on identifying failure modes by comparing predictions to ground-truth labels ~\cite[e.g.][]{zhang2022sliceteller, chung2020automated}, which does not give practitioners insight into failure modes that impact end-users but are not yet represented in offline, labeled datasets.

Our goal in this work is to understand how practitioners prioritize their resources across many potential failure modes, and what information they need to do so. We generate example challenge sets to guide this exploration using pattern-matching rules (the current state-of-the-art in MT) and topic modeling on areas of low coverage. However, further research is needed to evaluate and extend these methods. While we did not conduct a formal evaluation of our challenge sets in this work, both kinds of sets are certainly imperfect in terms of error identification -- there are perfect translations included in the challenge sets, and there are translation errors in the larger data that are not included in any challenge set. Given the large space of possible inputs, and probabilistic nature of machine learning, we cannot expect to ever have methods to identify all possible failures with perfect accuracy. Thus, there is a need for interactive visualization tools that support practitioners to explore and make sense of \textit{potential} failure modes and prioritize development and annotation resources under uncertainty.

\begin{figure*}[h!tb]
  \includegraphics[width=\linewidth]{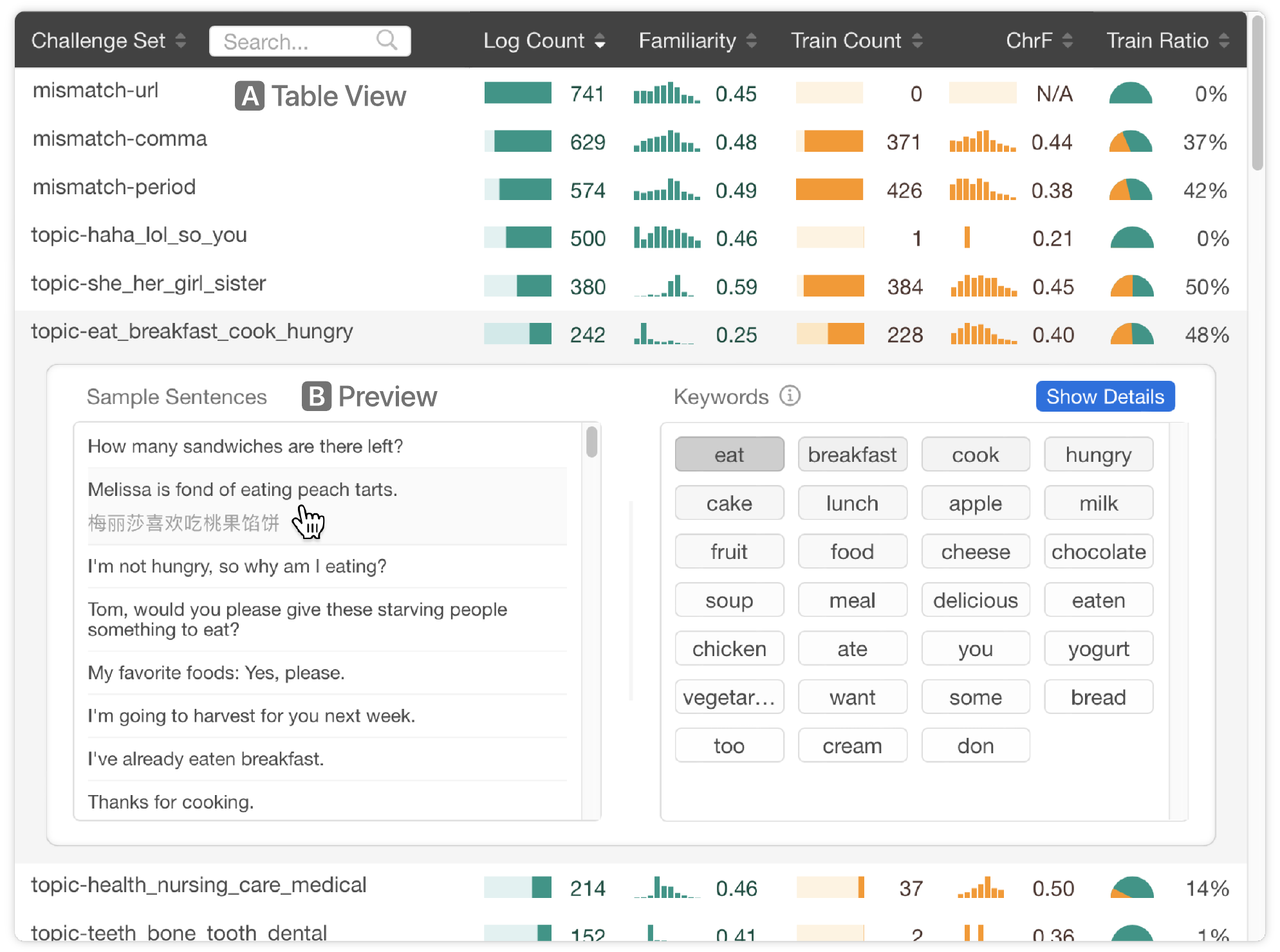}
  \Description{Screenshot of the table view}
  \caption[]{
  \textbf{(A) \textit{The Table View}} summarizes all challenge sets in a table, where users can compare the model's performance and familiarity across these sets.
  To help users interpret aggregate metrics, the \tableview{} visualizes the distributions of metrics and sets' compositions as sparkline-like charts.
  \textbf{(B) \textit{The Preview}} presents more details of a challenge set after a user clicks a row.
  The \textit{Sample Sentences} (left) lists 100 randomly selected source sentences from this set with translations.
  The \textit{Keywords} (right) visualizes the most representative words in this set, where a darker background indicates higher representativeness.
  }
  \Description{
    A screenshot of Angler's Table View.
    The Table View contains a sortable table with column headers: Challenge Set, Log Count, Familiarity, Train Count, ChrF, and Train Ratio. Each row in the table represents a challenge set.
    A challenge set named ''topic-eat_breakfast_cook_hungry'' is extended, where sample sentences and representative keywords are shown below that row.
    Each sample sentence is an English input.
    A clicked sentence has its Chinese translation below the input.
  }
  \label{fig:table}
\end{figure*}

\subsection{Table View}
\label{sec:ui:table}

\setlength{\columnsep}{13pt}%
\setlength{\intextsep}{0pt}%
\begin{wrapfigure}{R}{0.3\textwidth}
  \vspace{0pt}
  \centering
  \includegraphics[width=0.3\textwidth]{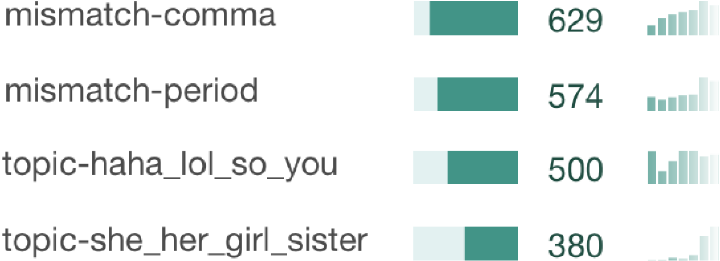}
  \vspace{-15pt}
  \caption[]{
    \tool{} distinguishes challenge sets created by unit test failures and unfamiliar topics by their names.
  }
  \Description{
    A screenshot of the name of the challenge sets and their log counts in the table view.
    From top to bottom, the names are ``mismatch-comma'', ``mismatch-period'', ``topic-haha_lol_so_you'' and ``topic-she_her_girl_sister''.
    Their log counts are 629, 574, 500, and 380.
  }
  \vspace{0pt}
  \label{fig:set-naming}
\end{wrapfigure}
When users launch \tool{}, they first see the \tableview{} listing all pre-computed challenge sets in a table~(\autoref{fig:table}\figpart{A}).
Each challenge set can contain samples from the \textcolor[HTML]{FA8231}{training data} and \textcolor[HTML]{009688}{usage logs}, color coded as \textcolor[HTML]{FA8231}{orange} and \textcolor[HTML]{009688}{green} respectively throughout \tool{}.
We name challenge sets based on their construction methods~(\autoref{fig:set-naming}).
For challenge sets created by unit test failures, we name them ``mismatch-[\textit{unit test name}].''
For challenge sets created by unfamiliar topics, we name them ``topic-[\textit{top-4 keywords}].''
These keywords are the same as keywords shown in the \preview{}~(\autoref{fig:table}\figpart{-B}).
In addition to the names of challenge sets, the \tableview{} view provides five metrics associated with each set:

\begin{itemize}
  \item \textcolor[HTML]{FA8231}{\textit{Train Count}} and \textcolor[HTML]{009688}{\textit{Log Count}}: the number of \textcolor[HTML]{FA8231}{training} and \textcolor[HTML]{009688}{usage log} samples in the set.
  \item \textcolor[HTML]{FA8231}{\textit{ChrF}}: a measure of the model's performance on the \textcolor[HTML]{FA8231}{training samples} in the set. ChrF is the F-score based on character \textit{n}-gram overlap between the hypothesis translation produced by the model and a validated reference translation~\cite{popovicChrFCharacterNgram2015}. We use the open-source SacreBLEU implementation of ChrF~\cite{postCallClarityReporting2018}.
  \item \textcolor[HTML]{009688}{\textit{Familiarity}}: a measure of how familiar the \textcolor[HTML]{009688}{usage log samples} in the set are to the model, by reference to the training data distribution~(\autoref{sec:ui:topic}).
  \item \textcolor[HTML]{FA8231}{\textit{Train Ratio}}: the percentage of samples in the set that are \textcolor[HTML]{FA8231}{training samples}.
\end{itemize}

Users can sort challenge sets by any of these metrics by clicking the sorting button~\vcenteredhbox{\includegraphics[height=10pt]{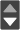}} in the table header.
To help users quickly compare these metrics across challenge sets, the \tableview{} also provides sparkline-like visualizations~\cite{tufteVisualDisplayQuantitative2013} in each row.
For each challenge set, the \tableview{} visualizes its sample counts as in-line bar charts, \textcolor[HTML]{FA8231}{ChrF} and \textcolor[HTML]{009688}{Familiarity} distributions as histograms, and the \textcolor[HTML]{FA8231}{training sample ratio} as a semi-circle pie chart.

After identifying an interesting challenge set, users can click the row to open a \preview{}~(\autoref{fig:table}\figpart{B}) in the table to see a preview of that set.
This view provides users with a quick summary of the set on demand.
On the left, users can browse 100 randomly sampled sentences from this challenge set; users can also click on each sentence to see the model's output translation.
The number of sentences in each challenge set varies from about 100 to 1000; challenge sets constructed from unit tests tend to have more sentences than ones constructed from unfamiliar topics.
We choose the number 100 because it gives a fair coverage of all sentences in the set and users can have a smooth experience in quickly browsing sentences from different sets.
If they are interested in one particular challenge set, they can view all sentences in that set's \detailview{}~(\autoref{sec:ui:detail}).
On the right, users can inspect the most representative keywords from this set.
Keywords are extracted and sorted by their class-based TF-IDF scores~\cite{grootendorstBERTopicNeuralTopic2022}.
Intuitively, these keywords are words that appear more frequently in this set than in all other sets.
In \tool{}, we list all keywords returned from BERTopic; future researchers and developers can determine a class-based TF-IDF score threshold to only display more frequent keywords (keywords with a darker background).\looseness=-1

\begin{figure*}[tb]
  \includegraphics[width=\linewidth]{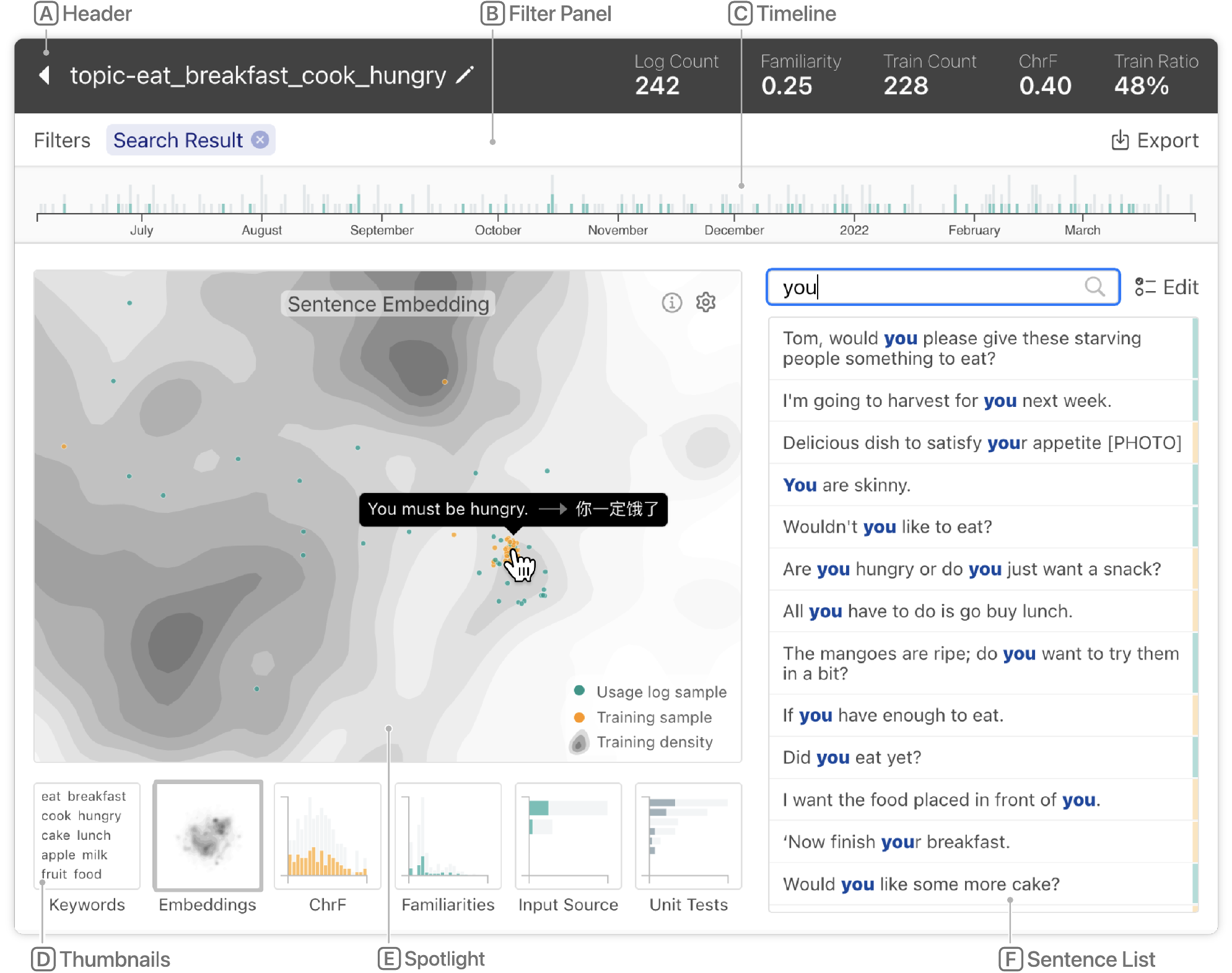}
  \caption[]{
  After a user selects a challenge set from the \tableview{}, \tool{} presents the \detailview{} to help the user further analyze this set from diverse perspectives.
  \textbf{(A) The Header} shows the name and statistics associated with this challenge set.
  \textbf{(B) The Filter Panel} helps users keep track of the currently active filters.
  \textbf{(C) The Timeline} visualizes the usage log count over time, allowing users to focus on traffic data from a particular time window.
  \textbf{(D) The Thumbnails} and \textbf{(E) Spotlight} visualize diverse representations of sentences in this set---users can click different \textit{Thumbnails} to switch the \textit{Spotlight}, on which users can further filter sentences with particular attributes.
  \textbf{(F) The Sentence List} displays all sentences that meet the active filters, where users can inspect translations, search words, and remove sentences from this set.
  }
  \Description{
    A screenshot of Angler's Detail View.
    At the top of the Detail View, there is a header describing the Name, Log count, Familiarity, Train Count, ChrF, and Train Ratio of the challenge set ``topic-eat_breakfast_cook_hungry''.
    Below the header, there is a filter bar with the filter tag ``Search Result''.
    Below the filter bar, there is a timeline histogram from June to April.
    There are two components below the timeline.
    On the left, there is a large embedding view visualizing the embedding spaces of the training data and usage logs.
    The cursor is hovering over a dot on the embedding view with a tooltip writing ``You must be hungry. → 你一定饿了。''
    Below the embedding view, there are six thumbnails labeled as ``Keywords'', ``Embeddings'', ``ChrF'', ``Familiarities'', ``Input Source'', and ``Unit Tests''.
    On the right, there is a list of sentences with a search bar and an edit button.
    The search bar is filled with a search query ``you.''
    Each sentence is an English input with the word ``you'' highlighted in blue.
  }
  \label{fig:detail}
\end{figure*}

\subsection{Detail View}
\label{sec:ui:detail}

\begin{figure*}[tb]
  \includegraphics[width=\linewidth]{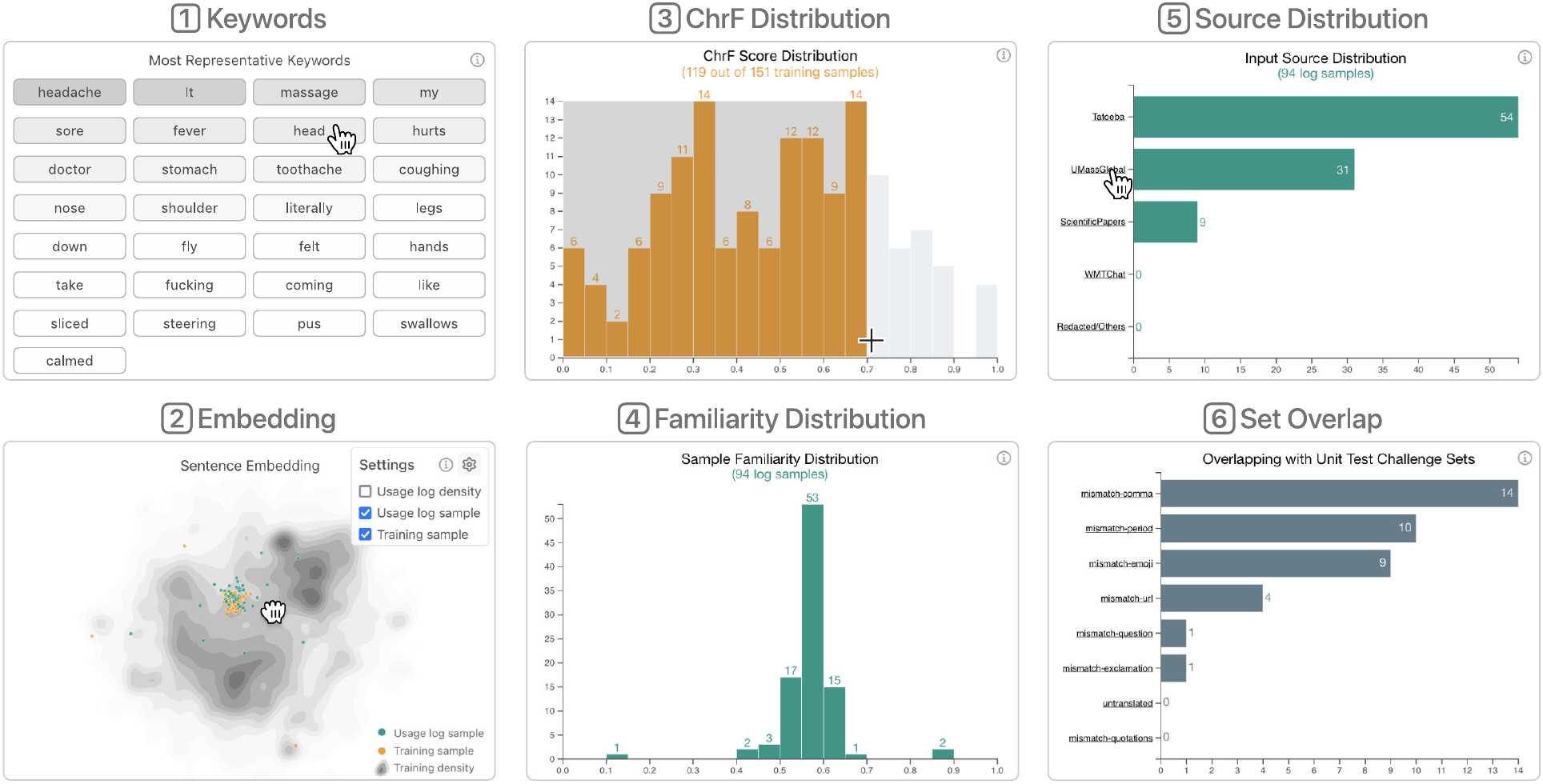}
  \caption[]{
  The \detailview{} presents six options for the \spotlight{} to help users explore a challenge set from diverse perspectives.
  \textbf{(1) The \textit{Keywords}} shows the most representative words in a set.
  \textbf{(2) The \textit{Embedding}} uses a zoomable scatter plot with contour backgrounds to help users explore the high-dimensional representations of sentences in a set.
  \textbf{(3) The \textit{ChrF Distribution}} allows users to inspect and filter training sentences by their ChrF scores.
  \textbf{(4) The \textit{Familiarity}} helps users filter usage logs by the models' familiarity scores.
  \textbf{(5) The \textit{Source Distribution}} visualizes the usage log source as a horizontal histogram where users can filter usage logs from particular sources.
  \textbf{(6) The \textit{Set Overlap}} allows users to see sentences that are also in other challenge sets.
  }
  \label{fig:spotlight}
  \Description{
    Six screenshots of different Spotlight View variations.
    (1) Keywords Spotlight is a grid of the most representative keywords.
    (2) Embedding Spotlight shows the embedding of all training data as well as colored dots representing training and usage log samples from this challenge set.
    (3) ChrF Distribution Spotlight is an orange bar chart where the bars in the range from 0 to 0.7 are selected by brushing.
    (4) Familiarity Distribution Spotlight is a green bar chart.
    (5) Source Distribution Spotlight is a green horizontal bar chart.
    (6) Set Overlap Spotlight is a gray horizontal bar char.
  }
\end{figure*}

To help users further analyze individual challenge sets~(\ref{item:different-info}), \tool{} presents the \detailview{}~(\autoref{fig:detail}) when a user clicks the \vcenteredhbox{\includegraphics[height=9pt]{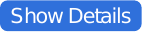}} button under a challenge set in the \tableview{}~(\autoref{fig:table}).
In the header of the \detailview{}~(\autoref{fig:detail}\figpart{A}), users can inspect the metrics associated with this challenge set and edit the set's name.
To explore sentences in this set through different perspectives, users can use the \timeline{}~(\autoref{fig:detail}\figpart{C}) and \spotlight{}~(\autoref{fig:detail}\figpart{E}) to filter sentences by different attributes.
The \filter{}~(\autoref{fig:detail}\figpart{B}) displays the currently applied filters, and the \samplelist{}~(\autoref{fig:detail}\figpart{F}) only shows sentences that satisfy these filters.
There are six visualization variations of the \spotlight{}~(\autoref{fig:spotlight}).
Users can switch between them to fit their exploration needs~(\ref{item:no-right-answer}) by clicking the corresponding \thumbnails{}~(\autoref{fig:detail}\figpart{D}).
Each \thumbnail{} is a simplified version of a \spotlight{} variation, where the visualization also updates in real-time when users add or remove filters.\looseness=-1

\subsubsection{Timeline}
\label{sec:ui:timeline}
To help users investigate how \textcolor[HTML]{009688}{usage logs} change over time, the \detailview{} provides a \timeline{}~(\autoref{fig:detail}\figpart{C}) panel on top of the window.
The \timeline{} visualizes the number of \textcolor[HTML]{009688}{user requests} in this set over time as a histogram, where the x-axis represents the time and the y-axis represents the \textcolor[HTML]{009688}{request count}.
Users can zoom and pan to inspect different periods.
Users can also brush the histogram to filter \textcolor[HTML]{009688}{usage logs} that are from a particular time window.\looseness=-1

\subsubsection{Keyword Spotlight}
\label{sec:ui:keyword}
Similar to the \textit{Keyword} panel in the \preview{}~(\autoref{sec:ui:table}), the \textit{Keyword Spotlight}~(\autoref{fig:spotlight}\figpart{-1}) displays the most representative words in a challenge set.
It sorts keywords by their representativeness, which is measured by the class-based TF-IDF scores~\cite{grootendorstBERTopicNeuralTopic2022}.
This view uses the darkness of the background color to encode a word's representativeness.
Users can click keywords to filter sentences that contain selected keywords.

\subsubsection{Embedding Spotlight}
\label{sec:ui:embedding}
To help users explore the semantic similarity of sentences in a challenge set, the \textit{Embedding Spotlight}~(\autoref{fig:spotlight}\figpart{-2}) visualizes a 2D projection of the sentences' high-dimensional representations~(\autoref{sec:ui:topic}) in a scatter plot.
Each dot in the scatter plot represents a sentence, and it is positioned by its UMAP coordinates.
Furthermore, we visualize the KDE density distributions~(\autoref{sec:ui:topic}) of all \textcolor[HTML]{FA8231}{training data} and all \textcolor[HTML]{009688}{usage logs} as contour plots.
Augmenting the scatter plot with density distributions of overall \textcolor[HTML]{FA8231}{training data} and \textcolor[HTML]{009688}{usage logs} allows users to discover use cases that are not well supported by existing \textcolor[HTML]{FA8231}{training data}~(\ref{item:explore-suspected}).

\subsubsection{ChrF Spotlight}
\label{sec:ui:chrf}
The \textit{ChrF Spotlight}~(\autoref{fig:spotlight}\figpart{-3}) visualizes the model's \textcolor[HTML]{FA8231}{ChrF score distribution} on the \textcolor[HTML]{FA8231}{training data} in this set as a histogram, allowing users to gain more insights regarding the model's performance on a particular set.
The x-axis encodes the \textcolor[HTML]{FA8231}{ChrF scores}, and the y-axis encodes the distribution frequency of \textcolor[HTML]{FA8231}{training data} in the set.
In addition, users can brush to select bins in the histogram, which would filter sentences with a \textcolor[HTML]{FA8231}{ChrF score} in the specified range.

\subsubsection{Familiarity Spotlight}
\label{sec:ui:familiarity}
The \textit{Familiarity Spotlight}~(\autoref{fig:spotlight}\figpart{-4}) is similar to the \textit{ChrF Spotlight}.
However, the x-axis here represents the model's \textcolor[HTML]{009688}{familiarity scores} on \textcolor[HTML]{009688}{usage logs} in the set.
The familiarity score is determined by the log-likelihood of observing \textcolor[HTML]{009688}{a user request} under the distribution of all \textcolor[HTML]{FA8231}{training data}~(\autoref{sec:ui:topic}).
Users can brush the histogram to filter sentences with particular \textcolor[HTML]{009688}{familiarity scores}.

\subsubsection{Source Spotlight}
\label{sec:ui:source}
To allow users to compare \textcolor[HTML]{009688}{usage logs} by the sources of these \textcolor[HTML]{009688}{user requests}, the \textit{Source Spotlight}~(\autoref{fig:spotlight}\figpart{-5}) visualizes \textcolor[HTML]{009688}{usage log} count as a horizontal bar chart.
The x-axis encodes the count of \textcolor[HTML]{009688}{usage logs} from one particular source, and the y-axis encodes the source category.
To focus on logs from particular sources, users can click the source names to create filters.
In our open-source demo, the \textit{Source Spotlight} shows the source dataset from which each sentence was sampled.
In the version of the tool developed for the MT team, the \textit{Source Spotlight} shows the source application from which a request was made (available for a sample of \textcolor[HTML]{009688}{usage logs}).\looseness=-1

\subsubsection{Overlap Spotlight}
\label{sec:ui:overlap}
The design of the \textit{Overlap Spotlight}~(\autoref{fig:spotlight}\figpart{-6}) is similar to \textit{Source Spotlight}.
Instead of encoding the source category, the y-axis here represents other challenge sets.
For example, for a challenge set created by unit test failures, the y-axis in its \textit{Overlap Spotlight} represents other challenge sets created by unfamiliar topics.
As unfamiliar topics are strictly non-overlapping, this view only shows overlap with challenge sets of the other type.
By cross-referencing two challenge set types (unit test failures and unfamiliar topics), this visualization can help users explore syntactic errors within semantic topics, and vice versa.

\subsubsection{\samplelist{}}
\label{sec:ui:list}
The \samplelist{}~(\autoref{fig:detail}\figpart{F}) shows all sentences in the challenge set that satisfy the currently applied filters.
Users can click a sentence to see the model's translation.
To further fine-tune a challenge set, users can click the \vcenteredhbox{\includegraphics[height=9pt]{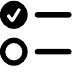}}~Edit button and remove unhelpful sentences from the set.
Finally, users can click the \vcenteredhbox{\includegraphics[height=9pt]{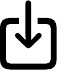}}~Export button to export sentences shown in the list along with their translations and attributes; users can then easily share these sentences with colleagues and human annotators~(\ref{item:no-right-answer}).

\subsection{Open-source Implementation}
\label{sec:ui:implementation}

\tool{} is an open-source interactive visualization system built with \textit{D3.js}~\cite{bostockDataDrivenDocuments2011}: users with diverse backgrounds~(\ref{item:no-right-answer}) can easily access our tool directly in their web browsers without installing any software or writing code.
We use the standard NLP suite for data processing (\eg \textit{NLTK}~\cite{loperNLTKNaturalLanguage2002}, \textit{Scikit-learn}~\cite{pedregosaScikitlearnMachineLearning2011}) and topic modeling (\eg \textit{BERTopic}~\cite{grootendorstBERTopicNeuralTopic2022}, \textit{UMAP}~\cite{mcinnesUMAPUniformManifold2018}).
We first implemented \tool{} with an industry-scale MT model (English $\rightarrow$ Spanish) and real \textcolor[HTML]{FA8231}{training data} and \textcolor[HTML]{009688}{usage logs}.
For demonstrations in this paper, we use the public MT model \textit{OPUS-MT} (English $\rightarrow$ Simplified Chinese)~\cite{tiedemannOPUSMTBuildingOpen2020} and its \textcolor[HTML]{FA8231}{training data}~\cite{tiedemann-2020-tatoeba}. To simulate \textcolor[HTML]{009688}{usage logs} we augment a sample of the model's test data~\cite{tiedemann-2020-tatoeba} with publicly available sources to emulate realistic use cases that can be difficult for MT models: social media \cite{blodgett2017dataset}, conversation \cite{wmtchat}, and scientific articles~\cite{data-scientific}.

\subsection{Usage Scenario}
\label{sec:ui:scenario}
We present a hypothetical usage scenario to illustrate how \tool{} can help Priya, an MT engineer, explore \textcolor[HTML]{009688}{usage logs} and guide new training data acquisition.
The first part of this usage scenario, in which the user explores and selects challenge sets of interest, is informed by real user interactions that we observed in the user study~(\autoref{sec:study}).
The second phase of the scenario describes how we envision extending \tool{} in future work to help practitioners use the datasets they collect with \tool{} to improve model performance.

Priya works on improving an English–Chinese translation model, and she only speaks English.
Priya first applies the challenge set extraction pipeline~(\autoref{sec:ui:creation}) to the \textcolor[HTML]{FA8231}{training data} and \textcolor[HTML]{009688}{usage logs} from the past 6 months.
The pipeline yields 100 challenge sets from unfamiliar topics and 6 challenge sets from unit test failures.
After Priya opens \tool{} to load extracted challenge sets in a web browser, she sees the \tableview{}~(\autoref{fig:teaser}\figpart{A}) summarize all 106 sets in a table with a variety of statistics.
Priya wants to prioritize subsets of data where the model may not perform well, but which are important to the end-users of the model, e.g., because they occur frequently in the \textcolor[HTML]{009688}{usage logs}, or represent a high-stakes use case.
To focus on data on which the current MT model may not perform well, Priya sorts challenge sets in ascending order by their mean \textcolor[HTML]{FA8231}{ChrF scores} by clicking the sort button.
After inspecting the top rows and their \previews{}, the \texttt{topic-headache} set draws Priya's attention---the MT model performs poorly on this set (mean \textcolor[HTML]{FA8231}{ChrF score} is only 0.39), and this set involves high-stakes medical topics where the MT quality is critical~(observed from the \textit{Keywords} in the \preview{}).

To learn more about this challenge set, Priya clicks the \vcenteredhbox{\includegraphics[height=9pt]{figures/icon-detail}} button to open the \detailview{}~(\autoref{fig:teaser}\figpart{B}).
Priya notices that the number of usage logs is consistent across the past nine months (from July 2021 to March 2022) in the \timeline{}~(\autoref{fig:teaser}\figpart{-B1}).
She then clicks the \textit{Embedding} \thumbnail{}~(\autoref{fig:detail}\figpart{D}) to switch the \spotlight{} from the default \textit{Keywords} view~(\autoref{fig:spotlight}\figpart{-1}) to the \textit{Embedding} view~(\autoref{fig:spotlight}\figpart{-2}).
Through zooming and hovering over the scatter plot, Priya finds that most sentences from this set form a cluster in the high-dimensional representation space, and all these sentences are about health issues.
She is surprised to see that people are using the model to communicate about health concerns, and wonders whether the training data covers this use case.
To explore this, Priya opens the \textit{Familiarity Distribution} \spotlight{}~(\autoref{fig:spotlight}\figpart{-4}) and brushes the histogram to select the region with low \textcolor[HTML]{009688}{familiarity scores}.
The \timeline{}, charts in \thumbnails{}, and the \samplelist{} update in-real time to focus on the \textcolor[HTML]{009688}{usage logs} with a \textcolor[HTML]{009688}{familiarity score} in the selected range.
Browsing the sentences in the \samplelist{}, Priya realizes that many of these unfamiliar sentences are about fever.
She worries that wrong translations about fevers could pose a health risk to users.
Therefore, Priya decides to prioritize improving her model's performance on this challenge set; she clicks the \vcenteredhbox{\includegraphics[height=9pt]{figures/icon-export}}~Export button to save all sentences along with their translations from this challenge set.\looseness=-1

The current \tool{} prototype was designed to explore what information practitioners need to prioritize subsets of data to send to annotators.
Priya follows a similar process to identify and export a few other challenge sets of high importance and sends all of this data to human annotators to acquire additional training data.
Human annotators can speak both English and Chinese.
They write reference Chinese translations for given English sentences by directly editing the translations produced by the model.
In the future, \tool{} could be extended to allow Priya to continue her analysis after the data has been annotated.

For example, Priya could check a few sentences from the health-related challenge set whose reference translations are significantly different from their translations produced by Priya's model.
At this point, Priya might find that her model has made several serious translation errors.
For example, her model translates the input sentence ``The fever burns you out'' to ``发烧会烧死你'' in Chinese~(\autoref{fig:teaser}\figpart{-B3}), which means ``fever will burn you to death.''
After retraining the MT model with the newly annotated data, Priya would hope to see that the model's \textcolor[HTML]{FA8231}{ChrF scores} and \textcolor[HTML]{009688}{familiarity scores} on the original challenge sets have significantly improved.
In this case, Priya would schedule to deploy her new MT model in the next software release cycle, now with better support for a safety-critical use case.

\section{User Study}
\label{sec:study}

We used \tool{} in a user study with seven people who contribute to machine translation development at Apple as ML engineers (E1--3), and in user experience-focused roles, such as product management, design, or analytics (UX1--4). Our goal in this study was to understand how users with different expertise would use \tool{} to explore and prioritize challenge sets. We were also interested in whether exploring challenge sets using \tool{} could help practitioners to uncover new insights about their models, and identify new ways to improve their models in line with users' needs. Our goal was not to measure whether \tool{} can support prioritization more effectively than another tool, e.g., by finding more translation errors, but rather to explore what kind of information is useful to practitioners, and how the process of exploring challenge sets could shape future evaluation practices.

The study was approved by our internal IRB. Each session was conducted over video chat and lasted between 45 minutes and one hour. With participants' consent, each session was recorded and transcribed for analysis. During each study session, we introduced \tool{} with a tutorial demonstrating each view (\autoref{fig:table}--\autoref{fig:spotlight}). \tool{} showed training and usage data from the team's own translation model, for the language pair English $\rightarrow$ Spanish. Next, we sent the participant a one-time secure link that allowed them to access the tool in their own browser, and asked them to share their screen while they explored the tool. For the remainder of the session, we asked the participant to think aloud as they completed three tasks:\looseness=-1

\begin{enumerate}
    \item[\textbf{T1.}] First, we asked the participant to navigate to the \detailview{} of a unit test challenge set that targeted mismatches in numbers between source and output translations, and to discuss what they saw.
    \item[\textbf{T2.}] Second, we asked the participant to choose a topic-based challenge set that was interesting to them, explain their choice, and again explore the \detailview{} to learn more.
    \item[\textbf{T3.}] Finally, we gave the participant a hypothetical budget of 2,000 sentences that they could choose to get evaluated by expert human translators, and asked them to explain how they would allocate that budget. The evaluation by professional translators could involve rating the quality of model-produced translations and/or correcting the translations to create gold standard reference translations that could be used for future model training.
\end{enumerate}

We analyzed the transcripts following a similar qualitative data analysis procedure to that of the formative interview study~(\autoref{sec:interview-methods}).
One author conducted two rounds of open coding, synthesizing and combining codes each round \cite{qualresearch}.
Next, a second author took the code book in development and independently coded all of the transcripts, adding new codes where relevant and noting disagreements.
These two authors then discussed and resolved disagreements, and converged on final coding scheme.

We found that \textbf{T1} and \textbf{T2} mainly served as a way for participants to acclimate to \tool{}'s interface, and understand the two types of challenge sets.
Although participants confirmed that \textbf{T3} was a realistic task for the team, most participants did not do the task as we had originally planned. We report our findings regarding how participants picked which challenge sets they deemed \textit{important} for model improvements, but we do not report on their fictional budget allocations because the majority of participants were resistant to allocating concrete (even completely hypothetical) numbers. We discuss this tension more in limitations \autoref{sec:evallimitations}. All three tasks required participants to prioritize among the available challenge sets, but our findings focus largely on participants' judgments during \textbf{T3}, where they spent the majority of the study time.

As discussed in Section \ref{sec:interview}, the team's existing approaches to model debugging and improvement were either one-off, focused analyses, which do not require prioritization between issues, or random samples of usage logs, which implicitly prioritize use cases based on frequency of requests.
Participants informally compared what they could do with \tool{} to these existing practices. Our goal in this study was to explore the space of possibilities for visual analytics tools to support prioritization, rather than quantify the relative benefit of our specific prototype compared to existing practices.\looseness=-1

\subsection{Results: Prioritizing under Uncertainty}
Participants had to rely on imperfect and incomplete metrics to estimate the quality of translations.
All participants knew a little Spanish, but not enough to spot-check the quality of a translation in most cases.
Though \tool{} shows sentences from usage data and model training data, only the training data had \textit{reference translations} certified as correct by human translators.
Sentences from the usage logs have no quality annotations~(\autoref{sec:ui:creation}).
Thus the ChrF quality estimation for any challenge set was based on the limited data with reference translations.
This evaluation setup is far from ideal, yet realistic to what MT practitioners ordinarily encounter.

Participants knew that no one metric was a reliable source of quality information, so they weighed multiple signals and still knew that human annotation would generally be required to get a reliable measure of quality. Three participants discussed how they incorporated uncertainty when interpreting metrics like average ChrF to ensure they were getting meaningful estimates of quality [\userTwo, \userThree, \userSeven{}]:

\begin{quote}
    \textit{``You might get a low metric or a low familiarity score, but the smaller the sample is the more likely it is [that] there's gonna be some noise in there that's kind of moving the metric.''} --- \userSeven
\end{quote}

Future iterations of the tool could use re-sampling methods to estimate confidence intervals for challenge set summary statistics to make this uncertainty more explicit.

Despite available metrics in \tool{} being uncertain proxies for model performance, participants nonetheless used metrics to judge the relative importance of a challenge set:

\begin{quote}
    \textit{``It's better to have a statistical way, I mean, \textup{[rather than]} just by what I'm thinking, right?''} --- \userThree
\end{quote}

All participants tended to rank the challenge sets by potential risk of model failure by combining low ChrF, low familiarity, \text{and} low train-ratio. Low ChrF indicates that the limited \textit{training data} that falls within the challenge set \textit{might be poorly translated}~\cite{popovicChrFCharacterNgram2015}. Low familiarity and low train-ratio are proxies for where the training data set \textit{might lack coverage} of the usage data. Low familiarity suggests that the user request data that falls within the challenge set is semantically different from the overall training data. Low train-ratio indicates that this challenge set represents a subgroup that is much more represented in the user requests than in the training data. Familiarity and train-ratio were calculated in a way that correlates (\autoref{sec:ui:creation}).

As a first pass, most participants used the sorting feature in the \tableview{}~(\autoref{fig:table}) to rank which challenge sets scored the lowest on one or more of these three metrics [\userTwo, \userThree, \userOne, \userFive, \userSeven{}]. Four of these five participants additionally considered set size, with a preference for larger sets. While participants could have stopped at this point, given the opportunity to explore further, none of the practitioners relied solely on the available metrics. Next, we discuss other ways that participants used \tool{} to explore the data beyond aggregate metrics and decide which sets to annotate.

\subsubsection{Estimating Meaningful Use Cases}
\label{sec:study:meaningful}

From the list of challenge sets in the \textit{Table View} sorted by metrics, three participants chose to prioritize topics that appeared to represent a ``meaningful,'' coherent use case [\userOne, \userThree, \userFour{}]. Partially, this is because the BERTopic~\cite{grootendorstBERTopicNeuralTopic2022} model tends to generate some ``topics'' with little meaning, \eg the topic \texttt{topic-haha\_lol\_so\_you} versus the more meaningful \texttt{topic-health\_nursing\_care\_medical} in \autoref{fig:table}.

Partially, participants needed to make judgments on the value of improving the model on various sentence types, since some challenge sets mostly contained data that appeared to be fraud, spam, or automated messages. Participants demonstrated the value of being able to directly read sentences in the \detailview{}~(\autoref{fig:detail}) to make these judgments:

\begin{quote}
   \textit{``Yeah, a lot of these are spam. [\ldots] As I'm kind of going through them, it's like a lot of spam, a lot of porn and a lot of things that are like, automated messages. So I would use my discretion, of course, and wouldn't just use the numbers.''} --- \userOne
\end{quote}

\userTwo knew from prior experience that \textit{``it's always good to look at what the data actually are [\ldots] besides looking at the high level statistics.''} They had seen in the past that even keyword summaries can be misleading and obscure complexity that is apparent when directly inspecting the data:

\begin{quote}
    \textit{``From my past experience, sometimes we have seen some data contain some keywords and we imagine them to be, for example, articles, but looking at the actual example, they are kind of fraud messaging. [\ldots] Combining them together as a single dedicated targeted test would not make too much sense for us to understand the performance on it.''} --- \userTwo
\end{quote}

Some of the dataset contained explicit sexual content and profanity, to which participants ascribed different value. \userOne argued for prioritizing model improvement resources towards use cases that aligned with their organization's values (such as supporting small business owners), \textit{over} explicit content. \userSeven was far more accepting of explicit content, arguing that if users were translating that content, there was no reason to treat it differently.

\subsubsection{Estimating Impact on Users}

Participants assessed how severe the consequences of specific model failures would be for end users. They considered the stakes of the interaction mediated by the model [\userOne, \userFour, \userFive{}], whether a failure was especially sensitive, \eg offensive outputs, and how likely a user was to be misled if they were to receive an erroneous translation of this nature, such as an incorrect date [\userFour{}].

In task \textbf{T1} all participants looked at number mismatch translations (\autoref{fig:unit-test}). All participants skimmed the raw translations to focus on specific sub-cases of number translation, for instance how the model converts roman numerals, dates, or currency. Even though they could not read the Spanish translation, \userThree, \userFour, and \userSeven talked about wanting to find ``obvious'' errors where they could clearly see numbers changing from English to Spanish. \eg 1,100 dollars to 100 dollars. An obvious error may not mislead a user, but could degrade their trust in the translation product.
Participants dug into specific sub-cases through filtering and search in \tool{} to get a sense for the severity of an error:

\begin{quote}
    \textit{``It's a really nice way of quickly getting into the patterns to see whether or not we're looking at something like a serious problem with translation or if it's just kind of surface level formatting issues.''} --- \userSeven{}
\end{quote}

\subsubsection{Estimating Complexity of the Error}

Practitioners wanted to prioritize annotation resources on more complex kinds of failures, rather than those that could be solved internally without additional annotations [\userOne, \userTwo{}]. For example, issues with translating numbers could be identified within the team by using regular expressions to match source and translations. For pattern-based failures, such as translating numbers or translating automated messages, \userTwo proposed trying data augmentation techniques first. Data augmentation includes increasing the samples and variations on existing data, \eg the sentence ``I ate breakfast on Sunday'' can be duplicated to create a sentence for all the weekdays ``I ate breakfast on Monday'':\looseness=-1

\begin{quote}
    \textit{``If it's a lack of data issue, it should be very easy to augment the data for this particular example.''} --- \userTwo
\end{quote}

While they used train-ratio and familiarity metrics to identify potential coverage issues, directly inspecting the data gave them insight into whether a problem was complex enough to warrant annotation.
\userSix, \userThree, and \userFour used the \textit{Embedding} view to develop more nuanced hypotheses about how customers' use of the model differs from the training data. Even within an apparently similar topic, participants used clusters of usage data with comparatively less training data to estimate subtopics that may need better coverage [\userOne, \userThree, \userSix{}]. They used their experience to hypothesize why or why not an area of low-coverage might be difficult for the model. For instance a cluster with a lot of domain-specific language may be best improved by paying for additional annotations [\userThree{}].

While we had initially prompted practitioners to budget annotation resources between the challenge sets we gave them, more often we found that they wanted to prioritize subgroups \textit{within those sets} to optimize annotation for the most complex and impactful issues.

We observed that practitioners were able to form more interesting and user-focused hypotheses for prioritization when they combined summary statistics with qualitative assessment of the data by reading sentences.
Their use of \tool{} was promising evidence for the strength of a visual analytics approach in MT prioritization. At the same time, practitioners demanded more features for flexibly creating challenge sets and exploring more analytics lenses than the \tool{} prototype supported. We next discuss strengths and limitations of the tool to inform next steps.

\subsection{Results: \tool{} Strengths and Usefulness}

\subsubsection{Develop Intuition for Model Behavior}

Neural machine translation models are large language models that are not easily understandable even to those who have developed them \cite{johnson2017multilingual}. We found that exploring data with \tool{} helped practitioners develop a deeper understanding of how their models work. \userSeven said that \textit{``a lot of what I like to do is just develop my like, mental model of how our translation models are working.''} Exploring challenge sets by the quality metrics gave \userThree \textit{``some insights into the weaknesses of the model.''} Given that practitioners' intuitions about model weaknesses guide their future debugging effort ~(\autoref{sec:interview-sourcing:brainstorming}), this can bring value beyond identifying specific failures in the moment. %

\subsubsection{Develop Intuition for Translation Usage}
\label{sec:study:intuition}

Participants used the topic-based challenge sets to improve their understanding of how customers use their translation products.
The \textit{Keyword Spotlight}, \textit{Sentence List}, and \textit{Source Spotlight} were especially useful for participants to develop hypotheses about how people use the model and the context where they are using it [\userOne, \userFour, \userFive, \userSix{}]. \userOne works in a user experience focused role and said that, \textit{``it helps us inform feature development when we understand the conversation topics that people are using \textup{[the model for]}.''} While browsing the unit test challenge set based on mismatches in numbers between source and output, \userSix imagined potential future features that could give users greater control over how different number formats are handled in translation, \eg when to use Roman numerals or convert date formats.\looseness=-1

\subsubsection{Develop a Shared Interdisciplinary Understanding}

UX-focused participants expressed excitement for using a visual analytics tool like \tool{} to broadly explore the use cases surrounding potential failure-modes---as opposed to a purely metrics-driven report that does not allow them to develop their sense of the use context.
\userOne and \userSix pointed out that being able to describe specific use cases makes it easier to engage cross-functional teams in planning and prioritizing product improvements and developments. \userOne wanted to spend time using \tool{} to \textit{``generate some insights about each of [the topics] in human understandable terms''} that they could then present to other team members.

\begin{quote}
    \textit{``Our [team members] come from a variety of backgrounds [\ldots] not all of them are engineers. So it's like, could I translate the high level findings here into something that they could understand in a brief?''} --- \userOne
\end{quote}

Using \tool{}, \userFour and \userFive even learned new insights about where the model performs unexpectedly well on specific kinds of inputs: \textit{``the date formatting changed. I didn't even know if that's something that we do''} --- \userFour{}.

Discussing improvements in terms of use cases and specific customer needs not only supports internal cross-functional collaboration, but also makes it easier to acquire new data targeting specific topics from external vendors [\userOne{}].

\subsection{Results: \tool{} Limitations and Usability Issues}
\label{sec:evallimitations}

As we mentioned in \autoref{sec:study}, most participants did not assign concrete numbers in the annotation budgeting task \textbf{T3}. Partially this was a limitation of the study setup, since we provided participants with a long list of pre-made challenge sets in \tool{}, and there was not enough time for a participant to closely examine all of them. However, in many cases participants also wanted more information and analytics features to drive their prioritization than \tool{} provides ---or else wanted to refine challenge sets from those we pre-made. A major design takeaway for future work is that although \tool{} is \textit{already} a relatively complex visualization tool, far more lenses and interactions were desired to complete the MT annotation prioritization task. We discuss these key areas for improvement next.\looseness=-1

\subsubsection{Provide More Context and Comparison}

Participants wanted to contextualize the data they were seeing in each challenge set with reference to the overall distribution of data in the training data and usage logs. For example, \userFive and \userSeven wanted to know what the average familiarity score was over all of the usage logs to interpret familiarity scores on specific challenge sets. Other participants suggested additional useful reference points, for instance, understanding overall ChrF score distribution by language pair [\userThree, \userFive{}] or understanding the overall distribution of topics in the training data and usage logs [\userFour, \userSeven{}]. \userSix suggested that it would be helpful to be able to more easily compare challenge sets:

\begin{quote}
    \textit{``It's not so easy for me to compare them [challenge sets]. So it would be great if I can somehow select a cluster and compare them side by side.''} --- \userSix
\end{quote}

\subsubsection{Support Authoring and Refining Challenge Sets}

Our focus in this work was on how visual analytics tools could support the process of prioritizing specific areas for improving MT models. Therefore, we chose two plausible methods for constructing challenge sets to use as examples in the study. While participants found those sets interesting, there was a clear need for future tooling to support challenge set \textit{creation} in addition to exploration.

Participants wanted to search all of the data by specific terms [\userFive, \userSix{}] and refine the unit test logic to better capture specific types of errors [\userTwo, \userThree, \userFour, \userSix, \userSeven{}]. \userSeven even asked if they could onboard their own datasets that they already had available to explore them with \tool{}'s visualizations.

\subsubsection{Offer Advanced Export Options for Custom Analysis}

Two participants [\userOne, \userThree{}] expressed a desire to conduct their own analyses on the data, \eg conduct a custom analysis over time [\userOne{}] or experiment with the underlying topic model [\userThree{}]. While \tool{} offers a simple export option, these options could be made more sophisticated to support users with advanced skills to build on the default visualizations.

\subsubsection{Expand Filtering and Sorting Capabilities}

Several participants found issues with the data filtering and sorting features that made it difficult to organize and prioritize data in the way they wished [\userOne, \userTwo, \userFive, \userSix{}]. For example, two participants expected to be able to filter to all sentences containing \textit{all} of the keywords selected, but the tool returned \textit{any} such sentences [\userFive, \userSix{}]. Two participants also struggled to keep track of which filters they had applied when navigating between views [\userTwo, \userFive{}], suggesting potentials to make the ability to view and remove filters more prominent in the interface. Two participants wanted to sort the \tableview{} by multiple columns to be able to organize and prioritize sets by multiple metrics or factors, \eg find large sets within the sets with the lowest average ChrF score [\userOne, \userTwo{}].

\subsubsection{Validate and Extend Challenge Set Creation Methods}

Grouping data using a topic model was a useful way for practitioners to explore data and better understand use cases for the model. However, it is not clear whether a group of sentences that are close together in a latent embedding space that was trained to group together sentences with similar meaning are also likely to be similarly difficult for an MT model.

There are certainly other dimensions by which to group data. As \userFive described,

\begin{quote}
    \textit{``When I think about trying to determine where we're doing poorly, there are a lot of dimensions you can look at. Topic is one, right? But there could be other dimensions, like how long the sentence is.''} --- \userFive
\end{quote}

In this paper, we found that exploring challenge sets has the potential to help practitioners prioritize their model evaluation and development resources on issues that are important to end-users. An important direction for future work is to validate that the challenge sets presented indeed represent areas where the translation model performance is relatively weaker. For instance, rather than asking practitioners to allocate hypothetical budgets, they could allocate real budgets and have professional translators evaluate the challenge sets selected to see whether they were able to identify new model failure cases using \tool{}.

In general, we designed \tool{} to be agnostic to the method of generating challenge sets. Thus, research developing and evaluating methods for generating challenge sets can proceed in parallel to efforts to improve visual analytics support for exploring, comparing, and prioritizing them. %
\section{Discussion and Future Work}

To conclude, we discuss our findings in the broader context of tooling for ML evaluation and debugging and highlight directions for future work.

\subsection{Trade-offs Between Automation and Human Curation}

\tool{} encourages MT practitioners to inspect training data and usage logs, so that they can better understand how end-users use MT models and detect model failures.
Practitioners can then annotate related data samples and retrain the model to address detected failures.
Since exploring raw data is a manual and tedious process, we introduce an approach that uses unit tests and topic modeling to automatically surface interesting challenge sets~(\autoref{sec:ui:creation}).
Our approach yields many challenge sets, but it still takes time for MT practitioners to inspect and fine-tune these sets.
One might argue that we should automate the whole pipeline and have human raters annotate all extracted challenge sets.
However, annotating MT data is expensive~\cite{freitagExpertsErrorsContext2021, moreMachineTranslationnessMachinelikeness2014}.
In our study, some challenge sets reveal MT errors that are trivial, where MT experts hesitate to spend the budget to annotate the challenge sets~(\autoref{sec:study:meaningful}).
Besides data acquisition prioritization, our mixed-initiative approach can also help users interact with raw usage logs and gain insights into the real-life use cases of MT models.
Future researchers can use \tool{} as a research instrument to further study the trade-offs between automation and human curation for challenge set creation.
To further reduce human effort, researchers can surface challenge sets more precisely before presenting them to MT practitioners.

\subsection{Generalization to Different Model Types}

We situate \tool{} in the MT context, as it is particularly challenging to discover failures for MT models due to the scarcity of ground truth and the high cost of human annotators~(\autoref{sec:related:mt-evaluation}).
However, our method is generalizable to different model types.
In our formative interview study, we find that it is a common practice to use challenge sets to test and monitor NLP, computer vision, and time-series models~(\autoref{sec:interview-challenge-set}).
ML practitioners can adapt our \textit{unit tests} and \textit{topic modeling} (clustering) approach to surface challenge sets for other model types.
Consider an image classification model; practitioners could define perturbation-based unit tests to detect model weaknesses.
For example, we would expect a model to give the same prediction when the input image is rotated, resized, or with different lighting~\cite{rebuffiDataAugmentationCan2021, joshiFairSASensitivity2022}.
Then, we can create challenge sets by collecting images where the model's prediction changes after adding image perturbations.
Similar to topic modeling, practitioners could use embedding clustering~\cite{baktashmotlaghDistributionmatchingEmbeddingVisual2016, caiHumanCenteredToolsCoping2019} and sub-group analysis~\cite{krishnakumarUDISUnsupervisedDiscovery2021, leeVisCUITVisualAuditor2022} to identify unfamiliar images from both usage logs and training data.
For example, if an image classifier team receives a user's complaint about a misclassification, they can use embedding-based image search~\cite{radfordLearningTransferableVisual2021} to identify similar images and create a challenge set.
Finally, researchers can adapt \tool{}'s \textit{overview+detail} design~(\autoref{sec:ui}) and open-source implementation~(\autoref{sec:ui:implementation}) to summarize extracted challenge sets and allow practitioners to explore and curate potentially error-prone images through different perspectives.

\subsection{Unit Tests for Machine Learning}

Researchers have argued that unit tests can help pay down the technical debt in ML systems~\cite{sculleyHiddenTechnicalDebt2015,sculleyDataCentricViewTechnical2022}.
There are many different ways to apply unit tests to an ML system.
For example, practitioners can write unit tests to validate the data quality~\cite{polyzotisDataValidationMachine2019, schelterUnitTestingData2019}, verify a model's behavior~\cite{ribeiroAccuracyBehavioralTesting2020, loveringUnitTestingConcepts2022}, and maintain ML operations (MLOps)~\cite{gargContinuousIntegrationContinuous2021, huyenDesigningMachineLearning2022}.
In \tool{}, we design simple rule-based unit tests, such as if the source does not contain offensive words, then the translation should not either.
We then apply these tests to the training data and usage logs to surface challenge sets~(\autoref{sec:ui:unit-test}).
Since they were intended to be a proof of concept, our unit tests were blunt and imperfect.
Still, MT practitioners in the evaluation study especially appreciated the unit tests, as they are powerful to detect glaring translation mistakes and yet are easy to adopt in the current model development workflow~(\autoref{sec:study:intuition}).
Therefore, we see rich research opportunities to study unit tests for ML systems.
For example, future researchers could extend our unit tests to support MT data validation and MLOps in general.
Researchers could also adapt \tool{} to design future interactive tools that allow ML practitioners to easily write, organize, and maintain unit tests for ML systems.

\subsection{Broader Impact}

To overcome the limitations of aggregate metrics on held-out test sets~(\autoref{sec:related:auditing}, \autoref{sec:related:mt-evaluation}), \tool{} uses \textit{real usage logs} to help MT practitioners gain a better understanding of how their models are used and prioritize model failures.
Drawing on usage data raises privacy and security concerns.
All of the authors have received training and license to use usage logs from their institution for this research.
Researchers adapting \tool{} should carefully consider the ethical implications of their choice of data source~\cite{barocasDesigningDisaggregatedEvaluations2021}.
Before collecting usage logs, researchers need to obtain consent from the users~\cite{rajiSavingFaceInvestigating2020} and compensate them when applicable~\cite{arrieta-ibarraShouldWeTreat2018}.
Usage logs must be de-identified before viewing them with \tool{}.
Finally, we encourage researchers and developers to thoroughly document their process of adopting \tool{} with new models and datasets~\cite{hegerUnderstandingMachineLearning2022}, including how they approach these ethical considerations.
\section{Conclusion}

In this work, we present \tool{}, an open-source interactive visualization system that empowers MT practitioners to prioritize model improvements by exploring and curating challenge sets.
To inform the design of the system, we conducted a formative interview study with \numParticipants ML practitioners to explore current practices in evaluating ML models and prioritizing evaluation resources.
Through a user study with \numParticipantsObs MT stakeholders across engineering and user experience-focused roles, we revealed how practitioners prioritize their efforts based on an understanding of how problems could impact end users.
We hope our work can inspire future researchers to design human-centered interactive tools that help ML practitioners improve their models in ways that enrich and improve the user experience. 

\begin{acks}
We thank our colleagues at Apple for their time during the interview study and evaluating our system.
We especially thank Danielle Olson who sparked early ideas and connections for pursuing this line of research, and Yanchao Ni for lending his expertise in machine translation for our work.
\end{acks}

\bibliographystyle{ACM-Reference-Format}

\bibliography{23-translate-chi}%


\begin{thebibliography}{137}


\ifx \showCODEN    \undefined \def \showCODEN     #1{\unskip}     \fi
\ifx \showDOI      \undefined \def \showDOI       #1{#1}\fi
\ifx \showISBNx    \undefined \def \showISBNx     #1{\unskip}     \fi
\ifx \showISBNxiii \undefined \def \showISBNxiii  #1{\unskip}     \fi
\ifx \showISSN     \undefined \def \showISSN      #1{\unskip}     \fi
\ifx \showLCCN     \undefined \def \showLCCN      #1{\unskip}     \fi
\ifx \shownote     \undefined \def \shownote      #1{#1}          \fi
\ifx \showarticletitle \undefined \def \showarticletitle #1{#1}   \fi
\ifx \showURL      \undefined \def \showURL       {\relax}        \fi
\providecommand\bibfield[2]{#2}
\providecommand\bibinfo[2]{#2}
\providecommand\natexlab[1]{#1}
\providecommand\showeprint[2][]{arXiv:#2}

\bibitem[Albrecht et~al\mbox{.}(2009)]%
        {albrechtChineseRoomVisualization2009}
\bibfield{author}{\bibinfo{person}{Joshua Albrecht}, \bibinfo{person}{Rebecca
  Hwa}, {and} \bibinfo{person}{G.~Elisabeta Marai}.}
  \bibinfo{year}{2009}\natexlab{}.
\newblock \showarticletitle{The {{Chinese Room}}: {{Visualization}} and
  {{Interaction}} to {{Understand}} and {{Correct Ambiguous Machine
  Translation}}}.
\newblock \bibinfo{journal}{\emph{Computer Graphics Forum}}
  \bibinfo{volume}{28} (\bibinfo{date}{June} \bibinfo{year}{2009}),
  \bibinfo{pages}{1047--1054}.
\newblock
\urldef\tempurl%
\url{https://doi.org/10.1111/j.1467-8659.2009.01443.x}
\showDOI{\tempurl}


\bibitem[Amershi et~al\mbox{.}(2019)]%
        {amershiSoftwareEngineeringMachine2019}
\bibfield{author}{\bibinfo{person}{Saleema Amershi}, \bibinfo{person}{Andrew
  Begel}, \bibinfo{person}{Christian Bird}, \bibinfo{person}{Robert DeLine},
  \bibinfo{person}{Harald Gall}, \bibinfo{person}{Ece Kamar},
  \bibinfo{person}{Nachiappan Nagappan}, \bibinfo{person}{Besmira Nushi}, {and}
  \bibinfo{person}{Thomas Zimmermann}.} \bibinfo{year}{2019}\natexlab{}.
\newblock \showarticletitle{Software Engineering for Machine Learning: A Case
  Study}. In \bibinfo{booktitle}{\emph{2019 IEEE/ACM 41st International
  Conference on Software Engineering: Software Engineering in Practice
  (ICSE-SEIP)}}. \bibinfo{publisher}{IEEE}, \bibinfo{address}{Montreal, QC,
  Canada}, \bibinfo{pages}{291--300}.
\newblock
\urldef\tempurl%
\url{https://doi.org/10.1109/ICSE-SEIP.2019.00042}
\showDOI{\tempurl}


\bibitem[{Arrieta-Ibarra} et~al\mbox{.}(2018)]%
        {arrieta-ibarraShouldWeTreat2018}
\bibfield{author}{\bibinfo{person}{Imanol {Arrieta-Ibarra}},
  \bibinfo{person}{Leonard Goff}, \bibinfo{person}{Diego
  {Jim{\'e}nez-Hern{\'a}ndez}}, \bibinfo{person}{Jaron Lanier}, {and}
  \bibinfo{person}{E.~Glen Weyl}.} \bibinfo{year}{2018}\natexlab{}.
\newblock \showarticletitle{Should We Treat Data as Labor? {{Moving}} beyond
  "{{Free}}"}.
\newblock \bibinfo{journal}{\emph{AEA Papers and Proceedings}}
  \bibinfo{volume}{108} (\bibinfo{date}{May} \bibinfo{year}{2018}),
  \bibinfo{pages}{38--42}.
\newblock
\urldef\tempurl%
\url{https://doi.org/10.1257/pandp.20181003}
\showDOI{\tempurl}


\bibitem[Avramidis et~al\mbox{.}(2019)]%
        {avramidisLinguisticEvaluationGermanEnglish2019}
\bibfield{author}{\bibinfo{person}{Eleftherios Avramidis},
  \bibinfo{person}{Vivien Macketanz}, \bibinfo{person}{Ursula Strohriegel},
  {and} \bibinfo{person}{Hans Uszkoreit}.} \bibinfo{year}{2019}\natexlab{}.
\newblock \showarticletitle{Linguistic {{Evaluation}} of {{German-English
  Machine Translation Using}} a {{Test Suite}}}. In
  \bibinfo{booktitle}{\emph{Proceedings of the {{Fourth Conference}} on
  {{Machine Translation}} ({{Volume}} 2: {{Shared Task Papers}}, {{Day}} 1)}}.
  \bibinfo{publisher}{Association for Computational Linguistics},
  \bibinfo{address}{Florence, Italy}, \bibinfo{pages}{445--454}.
\newblock
\urldef\tempurl%
\url{https://doi.org/10.18653/v1/w19-5351}
\showDOI{\tempurl}


\bibitem[Baktashmotlagh et~al\mbox{.}(2016)]%
        {baktashmotlaghDistributionmatchingEmbeddingVisual2016}
\bibfield{author}{\bibinfo{person}{Mahsa Baktashmotlagh},
  \bibinfo{person}{Mehrtash Har}, \bibinfo{person}{{i}}, {and}
  \bibinfo{person}{Mathieu Salzmann}.} \bibinfo{year}{2016}\natexlab{}.
\newblock \showarticletitle{Distribution-Matching Embedding for Visual Domain
  Adaptation}.
\newblock \bibinfo{journal}{\emph{Journal of Machine Learning Research}}
  \bibinfo{volume}{17}, \bibinfo{number}{108} (\bibinfo{year}{2016}),
  \bibinfo{pages}{1--30}.
\newblock
\urldef\tempurl%
\url{http://jmlr.org/papers/v17/15-207.html}
\showURL{%
\tempurl}


\bibitem[Banerjee and Lavie(2005)]%
        {banerjeeMETEORAutomaticMetric2005}
\bibfield{author}{\bibinfo{person}{Satanjeev Banerjee} {and}
  \bibinfo{person}{Alon Lavie}.} \bibinfo{year}{2005}\natexlab{}.
\newblock \showarticletitle{{{METEOR}}: {{An}} Automatic Metric for {{MT}}
  Evaluation with Improved Correlation with Human Judgments}. In
  \bibinfo{booktitle}{\emph{Proceedings of the Acl Workshop on Intrinsic and
  Extrinsic Evaluation Measures for Machine Translation and/or Summarization}}.
  \bibinfo{publisher}{Association for Computational Linguistics},
  \bibinfo{address}{Ann Arbor, Michigan}, \bibinfo{pages}{65--72}.
\newblock


\bibitem[Barocas et~al\mbox{.}(2021)]%
        {barocasDesigningDisaggregatedEvaluations2021}
\bibfield{author}{\bibinfo{person}{Solon Barocas}, \bibinfo{person}{Anhong
  Guo}, \bibinfo{person}{Ece Kamar}, \bibinfo{person}{Jacquelyn Krones},
  \bibinfo{person}{Meredith~Ringel Morris}, \bibinfo{person}{Jennifer~Wortman
  Vaughan}, \bibinfo{person}{W.~Duncan Wadsworth}, {and} \bibinfo{person}{Hanna
  Wallach}.} \bibinfo{year}{2021}\natexlab{}.
\newblock \showarticletitle{Designing Disaggregated Evaluations of AI Systems:
  Choices, Considerations, and Tradeoffs}. In
  \bibinfo{booktitle}{\emph{Proceedings of the 2021 AAAI/ACM Conference on AI,
  Ethics, and Society}} (Virtual Event, USA) \emph{(\bibinfo{series}{AIES
  '21})}. \bibinfo{publisher}{Association for Computing Machinery},
  \bibinfo{address}{New York, NY, USA}, \bibinfo{pages}{368–378}.
\newblock
\showISBNx{9781450384735}
\urldef\tempurl%
\url{https://doi.org/10.1145/3461702.3462610}
\showDOI{\tempurl}


\bibitem[{Beauxis-Aussalet} et~al\mbox{.}(2021)]%
        {beauxis-aussaletRoleInteractiveVisualization2021}
\bibfield{author}{\bibinfo{person}{Emma {Beauxis-Aussalet}},
  \bibinfo{person}{Michael Behrisch}, \bibinfo{person}{Rita Borgo},
  \bibinfo{person}{Duen~Horng Chau}, \bibinfo{person}{Christopher Collins},
  \bibinfo{person}{David Ebert}, \bibinfo{person}{Mennatallah {El-Assady}},
  \bibinfo{person}{Alex Endert}, \bibinfo{person}{Daniel~A. Keim},
  \bibinfo{person}{Jorn Kohlhammer}, \bibinfo{person}{Daniela Oelke},
  \bibinfo{person}{Jaakko Peltonen}, \bibinfo{person}{Maria Riveiro},
  \bibinfo{person}{Tobias Schreck}, \bibinfo{person}{Hendrik Strobelt}, {and}
  \bibinfo{person}{Jarke~J. {van Wijk}}.} \bibinfo{year}{2021}\natexlab{}.
\newblock \showarticletitle{The {{Role}} of {{Interactive Visualization}} in
  {{Fostering Trust}} in {{AI}}}.
\newblock \bibinfo{journal}{\emph{IEEE Computer Graphics and Applications}}
  \bibinfo{volume}{41} (\bibinfo{date}{Nov.} \bibinfo{year}{2021}),
  \bibinfo{pages}{7--12}.
\newblock
\urldef\tempurl%
\url{https://doi.org/10.1109/mcg.2021.3107875}
\showDOI{\tempurl}


\bibitem[Bentley(1975)]%
        {bentleyMultidimensionalBinarySearch1975}
\bibfield{author}{\bibinfo{person}{Jon~Louis Bentley}.}
  \bibinfo{year}{1975}\natexlab{}.
\newblock \showarticletitle{Multidimensional Binary Search Trees Used for
  Associative Searching}.
\newblock \bibinfo{journal}{\emph{Commun. ACM}}  \bibinfo{volume}{18}
  (\bibinfo{year}{1975}).
\newblock
\urldef\tempurl%
\url{https://doi.org/10.1145/361002.361007}
\showDOI{\tempurl}


\bibitem[Bhatt et~al\mbox{.}(2021)]%
        {bhatt2021case}
\bibfield{author}{\bibinfo{person}{Shaily Bhatt}, \bibinfo{person}{Rahul Jain},
  \bibinfo{person}{Sandipan Dandapat}, {and} \bibinfo{person}{Sunayana
  Sitaram}.} \bibinfo{year}{2021}\natexlab{}.
\newblock \showarticletitle{A Case Study of Efficacy and Challenges in
  Practical Human-in-Loop Evaluation of {{NLP}} Systems Using Checklist}. In
  \bibinfo{booktitle}{\emph{Workshop on Human Evaluation of {{NLP}} Systems}}.
\newblock


\bibitem[Blodgett et~al\mbox{.}(2017)]%
        {blodgett2017dataset}
\bibfield{author}{\bibinfo{person}{Su~Lin Blodgett}, \bibinfo{person}{Johnny
  Wei}, {and} \bibinfo{person}{Brendan O{'}Connor}.}
  \bibinfo{year}{2017}\natexlab{}.
\newblock \showarticletitle{A Dataset and Classifier for Recognizing Social
  Media {E}nglish}. In \bibinfo{booktitle}{\emph{Proceedings of the 3rd
  Workshop on Noisy User-generated Text}}. \bibinfo{publisher}{Association for
  Computational Linguistics}, \bibinfo{address}{Copenhagen, Denmark},
  \bibinfo{pages}{56--61}.
\newblock
\urldef\tempurl%
\url{https://doi.org/10.18653/v1/W17-4408}
\showDOI{\tempurl}


\bibitem[Bolukbasi et~al\mbox{.}(2016)]%
        {bolukbasiManComputerProgrammer2016}
\bibfield{author}{\bibinfo{person}{Tolga Bolukbasi}, \bibinfo{person}{Kai-Wei
  Chang}, \bibinfo{person}{James~Y Zou}, \bibinfo{person}{Venkatesh Saligrama},
  {and} \bibinfo{person}{Adam~T Kalai}.} \bibinfo{year}{2016}\natexlab{}.
\newblock \showarticletitle{Man Is to Computer Programmer as Woman Is to
  Homemaker? {{Debiasing}} Word Embeddings}. In
  \bibinfo{booktitle}{\emph{NeurIPS}}, Vol.~\bibinfo{volume}{29}.
\newblock


\bibitem[Bostock et~al\mbox{.}(2011)]%
        {bostockDataDrivenDocuments2011}
\bibfield{author}{\bibinfo{person}{M. Bostock}, \bibinfo{person}{V.
  Ogievetsky}, {and} \bibinfo{person}{J. Heer}.}
  \bibinfo{year}{2011}\natexlab{}.
\newblock \showarticletitle{{{D}}{$^3$} {{Data-Driven Documents}}}.
\newblock \bibinfo{journal}{\emph{IEEE TVCG}}  \bibinfo{volume}{17}
  (\bibinfo{date}{Dec.} \bibinfo{year}{2011}).
\newblock
\urldef\tempurl%
\url{https://doi.org/10.1109/tvcg.2011.185}
\showDOI{\tempurl}


\bibitem[Buolamwini and Gebru(2018)]%
        {buolamwiniGenderShadesIntersectional2018}
\bibfield{author}{\bibinfo{person}{Joy Buolamwini} {and}
  \bibinfo{person}{Timnit Gebru}.} \bibinfo{year}{2018}\natexlab{}.
\newblock \showarticletitle{Gender Shades: {{Intersectional}} Accuracy
  Disparities in Commercial Gender Classification}. In
  \bibinfo{booktitle}{\emph{Proceedings of the 1st Conference on Fairness,
  Accountability and Transparency}} \emph{(\bibinfo{series}{Proceedings of
  Machine Learning Research}, Vol.~\bibinfo{volume}{81})}.
\newblock


\bibitem[Burchardt et~al\mbox{.}(2017)]%
        {burchardtLinguisticEvaluationRulebased2017}
\bibfield{author}{\bibinfo{person}{Aljoscha Burchardt}, \bibinfo{person}{Vivien
  Macketanz}, \bibinfo{person}{Jon Dehdari}, \bibinfo{person}{Georg Heigold},
  \bibinfo{person}{Peter {Jan-Thorsten}}, {and} \bibinfo{person}{Philip
  Williams}.} \bibinfo{year}{2017}\natexlab{}.
\newblock \showarticletitle{A Linguistic Evaluation of Rule-Based,
  Phrase-Based, and Neural {{MT}} Engines}.
\newblock \bibinfo{journal}{\emph{The Prague Bulletin of Mathematical
  Linguistics}}  \bibinfo{volume}{108} (\bibinfo{year}{2017}).
\newblock
\urldef\tempurl%
\url{https://doi.org/10.1515/pralin-2017-0017}
\showDOI{\tempurl}


\bibitem[Burlot et~al\mbox{.}(2018)]%
        {burlotWMT18Morpheval2018}
\bibfield{author}{\bibinfo{person}{Franck Burlot}, \bibinfo{person}{Yves
  Scherrer}, \bibinfo{person}{Vinit Ravishankar}, \bibinfo{person}{Ond{\v r}ej
  Bojar}, \bibinfo{person}{Stig-Arne Gr{\"o}nroos}, \bibinfo{person}{Maarit
  Koponen}, \bibinfo{person}{Tommi Nieminen}, {and} \bibinfo{person}{Fran{\c
  c}ois Yvon}.} \bibinfo{year}{2018}\natexlab{}.
\newblock \showarticletitle{The {{WMT}}'18 {{Morpheval}} Test Suites for
  {{English-Czech}}, {{English-German}}, {{English-Finnish}} and
  {{Turkish-English}}}. In \bibinfo{booktitle}{\emph{Proceedings of the Third
  Conference on Machine Translation: {{Shared}} Task Papers}}.
\newblock
\urldef\tempurl%
\url{https://doi.org/10.18653/v1/w18-6433}
\showDOI{\tempurl}


\bibitem[Burlot and Yvon(2017)]%
        {burlot2017evaluating}
\bibfield{author}{\bibinfo{person}{Franck Burlot} {and}
  \bibinfo{person}{Fran{\c c}ois Yvon}.} \bibinfo{year}{2017}\natexlab{}.
\newblock \showarticletitle{Evaluating the Morphological Competence of Machine
  Translation Systems}. In \bibinfo{booktitle}{\emph{Proceedings of the Second
  Conference on Machine Translation}}.
\newblock
\urldef\tempurl%
\url{https://doi.org/10.18653/v1/w17-4705}
\showDOI{\tempurl}


\bibitem[Cabrera et~al\mbox{.}(2021)]%
        {cabreraDiscoveringValidatingAI2021}
\bibfield{author}{\bibinfo{person}{{\'A}ngel~Alexander Cabrera},
  \bibinfo{person}{Abraham~J. Druck}, \bibinfo{person}{Jason~I. Hong}, {and}
  \bibinfo{person}{Adam Perer}.} \bibinfo{year}{2021}\natexlab{}.
\newblock \showarticletitle{Discovering and {{Validating AI Errors With
  Crowdsourced Failure Reports}}}.
\newblock \bibinfo{journal}{\emph{Proceedings of the ACM on Human-Computer
  Interaction}}  \bibinfo{volume}{5} (\bibinfo{date}{Oct.}
  \bibinfo{year}{2021}).
\newblock
\urldef\tempurl%
\url{https://doi.org/10.1145/3479569}
\showDOI{\tempurl}


\bibitem[Cabrera et~al\mbox{.}(2019)]%
        {cabreraFAIRVISVisualAnalytics2019}
\bibfield{author}{\bibinfo{person}{Angel~Alexander Cabrera},
  \bibinfo{person}{Will Epperson}, \bibinfo{person}{Fred Hohman},
  \bibinfo{person}{Minsuk Kahng}, \bibinfo{person}{Jamie Morgenstern}, {and}
  \bibinfo{person}{Duen~Horng Chau}.} \bibinfo{year}{2019}\natexlab{}.
\newblock \showarticletitle{{{FAIRVIS}}: {{Visual Analytics}} for {{Discovering
  Intersectional Bias}} in {{Machine Learning}}}. In
  \bibinfo{booktitle}{\emph{2019 {{IEEE Conference}} on {{Visual Analytics
  Science}} and {{Technology}} ({{VAST}})}}.
\newblock
\urldef\tempurl%
\url{https://doi.org/10.1109/vast47406.2019.8986948}
\showDOI{\tempurl}


\bibitem[Cai et~al\mbox{.}(2019)]%
        {caiHumanCenteredToolsCoping2019}
\bibfield{author}{\bibinfo{person}{Carrie~J. Cai}, \bibinfo{person}{Emily
  Reif}, \bibinfo{person}{Narayan Hegde}, \bibinfo{person}{Jason Hipp},
  \bibinfo{person}{Been Kim}, \bibinfo{person}{Daniel Smilkov},
  \bibinfo{person}{Martin Wattenberg}, \bibinfo{person}{Fernanda Viegas},
  \bibinfo{person}{Greg~S. Corrado}, \bibinfo{person}{Martin~C. Stumpe}, {and}
  \bibinfo{person}{Michael Terry}.} \bibinfo{year}{2019}\natexlab{}.
\newblock \showarticletitle{Human-{{Centered Tools}} for {{Coping}} with
  {{Imperfect Algorithms During Medical Decision-Making}}}. In
  \bibinfo{booktitle}{\emph{Proceedings of the 2019 {{CHI Conference}} on
  {{Human Factors}} in {{Computing Systems}}}}.
\newblock
\urldef\tempurl%
\url{https://doi.org/10.1145/3290605.3300234}
\showDOI{\tempurl}


\bibitem[Callison-Burch(2009)]%
        {callison-burch-2009-fast}
\bibfield{author}{\bibinfo{person}{Chris Callison-Burch}.}
  \bibinfo{year}{2009}\natexlab{}.
\newblock \showarticletitle{Fast, Cheap, and Creative: Evaluating Translation
  Quality Using {A}mazon{'}s {M}echanical {T}urk}. In
  \bibinfo{booktitle}{\emph{Proceedings of the 2009 Conference on Empirical
  Methods in Natural Language Processing}}. \bibinfo{publisher}{Association for
  Computational Linguistics}, \bibinfo{address}{Singapore},
  \bibinfo{pages}{286--295}.
\newblock
\urldef\tempurl%
\url{https://aclanthology.org/D09-1030}
\showURL{%
\tempurl}


\bibitem[Callison-Burch et~al\mbox{.}(2007)]%
        {callison-burch-etal-2007-meta}
\bibfield{author}{\bibinfo{person}{Chris Callison-Burch},
  \bibinfo{person}{Cameron Fordyce}, \bibinfo{person}{Philipp Koehn},
  \bibinfo{person}{Christof Monz}, {and} \bibinfo{person}{Josh Schroeder}.}
  \bibinfo{year}{2007}\natexlab{}.
\newblock \showarticletitle{(Meta-) Evaluation of Machine Translation}. In
  \bibinfo{booktitle}{\emph{Proceedings of the Second Workshop on Statistical
  Machine Translation}}. \bibinfo{publisher}{Association for Computational
  Linguistics}, \bibinfo{address}{Prague, Czech Republic},
  \bibinfo{pages}{136--158}.
\newblock
\urldef\tempurl%
\url{https://aclanthology.org/W07-0718}
\showURL{%
\tempurl}


\bibitem[{Callison-Burch} et~al\mbox{.}(2006)]%
        {callison-burchReevaluatingRoleBleu2006}
\bibfield{author}{\bibinfo{person}{Chris {Callison-Burch}},
  \bibinfo{person}{Miles Osborne}, {and} \bibinfo{person}{Philipp Koehn}.}
  \bibinfo{year}{2006}\natexlab{}.
\newblock \showarticletitle{Re-Evaluating the Role of {{Bleu}} in Machine
  Translation Research}. In \bibinfo{booktitle}{\emph{11th {{Conference}} of
  the {{European Chapter}} of the {{Association}} for {{Computational
  Linguistics}}}}.
\newblock


\bibitem[Choshen and Abend(2019)]%
        {choshenAutomaticallyExtractingChallenge2019}
\bibfield{author}{\bibinfo{person}{Leshem Choshen} {and} \bibinfo{person}{Omri
  Abend}.} \bibinfo{year}{2019}\natexlab{}.
\newblock \showarticletitle{Automatically Extracting Challenge Sets for
  Non-Local Phenomena in Neural Machine Translation}. In
  \bibinfo{booktitle}{\emph{{{CoNLL}}}}.
\newblock
\urldef\tempurl%
\url{https://doi.org/10.18653/v1/k19-1028}
\showDOI{\tempurl}


\bibitem[Chung et~al\mbox{.}(2020)]%
        {chung2020automated}
\bibfield{author}{\bibinfo{person}{Yeounoh Chung}, \bibinfo{person}{Tim
  Kraska}, \bibinfo{person}{Neoklis Polyzotis}, \bibinfo{person}{Ki~Hyun Tae},
  {and} \bibinfo{person}{Steven~Euijong Whang}.}
  \bibinfo{year}{2020}\natexlab{}.
\newblock \showarticletitle{Automated Data Slicing for Model Validation: A Big
  Data - AI Integration Approach}.
\newblock \bibinfo{journal}{\emph{IEEE Transactions on Knowledge and Data
  Engineering}} \bibinfo{volume}{32}, \bibinfo{number}{12}
  (\bibinfo{year}{2020}), \bibinfo{pages}{2284--2296}.
\newblock
\urldef\tempurl%
\url{https://doi.org/10.1109/TKDE.2019.2916074}
\showDOI{\tempurl}


\bibitem[Cockburn et~al\mbox{.}(2009)]%
        {cockburnReviewOverviewDetail2009}
\bibfield{author}{\bibinfo{person}{Andy Cockburn}, \bibinfo{person}{Amy
  Karlson}, {and} \bibinfo{person}{Benjamin~B. Bederson}.}
  \bibinfo{year}{2009}\natexlab{}.
\newblock \showarticletitle{A {{Review}} of {{Overview}}+{{Detail}},
  {{Zooming}}, and {{Focus}}+{{Context Interfaces}}}.
\newblock \bibinfo{journal}{\emph{ACM Comput. Surv.}}  \bibinfo{volume}{41}
  (\bibinfo{date}{Jan.} \bibinfo{year}{2009}).
\newblock
\urldef\tempurl%
\url{https://doi.org/10.1145/1456650.1456652}
\showDOI{\tempurl}


\bibitem[Coenen and Pearce(2019)]%
        {coenenUnderstandingUMAP2019}
\bibfield{author}{\bibinfo{person}{Andy Coenen} {and} \bibinfo{person}{Adam
  Pearce}.} \bibinfo{year}{2019}\natexlab{}.
\newblock \bibinfo{title}{Understanding {{UMAP}}}.
\newblock
\newblock
\urldef\tempurl%
\url{https://pair-code.github.io/understanding-umap/}
\showURL{%
\tempurl}


\bibitem[Das et~al\mbox{.}(2019)]%
        {dasDangersMachineTranslation2019}
\bibfield{author}{\bibinfo{person}{Prithwijit Das}, \bibinfo{person}{Anna
  Kuznetsova}, \bibinfo{person}{Meng'ou Zhu}, {and} \bibinfo{person}{Ruth
  Milanaik}.} \bibinfo{year}{2019}\natexlab{}.
\newblock \showarticletitle{Dangers of Machine Translation: {{The}} Need for
  Professionally Translated Anticipatory Guidance Resources for Limited English
  Proficiency Caregivers}.
\newblock \bibinfo{journal}{\emph{Clinical Pediatrics (Phila)}}
  \bibinfo{volume}{58} (\bibinfo{date}{Feb.} \bibinfo{year}{2019}).
\newblock


\bibitem[{de Vries} et~al\mbox{.}(2019)]%
        {devriesDoesObjectRecognition2019}
\bibfield{author}{\bibinfo{person}{Terrance {de Vries}}, \bibinfo{person}{Ishan
  Misra}, \bibinfo{person}{Changhan Wang}, {and} \bibinfo{person}{Laurens {van
  der Maaten}}.} \bibinfo{year}{2019}\natexlab{}.
\newblock \showarticletitle{Does Object Recognition Work for Everyone?}. In
  \bibinfo{booktitle}{\emph{Proceedings of the {{IEEE}}/{{CVF}} Conference on
  Computer Vision and Pattern Recognition ({{CVPR}}) Workshops}}.
\newblock


\bibitem[{d'Eon} et~al\mbox{.}(2022)]%
        {deonSpotlightGeneralMethod2022}
\bibfield{author}{\bibinfo{person}{Greg {d'Eon}}, \bibinfo{person}{Jason
  {d'Eon}}, \bibinfo{person}{James~R Wright}, {and} \bibinfo{person}{Kevin
  {Leyton-Brown}}.} \bibinfo{year}{2022}\natexlab{}.
\newblock \showarticletitle{The Spotlight: {{A}} General Method for Discovering
  Systematic Errors in Deep Learning Models}. In \bibinfo{booktitle}{\emph{2022
  {{ACM}} Conference on Fairness, Accountability, and Transparency}}.
\newblock
\urldef\tempurl%
\url{https://doi.org/10.1145/3531146.3533240}
\showDOI{\tempurl}


\bibitem[DeVos et~al\mbox{.}(2022)]%
        {devos2022toward}
\bibfield{author}{\bibinfo{person}{Alicia DeVos}, \bibinfo{person}{Aditi
  Dhabalia}, \bibinfo{person}{Hong Shen}, \bibinfo{person}{Kenneth Holstein},
  {and} \bibinfo{person}{Motahhare Eslami}.} \bibinfo{year}{2022}\natexlab{}.
\newblock \showarticletitle{Toward User-Driven Algorithm Auditing:
  {{Investigating}} Users' Strategies for Uncovering Harmful Algorithmic
  Behavior}. In \bibinfo{booktitle}{\emph{Proceedings of the 2022 {{CHI}}
  Conference on Human Factors in Computing Systems}}
  \emph{(\bibinfo{series}{{{CHI}} '22})}. Article \bibinfo{articleno}{626}.
\newblock
\urldef\tempurl%
\url{https://doi.org/10.1145/3491102.3517441}
\showDOI{\tempurl}


\bibitem[Dud{\'i}k et~al\mbox{.}(2020)]%
        {dudikFairlearnToolkitAssessing2020}
\bibfield{author}{\bibinfo{person}{Miro Dud{\'i}k}, \bibinfo{person}{Sarah
  Bird}, \bibinfo{person}{Hanna Wallach}, {and} \bibinfo{person}{Kathleen
  Walker}.} \bibinfo{year}{2020}\natexlab{}.
\newblock \showarticletitle{Fairlearn: {{A}} Toolkit for Assessing and
  Improving Fairness in {{AI}}}.
\newblock  (\bibinfo{date}{May} \bibinfo{year}{2020}).
\newblock


\bibitem[Farinha et~al\mbox{.}(2022)]%
        {wmtchat}
\bibfield{author}{\bibinfo{person}{Ana~C. Farinha}, \bibinfo{person}{M.~Amin
  Farajian}, \bibinfo{person}{Patrick Fernandes José Souza João
  Alves~António Lopes}, \bibinfo{person}{Helena Moniz}, {and}
  \bibinfo{person}{André Martins}.} \bibinfo{year}{2022}\natexlab{}.
\newblock \bibinfo{title}{WMT22 Chat Task}.
\newblock
\newblock
\urldef\tempurl%
\url{https://wmt-chat-task.github.io/}
\showURL{%
\tempurl}


\bibitem[Freitag et~al\mbox{.}(2021)]%
        {freitagExpertsErrorsContext2021}
\bibfield{author}{\bibinfo{person}{Markus Freitag}, \bibinfo{person}{George
  Foster}, \bibinfo{person}{David Grangier}, \bibinfo{person}{Viresh Ratnakar},
  \bibinfo{person}{Qijun Tan}, {and} \bibinfo{person}{Wolfgang Macherey}.}
  \bibinfo{year}{2021}\natexlab{}.
\newblock \showarticletitle{Experts, {{Errors}}, and {{Context}}: {{A
  Large-Scale Study}} of {{Human Evaluation}} for {{Machine Translation}}}.
\newblock \bibinfo{journal}{\emph{Transactions of the Association for
  Computational Linguistics}}  \bibinfo{volume}{9} (\bibinfo{date}{Dec.}
  \bibinfo{year}{2021}).
\newblock
\urldef\tempurl%
\url{https://doi.org/10.1162/tacl_a_00437}
\showDOI{\tempurl}


\bibitem[Gardner et~al\mbox{.}(2020)]%
        {gardner2020evaluating}
\bibfield{author}{\bibinfo{person}{Matt Gardner}, \bibinfo{person}{Yoav Artzi},
  \bibinfo{person}{Victoria Basmov}, \bibinfo{person}{Jonathan Berant},
  \bibinfo{person}{Ben Bogin}, \bibinfo{person}{Sihao Chen},
  \bibinfo{person}{Pradeep Dasigi}, \bibinfo{person}{Dheeru Dua},
  \bibinfo{person}{Yanai Elazar}, \bibinfo{person}{Ananth Gottumukkala},
  \bibinfo{person}{Nitish Gupta}, \bibinfo{person}{Hannaneh Hajishirzi},
  \bibinfo{person}{Gabriel Ilharco}, \bibinfo{person}{Daniel Khashabi},
  \bibinfo{person}{Kevin Lin}, \bibinfo{person}{Jiangming Liu},
  \bibinfo{person}{Nelson~F. Liu}, \bibinfo{person}{Phoebe Mulcaire},
  \bibinfo{person}{Qiang Ning}, \bibinfo{person}{Sameer Singh},
  \bibinfo{person}{Noah~A. Smith}, \bibinfo{person}{Sanjay Subramanian},
  \bibinfo{person}{Reut Tsarfaty}, \bibinfo{person}{Eric Wallace},
  \bibinfo{person}{Ally Zhang}, {and} \bibinfo{person}{Ben Zhou}.}
  \bibinfo{year}{2020}\natexlab{}.
\newblock \showarticletitle{Evaluating Models' Local Decision Boundaries via
  Contrast Sets}. In \bibinfo{booktitle}{\emph{Findings of the Association for
  Computational Linguistics: {{EMNLP}} 2020}}.
\newblock
\urldef\tempurl%
\url{https://doi.org/10.18653/v1/2020.findings-emnlp.117}
\showDOI{\tempurl}


\bibitem[Garg et~al\mbox{.}(2021)]%
        {gargContinuousIntegrationContinuous2021}
\bibfield{author}{\bibinfo{person}{Satvik Garg}, \bibinfo{person}{Pradyumn
  Pundir}, \bibinfo{person}{Geetanjali Rathee}, \bibinfo{person}{P.K. Gupta},
  \bibinfo{person}{Somya Garg}, {and} \bibinfo{person}{Saransh Ahlawat}.}
  \bibinfo{year}{2021}\natexlab{}.
\newblock \showarticletitle{On Continuous Integration / Continuous Delivery for
  Automated Deployment of Machine Learning Models Using {{MLOps}}}. In
  \bibinfo{booktitle}{\emph{2021 {{IEEE}} Fourth International Conference on
  Artificial Intelligence and Knowledge Engineering ({{AIKE}})}}.
\newblock
\urldef\tempurl%
\url{https://doi.org/10.1109/aike52691.2021.00010}
\showDOI{\tempurl}


\bibitem[Geva et~al\mbox{.}(2019)]%
        {gevaAreWeModeling2019}
\bibfield{author}{\bibinfo{person}{Mor Geva}, \bibinfo{person}{Yoav Goldberg},
  {and} \bibinfo{person}{Jonathan Berant}.} \bibinfo{year}{2019}\natexlab{}.
\newblock \showarticletitle{Are We Modeling the Task or the Annotator? {{An}}
  Investigation of Annotator Bias in Natural Language Understanding Datasets}.
  In \bibinfo{booktitle}{\emph{{{EMNLP-IJCNLP}}}}.
\newblock
\urldef\tempurl%
\url{https://doi.org/10.18653/v1/d19-1107}
\showDOI{\tempurl}


\bibitem[Grootendorst(2022)]%
        {grootendorstBERTopicNeuralTopic2022}
\bibfield{author}{\bibinfo{person}{Maarten Grootendorst}.}
  \bibinfo{year}{2022}\natexlab{}.
\newblock \showarticletitle{{{BERTopic}}: {{Neural}} Topic Modeling with a
  Class-Based {{TF-IDF}} Procedure}.
\newblock \bibinfo{journal}{\emph{arXiv preprint arXiv:2203.05794}}
  (\bibinfo{year}{2022}).
\newblock


\bibitem[Han(2018)]%
        {hanMachineTranslationEvaluation2018}
\bibfield{author}{\bibinfo{person}{Lifeng Han}.}
  \bibinfo{year}{2018}\natexlab{}.
\newblock \showarticletitle{Machine {{Translation Evaluation Resources}} and
  {{Methods}}: {{A Survey}}}.
\newblock \bibinfo{journal}{\emph{arxiv:1605.04515}} (\bibinfo{date}{Sept.}
  \bibinfo{year}{2018}).
\newblock


\bibitem[Heger et~al\mbox{.}(2022)]%
        {hegerUnderstandingMachineLearning2022}
\bibfield{author}{\bibinfo{person}{Amy~K. Heger}, \bibinfo{person}{Liz~B.
  Marquis}, \bibinfo{person}{Mihaela Vorvoreanu}, \bibinfo{person}{Hanna
  Wallach}, {and} \bibinfo{person}{Jennifer~Wortman Vaughan}.}
  \bibinfo{year}{2022}\natexlab{}.
\newblock \showarticletitle{Understanding {{Machine Learning Practitioners}}'
  {{Data Documentation Perceptions}}, {{Needs}}, {{Challenges}}, and
  {{Desiderata}}}.
\newblock \bibinfo{journal}{\emph{arxiv:2206.02923}} (\bibinfo{date}{Aug.}
  \bibinfo{year}{2022}).
\newblock


\bibitem[Helcl and Libovick{\'y}(2017)]%
        {helclNeuralMonkeyOpensource2017}
\bibfield{author}{\bibinfo{person}{Jind{\v r}ich Helcl} {and}
  \bibinfo{person}{Jind{\v r}ich Libovick{\'y}}.}
  \bibinfo{year}{2017}\natexlab{}.
\newblock \showarticletitle{Neural Monkey: {{An}} Open-Source Tool for Sequence
  Learning}.
\newblock \bibinfo{journal}{\emph{The Prague Bulletin of Mathematical
  Linguistics}} (\bibinfo{year}{2017}).
\newblock
\urldef\tempurl%
\url{https://doi.org/10.1515/pralin-2017-0001}
\showDOI{\tempurl}


\bibitem[Hohman et~al\mbox{.}(2019)]%
        {hohmanVisualAnalyticsDeep2019}
\bibfield{author}{\bibinfo{person}{Fred Hohman}, \bibinfo{person}{Minsuk
  Kahng}, \bibinfo{person}{Robert Pienta}, {and} \bibinfo{person}{Duen~Horng
  Chau}.} \bibinfo{year}{2019}\natexlab{}.
\newblock \showarticletitle{Visual {{Analytics}} in {{Deep Learning}}: {{An
  Interrogative Survey}} for the {{Next Frontiers}}}.
\newblock \bibinfo{journal}{\emph{IEEE Transactions on Visualization and
  Computer Graphics}}  \bibinfo{volume}{25} (\bibinfo{date}{Aug.}
  \bibinfo{year}{2019}).
\newblock
\urldef\tempurl%
\url{https://doi.org/10.1109/tvcg.2018.2843369}
\showDOI{\tempurl}


\bibitem[Holstein et~al\mbox{.}(2019)]%
        {holsteinImprovingFairnessMachine2019b}
\bibfield{author}{\bibinfo{person}{Kenneth Holstein}, \bibinfo{person}{Jennifer
  Wortman~Vaughan}, \bibinfo{person}{Hal Daum{\'e}~III}, \bibinfo{person}{Miro
  Dudik}, {and} \bibinfo{person}{Hanna Wallach}.}
  \bibinfo{year}{2019}\natexlab{}.
\newblock \showarticletitle{Improving Fairness in Machine Learning Systems:
  {{What}} Do Industry Practitioners Need?}. In
  \bibinfo{booktitle}{\emph{Proceedings of the 2019 {{CHI}} Conference on Human
  Factors in Computing Systems}}.
\newblock
\urldef\tempurl%
\url{https://doi.org/10.1145/3290605.3300830}
\showDOI{\tempurl}


\bibitem[Hooker et~al\mbox{.}(2020)]%
        {hookerCharacterisingBiasCompressed2020}
\bibfield{author}{\bibinfo{person}{Sara Hooker}, \bibinfo{person}{Nyalleng
  Moorosi}, \bibinfo{person}{Gregory Clark}, \bibinfo{person}{Samy Bengio},
  {and} \bibinfo{person}{Emily Denton}.} \bibinfo{year}{2020}\natexlab{}.
\newblock \showarticletitle{Characterising {{Bias}} in {{Compressed Models}}}.
\newblock \bibinfo{journal}{\emph{arxiv:2010.03058}} (\bibinfo{date}{Dec.}
  \bibinfo{year}{2020}).
\newblock


\bibitem[Hoover et~al\mbox{.}(2020)]%
        {hooverExBERTVisualAnalysis2020}
\bibfield{author}{\bibinfo{person}{Benjamin Hoover}, \bibinfo{person}{Hendrik
  Strobelt}, {and} \bibinfo{person}{Sebastian Gehrmann}.}
  \bibinfo{year}{2020}\natexlab{}.
\newblock \showarticletitle{{{exBERT}}: {{A Visual Analysis Tool}} to {{Explore
  Learned Representations}} in {{Transformer Models}}}. In
  \bibinfo{booktitle}{\emph{Proceedings of the 58th Annual Meeting of the
  Association for Computational Linguistics: {{System}} Demonstrations}}.
\newblock
\urldef\tempurl%
\url{https://doi.org/10.18653/v1/2020.acl-demos.22}
\showDOI{\tempurl}


\bibitem[Hopkins et~al\mbox{.}(2023)]%
        {hopkins2023designing}
\bibfield{author}{\bibinfo{person}{Aspen Hopkins}, \bibinfo{person}{Fred
  Hohman}, \bibinfo{person}{Luca Zappella}, \bibinfo{person}{Xavier~Suau
  Cuadros}, {and} \bibinfo{person}{Dominik Moritz}.}
  \bibinfo{year}{2023}\natexlab{}.
\newblock \showarticletitle{Designing data: Proactive data collection and
  iteration for machine learning}.
\newblock \bibinfo{journal}{\emph{arXiv preprint arXiv:2301.10319}}
  (\bibinfo{year}{2023}).
\newblock


\bibitem[Huyen(2022)]%
        {huyenDesigningMachineLearning2022}
\bibfield{author}{\bibinfo{person}{Chip Huyen}.}
  \bibinfo{year}{2022}\natexlab{}.
\newblock \bibinfo{booktitle}{\emph{Designing Machine Learning Systems: An
  Iterative Process for Production-Ready Applications} (\bibinfo{edition}{first
  edition} ed.)}.
\newblock


\bibitem[Ipeirotis et~al\mbox{.}(2010)]%
        {ipeirotisQualityManagementAmazon2010}
\bibfield{author}{\bibinfo{person}{Panagiotis~G. Ipeirotis},
  \bibinfo{person}{Foster Provost}, {and} \bibinfo{person}{Jing Wang}.}
  \bibinfo{year}{2010}\natexlab{}.
\newblock \showarticletitle{Quality Management on {{Amazon Mechanical Turk}}}.
  In \bibinfo{booktitle}{\emph{Proceedings of the {{ACM SIGKDD Workshop}} on
  {{Human Computation}} - {{HCOMP}} '10}}.
\newblock
\urldef\tempurl%
\url{https://doi.org/10.1145/1837885.1837906}
\showDOI{\tempurl}


\bibitem[Isabelle et~al\mbox{.}(2017)]%
        {isabelleChallengeSetApproach2017}
\bibfield{author}{\bibinfo{person}{Pierre Isabelle}, \bibinfo{person}{Colin
  Cherry}, {and} \bibinfo{person}{George Foster}.}
  \bibinfo{year}{2017}\natexlab{}.
\newblock \showarticletitle{A Challenge Set Approach to Evaluating Machine
  Translation}. In \bibinfo{booktitle}{\emph{Proceedings of the 2017 Conference
  on Empirical Methods in Natural Language Processing}}.
\newblock
\urldef\tempurl%
\url{https://doi.org/10.18653/v1/d17-1263}
\showDOI{\tempurl}


\bibitem[Johnson et~al\mbox{.}(2017)]%
        {johnson2017multilingual}
\bibfield{author}{\bibinfo{person}{Melvin Johnson}, \bibinfo{person}{Mike
  Schuster}, \bibinfo{person}{Quoc~V. Le}, \bibinfo{person}{Maxim Krikun},
  \bibinfo{person}{Yonghui Wu}, \bibinfo{person}{Zhifeng Chen},
  \bibinfo{person}{Nikhil Thorat}, \bibinfo{person}{Fernanda Viégas},
  \bibinfo{person}{Martin Wattenberg}, \bibinfo{person}{Greg Corrado},
  \bibinfo{person}{Macduff Hughes}, {and} \bibinfo{person}{Jeffrey Dean}.}
  \bibinfo{year}{2017}\natexlab{}.
\newblock \showarticletitle{Google’s Multilingual Neural Machine Translation
  System: Enabling Zero-Shot Translation}.
\newblock \bibinfo{journal}{\emph{Transactions of the Association for
  Computational Linguistics}} (\bibinfo{year}{2017}),
  \bibinfo{pages}{339–351}.
\newblock
\urldef\tempurl%
\url{https://doi.org/10.1162/tacl_a_00065}
\showDOI{\tempurl}


\bibitem[Joshi et~al\mbox{.}(2022)]%
        {joshiFairSASensitivity2022}
\bibfield{author}{\bibinfo{person}{Aparna~R Joshi},
  \bibinfo{person}{Xavier~Suau Cuadros}, \bibinfo{person}{Nivedha Sivakumar},
  \bibinfo{person}{Luca Zappella}, {and} \bibinfo{person}{Nicholas
  Apostoloff}.} \bibinfo{year}{2022}\natexlab{}.
\newblock \showarticletitle{Fair {{SA}}: {{Sensitivity}} Analysis for Fairness
  in Face Recognition}. In \bibinfo{booktitle}{\emph{Algorithmic Fairness
  through the Lens of Causality and Robustness Workshop}}. {PMLR}.
\newblock


\bibitem[Khoong et~al\mbox{.}(2019)]%
        {khoongAssessingUseGoogle2019}
\bibfield{author}{\bibinfo{person}{Elaine~C Khoong}, \bibinfo{person}{Eric
  Steinbrook}, \bibinfo{person}{Cortlyn Brown}, {and} \bibinfo{person}{Alicia
  Fernandez}.} \bibinfo{year}{2019}\natexlab{}.
\newblock \showarticletitle{Assessing the Use of {{Google Translate}} for
  {{Spanish}} and {{Chinese}} Translations of Emergency Department Discharge
  Instructions}.
\newblock \bibinfo{journal}{\emph{JAMA internal medicine}}
  \bibinfo{volume}{179} (\bibinfo{year}{2019}).
\newblock
\urldef\tempurl%
\url{https://doi.org/10.1001/jamainternmed.2018.7653}
\showDOI{\tempurl}


\bibitem[Kiela et~al\mbox{.}(2021)]%
        {kielaDynabenchRethinkingBenchmarking2021}
\bibfield{author}{\bibinfo{person}{Douwe Kiela}, \bibinfo{person}{Max Bartolo},
  \bibinfo{person}{Yixin Nie}, \bibinfo{person}{Divyansh Kaushik},
  \bibinfo{person}{Atticus Geiger}, \bibinfo{person}{Zhengxuan Wu},
  \bibinfo{person}{Bertie Vidgen}, \bibinfo{person}{Grusha Prasad},
  \bibinfo{person}{Amanpreet Singh}, \bibinfo{person}{Pratik Ringshia},
  \bibinfo{person}{Zhiyi Ma}, \bibinfo{person}{Tristan Thrush},
  \bibinfo{person}{Sebastian Riedel}, \bibinfo{person}{Zeerak Waseem},
  \bibinfo{person}{Pontus Stenetorp}, \bibinfo{person}{Robin Jia},
  \bibinfo{person}{Mohit Bansal}, \bibinfo{person}{Christopher Potts}, {and}
  \bibinfo{person}{Adina Williams}.} \bibinfo{year}{2021}\natexlab{}.
\newblock \showarticletitle{Dynabench: {{Rethinking Benchmarking}} in {{NLP}}}.
  In \bibinfo{booktitle}{\emph{Proceedings of the 2021 {{Conference}} of the
  {{North American Chapter}} of the {{Association}} for {{Computational
  Linguistics}}: {{Human Language Technologies}}}}.
\newblock
\urldef\tempurl%
\url{https://doi.org/10.18653/v1/2021.naacl-main.324}
\showDOI{\tempurl}


\bibitem[King and Maeggard(1998)]%
        {kingIssuesNaturalLanguage1998}
\bibfield{author}{\bibinfo{person}{Margaret King} {and} \bibinfo{person}{Bente
  Maeggard}.} \bibinfo{year}{1998}\natexlab{}.
\newblock \showarticletitle{Issues in Natural Language Systems Evaluation.}. In
  \bibinfo{booktitle}{\emph{{{LREC}}}}.
\newblock


\bibitem[Kit and Wong(2008)]%
        {kitComparativeEvaluationOnline2008}
\bibfield{author}{\bibinfo{person}{Chunyu Kit} {and} \bibinfo{person}{Tak~Ming
  Wong}.} \bibinfo{year}{2008}\natexlab{}.
\newblock \showarticletitle{Comparative Evaluation of Online Machine
  Translation Systems with Legal Texts}.
\newblock \bibinfo{journal}{\emph{Law Library Journal}}  \bibinfo{volume}{100}
  (\bibinfo{year}{2008}).
\newblock


\bibitem[Klejch et~al\mbox{.}(2015)]%
        {klejchMTComparEvalGraphicalEvaluation2015}
\bibfield{author}{\bibinfo{person}{Ondrej Klejch}, \bibinfo{person}{Eleftherios
  Avramidis}, \bibinfo{person}{Aljoscha Burchardt}, {and}
  \bibinfo{person}{Martin Popel}.} \bibinfo{year}{2015}\natexlab{}.
\newblock \showarticletitle{{{MT-ComparEval}}: {{Graphical}} Evaluation
  Interface for Machine Translation Development.}
\newblock \bibinfo{journal}{\emph{Prague Bull. Math. Linguistics}}
  \bibinfo{volume}{104} (\bibinfo{year}{2015}).
\newblock
\urldef\tempurl%
\url{https://doi.org/10.1515/pralin-2015-0014}
\showDOI{\tempurl}


\bibitem[Koehn(2007)]%
        {koehnEuroMatrixMachineTranslation2007}
\bibfield{author}{\bibinfo{person}{Philipp Koehn}.} \bibinfo{year}{09 10-14
  2007}\natexlab{}.
\newblock \showarticletitle{{{EuroMatrix}} \textendash{} Machine Translation
  for All {{European}} Languages}. In \bibinfo{booktitle}{\emph{Proceedings of
  Machine Translation Summit {{XI}}: {{Invited}} Papers}}.
\newblock


\bibitem[Koenecke et~al\mbox{.}(2020)]%
        {koeneckeRacialDisparitiesAutomated2020}
\bibfield{author}{\bibinfo{person}{Allison Koenecke}, \bibinfo{person}{Andrew
  Nam}, \bibinfo{person}{Emily Lake}, \bibinfo{person}{Joe Nudell},
  \bibinfo{person}{Minnie Quartey}, \bibinfo{person}{Zion Mengesha},
  \bibinfo{person}{Connor Toups}, \bibinfo{person}{John~R. Rickford},
  \bibinfo{person}{Dan Jurafsky}, {and} \bibinfo{person}{Sharad Goel}.}
  \bibinfo{year}{2020}\natexlab{}.
\newblock \showarticletitle{Racial Disparities in Automated Speech
  Recognition}.
\newblock \bibinfo{journal}{\emph{Proceedings of the National Academy of
  Sciences}}  \bibinfo{volume}{117} (\bibinfo{date}{April}
  \bibinfo{year}{2020}).
\newblock
\urldef\tempurl%
\url{https://doi.org/10.1073/pnas.1915768117}
\showDOI{\tempurl}


\bibitem[Koh et~al\mbox{.}(2021)]%
        {kohWILDSBenchmarkIntheWild2021}
\bibfield{author}{\bibinfo{person}{Pang~Wei Koh}, \bibinfo{person}{Shiori
  Sagawa}, \bibinfo{person}{Henrik Marklund}, \bibinfo{person}{Sang~Michael
  Xie}, \bibinfo{person}{Marvin Zhang}, \bibinfo{person}{Akshay Balsubramani},
  \bibinfo{person}{Weihua Hu}, \bibinfo{person}{Michihiro Yasunaga},
  \bibinfo{person}{Richard~Lanas Phillips}, \bibinfo{person}{Irena Gao},
  \bibinfo{person}{Tony Lee}, \bibinfo{person}{Etienne David},
  \bibinfo{person}{Ian Stavness}, \bibinfo{person}{Wei Guo},
  \bibinfo{person}{Berton Earnshaw}, \bibinfo{person}{Imran Haque},
  \bibinfo{person}{Sara~M Beery}, \bibinfo{person}{Jure Leskovec},
  \bibinfo{person}{Anshul Kundaje}, \bibinfo{person}{Emma Pierson},
  \bibinfo{person}{Sergey Levine}, \bibinfo{person}{Chelsea Finn}, {and}
  \bibinfo{person}{Percy Liang}.} \bibinfo{year}{2021}\natexlab{}.
\newblock \showarticletitle{{{WILDS}}: {{A}} Benchmark of in-the-{{Wild}}
  Distribution Shifts}. In \bibinfo{booktitle}{\emph{ICML}}
  \emph{(\bibinfo{series}{Proceedings of Machine Learning Research},
  Vol.~\bibinfo{volume}{139})}.
\newblock


\bibitem[Krishnakumar et~al\mbox{.}(2021)]%
        {krishnakumarUDISUnsupervisedDiscovery2021}
\bibfield{author}{\bibinfo{person}{Arvindkumar Krishnakumar},
  \bibinfo{person}{Viraj Prabhu}, \bibinfo{person}{Sruthi Sudhakar}, {and}
  \bibinfo{person}{Judy Hoffman}.} \bibinfo{year}{2021}\natexlab{}.
\newblock \showarticletitle{{{UDIS}}: {{Unsupervised Discovery}} of {{Bias}} in
  {{Deep Visual Recognition Models}}}.
\newblock \bibinfo{journal}{\emph{arxiv:2110.15499}} (\bibinfo{date}{Oct.}
  \bibinfo{year}{2021}).
\newblock


\bibitem[Lee et~al\mbox{.}(2017)]%
        {leeInteractiveVisualizationManipulation2017}
\bibfield{author}{\bibinfo{person}{Jaesong Lee}, \bibinfo{person}{Joong-Hwi
  Shin}, {and} \bibinfo{person}{Jun-Seok Kim}.}
  \bibinfo{year}{2017}\natexlab{}.
\newblock \showarticletitle{Interactive {{Visualization}} and {{Manipulation}}
  of {{Attention-based Neural Machine Translation}}}. In
  \bibinfo{booktitle}{\emph{Proceedings of the 2017 {{Conference}} on
  {{Empirical Methods}} in {{Natural}} {{Language Processing}}: {{System
  Demonstrations}}}}.
\newblock


\bibitem[Lee et~al\mbox{.}(2022)]%
        {leeVisCUITVisualAuditor2022}
\bibfield{author}{\bibinfo{person}{Seongmin Lee}, \bibinfo{person}{Zijie~J.
  Wang}, \bibinfo{person}{Judy Hoffman}, {and} \bibinfo{person}{Duen~Horng
  Chau}.} \bibinfo{year}{2022}\natexlab{}.
\newblock \showarticletitle{{{VisCUIT}}: {{Visual}} Auditor for Bias in {{CNN}}
  Image Classifier}. In \bibinfo{booktitle}{\emph{Proceedings of the
  {{IEEE}}/{{CVF}} Conference on Computer Vision and Pattern Recognition
  ({{CVPR}})}}.
\newblock


\bibitem[Li et~al\mbox{.}(2022)]%
        {liUnifiedUnderstandingDeep2022}
\bibfield{author}{\bibinfo{person}{Zhen Li}, \bibinfo{person}{Xiting Wang},
  \bibinfo{person}{Weikai Yang}, \bibinfo{person}{Jing Wu},
  \bibinfo{person}{Zhengyan Zhang}, \bibinfo{person}{Zhiyuan Liu},
  \bibinfo{person}{Maosong Sun}, \bibinfo{person}{Hui Zhang}, {and}
  \bibinfo{person}{Shixia Liu}.} \bibinfo{year}{2022}\natexlab{}.
\newblock \showarticletitle{A {{Unified Understanding}} of {{Deep NLP Models}}
  for {{Text Classification}}}.
\newblock \bibinfo{journal}{\emph{IEEE Transactions on Visualization and
  Computer Graphics}} (\bibinfo{year}{2022}).
\newblock
\urldef\tempurl%
\url{https://doi.org/10.1109/tvcg.2022.3184186}
\showDOI{\tempurl}


\bibitem[Liebling et~al\mbox{.}(2020)]%
        {liebling2020unmet}
\bibfield{author}{\bibinfo{person}{Daniel~J. Liebling}, \bibinfo{person}{Michal
  Lahav}, \bibinfo{person}{Abigail Evans}, \bibinfo{person}{Aaron Donsbach},
  \bibinfo{person}{Jess Holbrook}, \bibinfo{person}{Boris Smus}, {and}
  \bibinfo{person}{Lindsey Boran}.} \bibinfo{year}{2020}\natexlab{}.
\newblock \showarticletitle{Unmet Needs and Opportunities for Mobile
  Translation AI}. In \bibinfo{booktitle}{\emph{Proceedings of the 2020 CHI
  Conference on Human Factors in Computing Systems}} (Honolulu, HI, USA)
  \emph{(\bibinfo{series}{CHI '20})}. \bibinfo{publisher}{Association for
  Computing Machinery}, \bibinfo{address}{New York, NY, USA},
  \bibinfo{pages}{1–13}.
\newblock
\showISBNx{9781450367080}
\urldef\tempurl%
\url{https://doi.org/10.1145/3313831.3376261}
\showDOI{\tempurl}


\bibitem[Lin and Och(2004)]%
        {linORANGEMethodEvaluating2004}
\bibfield{author}{\bibinfo{person}{Chin-Yew Lin} {and}
  \bibinfo{person}{Franz~Josef Och}.} \bibinfo{year}{2004}\natexlab{}.
\newblock \showarticletitle{{{ORANGE}}: A Method for Evaluating Automatic
  Evaluation Metrics for Machine Translation}. In
  \bibinfo{booktitle}{\emph{{{COLING}} 2004: {{Proceedings}} of the 20th
  International Conference on Computational Linguistics}}.
\newblock
\urldef\tempurl%
\url{https://doi.org/10.3115/1220355.1220427}
\showDOI{\tempurl}


\bibitem[Liu et~al\mbox{.}(2018)]%
        {liuVisualDiagnosisTree2018}
\bibfield{author}{\bibinfo{person}{Shixia Liu}, \bibinfo{person}{Jiannan Xiao},
  \bibinfo{person}{Junlin Liu}, \bibinfo{person}{Xiting Wang},
  \bibinfo{person}{Jing Wu}, {and} \bibinfo{person}{Jun Zhu}.}
  \bibinfo{year}{2018}\natexlab{}.
\newblock \showarticletitle{Visual {{Diagnosis}} of {{Tree Boosting Methods}}}.
\newblock \bibinfo{journal}{\emph{IEEE TVCG}}  \bibinfo{volume}{24}
  (\bibinfo{date}{Jan.} \bibinfo{year}{2018}).
\newblock
\urldef\tempurl%
\url{https://doi.org/10.1109/tvcg.2017.2744378}
\showDOI{\tempurl}


\bibitem[Loper and Bird(2002)]%
        {loperNLTKNaturalLanguage2002}
\bibfield{author}{\bibinfo{person}{Edward Loper} {and} \bibinfo{person}{Steven
  Bird}.} \bibinfo{year}{2002}\natexlab{}.
\newblock \showarticletitle{{{NLTK}}: The {{Natural Language Toolkit}}}. In
  \bibinfo{booktitle}{\emph{Proceedings of the {{ACL-02 Workshop}} on
  {{Effective}} Tools and Methodologies for Teaching Natural Language
  Processing and Computational Linguistics -}}, Vol.~\bibinfo{volume}{1}.
\newblock
\urldef\tempurl%
\url{https://doi.org/10.3115/1219044.1219075}
\showDOI{\tempurl}


\bibitem[Lovering and Pavlick(2022)]%
        {loveringUnitTestingConcepts2022}
\bibfield{author}{\bibinfo{person}{Charles Lovering} {and}
  \bibinfo{person}{Ellie Pavlick}.} \bibinfo{year}{2022}\natexlab{}.
\newblock \showarticletitle{Unit {{Testing}} for {{Concepts}} in {{Neural
  Networks}}}.
\newblock \bibinfo{journal}{\emph{arxiv:2208.10244}} (\bibinfo{date}{July}
  \bibinfo{year}{2022}).
\newblock


\bibitem[Lundberg and Lee(2017)]%
        {lundbergUnifiedApproachInterpreting2017}
\bibfield{author}{\bibinfo{person}{Scott~M. Lundberg} {and}
  \bibinfo{person}{Su-In Lee}.} \bibinfo{year}{2017}\natexlab{}.
\newblock \showarticletitle{A Unified Approach to Interpreting Model
  Predictions}. In \bibinfo{booktitle}{\emph{Proceedings of the 31st
  International Conference on Neural Information Processing Systems}}
  \emph{(\bibinfo{series}{{{NIPS}}'17})}.
\newblock


\bibitem[M and M.N(2015)]%
        {mReviewEvaluationMetrics2015}
\bibfield{author}{\bibinfo{person}{Hossin M} {and} \bibinfo{person}{Sulaiman
  M.N}.} \bibinfo{year}{2015}\natexlab{}.
\newblock \showarticletitle{A {{Review}} on {{Evaluation Metrics}} for {{Data
  Classification Evaluations}}}.
\newblock \bibinfo{journal}{\emph{International Journal of Data Mining \&
  Knowledge Management Process}}  \bibinfo{volume}{5} (\bibinfo{date}{March}
  \bibinfo{year}{2015}).
\newblock
\urldef\tempurl%
\url{https://doi.org/10.5121/ijdkp.2015.5201}
\showDOI{\tempurl}


\bibitem[Macketanz et~al\mbox{.}(2018)]%
        {macketanzFinegrainedEvaluationGermanEnglish2018}
\bibfield{author}{\bibinfo{person}{Vivien Macketanz},
  \bibinfo{person}{Eleftherios Avramidis}, \bibinfo{person}{Aljoscha
  Burchardt}, {and} \bibinfo{person}{Hans Uszkoreit}.}
  \bibinfo{year}{2018}\natexlab{}.
\newblock \showarticletitle{Fine-Grained Evaluation of {{German-English}}
  Machine Translation Based on a Test Suite}. In
  \bibinfo{booktitle}{\emph{Proceedings of the Third Conference on Machine
  Translation: {{Shared}} Task Papers}}.
\newblock
\urldef\tempurl%
\url{https://doi.org/10.18653/v1/w18-6436}
\showDOI{\tempurl}


\bibitem[Madnani(2011)]%
        {madnaniIBLEUInteractivelyDebugging2011}
\bibfield{author}{\bibinfo{person}{Nitin Madnani}.}
  \bibinfo{year}{2011}\natexlab{}.
\newblock \showarticletitle{{{iBLEU}}: {{Interactively Debugging}} and
  {{Scoring Statistical Machine Translation Systems}}}. In
  \bibinfo{booktitle}{\emph{2011 {{IEEE Fifth International Conference}} on
  {{Semantic Computing}}}}.
\newblock
\urldef\tempurl%
\url{https://doi.org/10.1109/icsc.2011.36}
\showDOI{\tempurl}


\bibitem[Mathur et~al\mbox{.}(2020)]%
        {mathurTangledBLEUReevaluating2020}
\bibfield{author}{\bibinfo{person}{Nitika Mathur}, \bibinfo{person}{Timothy
  Baldwin}, {and} \bibinfo{person}{Trevor Cohn}.}
  \bibinfo{year}{2020}\natexlab{}.
\newblock \showarticletitle{Tangled up in {{BLEU}}: {{Reevaluating}} the
  Evaluation of Automatic Machine Translation Evaluation Metrics}. In
  \bibinfo{booktitle}{\emph{Proceedings of the 58th Annual Meeting of the
  Association for Computational Linguistics}}.
\newblock
\urldef\tempurl%
\url{https://doi.org/10.18653/v1/2020.acl-main.448}
\showDOI{\tempurl}


\bibitem[McInnes et~al\mbox{.}(2018)]%
        {mcinnesUMAPUniformManifold2018}
\bibfield{author}{\bibinfo{person}{L. McInnes}, \bibinfo{person}{J. Healy},
  {and} \bibinfo{person}{J. Melville}.} \bibinfo{year}{2018}\natexlab{}.
\newblock \showarticletitle{{{UMAP}}: {{Uniform}} Manifold Approximation and
  Projection for Dimension Reduction}.
\newblock \bibinfo{journal}{\emph{ArXiv e-prints}} (\bibinfo{date}{Feb.}
  \bibinfo{year}{2018}).
\newblock


\bibitem[Merriam and Associates(2002)]%
        {qualresearch}
\bibfield{author}{\bibinfo{person}{Sharan~B Merriam} {and}
  \bibinfo{person}{Associates}.} \bibinfo{year}{2002}\natexlab{}.
\newblock \showarticletitle{Introduction to qualitative research}.
\newblock In \bibinfo{booktitle}{\emph{Qualitative research in practice:
  Examples for discussion and analysis}}. \bibinfo{publisher}{Jossey-Bass},
  \bibinfo{address}{Hoboken, NJ, USA}, \bibinfo{pages}{1--17}.
\newblock


\bibitem[Mor{\'e} and Climent(2014)]%
        {moreMachineTranslationnessMachinelikeness2014}
\bibfield{author}{\bibinfo{person}{Joaquim Mor{\'e}} {and}
  \bibinfo{person}{Salvador Climent}.} \bibinfo{year}{2014}\natexlab{}.
\newblock \showarticletitle{Machine Translationness: {{Machine-likeness}} in
  Machine Translation Evaluation}. In \bibinfo{booktitle}{\emph{Proceedings of
  the Ninth International Conference on Language Resources and Evaluation
  ({{LREC}}'14)}}.
\newblock


\bibitem[Munechika et~al\mbox{.}(2022)]%
        {munechikaVisualAuditorInteractive2022}
\bibfield{author}{\bibinfo{person}{David Munechika}, \bibinfo{person}{Zijie~J.
  Wang}, \bibinfo{person}{Jack Reidy}, \bibinfo{person}{Josh Rubin},
  \bibinfo{person}{Krishna Gade}, \bibinfo{person}{Krishnaram Kenthapadi},
  {and} \bibinfo{person}{Duen~Horng Chau}.} \bibinfo{year}{2022}\natexlab{}.
\newblock \showarticletitle{Visual {{Auditor}}: {{Interactive Visualization}}
  for {{Detection}} and {{Summarization}} of {{Model Biases}}}. In
  \bibinfo{booktitle}{\emph{2022 {{IEEE Visualization Conference}} ({{VIS}})}}.
\newblock


\bibitem[Munz et~al\mbox{.}(2022)]%
        {munzVisualizationbasedImprovementNeural2022}
\bibfield{author}{\bibinfo{person}{Tanja Munz}, \bibinfo{person}{Dirk
  V{\"a}th}, \bibinfo{person}{Paul Kuznecov}, \bibinfo{person}{Ngoc~Thang Vu},
  {and} \bibinfo{person}{Daniel Weiskopf}.} \bibinfo{year}{2022}\natexlab{}.
\newblock \showarticletitle{Visualization-Based Improvement of Neural Machine
  Translation}.
\newblock \bibinfo{journal}{\emph{Computers \& Graphics}}
  \bibinfo{volume}{103} (\bibinfo{date}{April} \bibinfo{year}{2022}).
\newblock
\urldef\tempurl%
\url{https://doi.org/10.1016/j.cag.2021.12.003}
\showDOI{\tempurl}


\bibitem[Naik et~al\mbox{.}(2018)]%
        {naik2018stress}
\bibfield{author}{\bibinfo{person}{Aakanksha Naik}, \bibinfo{person}{Abhilasha
  Ravichander}, \bibinfo{person}{Norman Sadeh}, \bibinfo{person}{Carolyn Rose},
  {and} \bibinfo{person}{Graham Neubig}.} \bibinfo{year}{2018}\natexlab{}.
\newblock \showarticletitle{Stress Test Evaluation for Natural Language
  Inference}. In \bibinfo{booktitle}{\emph{The 27th International Conference on
  Computational Linguistics ({{COLING}})}}.
\newblock


\bibitem[Neubig et~al\mbox{.}(2019)]%
        {neubigComparemtToolHolistic2019}
\bibfield{author}{\bibinfo{person}{Graham Neubig}, \bibinfo{person}{Zi-Yi Dou},
  \bibinfo{person}{Junjie Hu}, \bibinfo{person}{Paul Michel},
  \bibinfo{person}{Danish Pruthi}, {and} \bibinfo{person}{Xinyi Wang}.}
  \bibinfo{year}{2019}\natexlab{}.
\newblock \showarticletitle{Compare-Mt: {{A Tool}} for {{Holistic Comparison}}
  of {{Language Generation Systems}}}. In \bibinfo{booktitle}{\emph{Proceedings
  of the 2019 {{Conference}} of the {{North}}}}.
\newblock
\urldef\tempurl%
\url{https://doi.org/10.18653/v1/n19-4007}
\showDOI{\tempurl}


\bibitem[Obermeyer et~al\mbox{.}(2019)]%
        {obermeyerDissectingRacialBias2019}
\bibfield{author}{\bibinfo{person}{Ziad Obermeyer}, \bibinfo{person}{Brian
  Powers}, \bibinfo{person}{Christine Vogeli}, {and} \bibinfo{person}{Sendhil
  Mullainathan}.} \bibinfo{year}{2019}\natexlab{}.
\newblock \showarticletitle{Dissecting Racial Bias in an Algorithm Used to
  Manage the Health of Populations}.
\newblock \bibinfo{journal}{\emph{Science}}  \bibinfo{volume}{366}
  (\bibinfo{date}{Oct.} \bibinfo{year}{2019}).
\newblock
\urldef\tempurl%
\url{https://doi.org/10.1126/science.aax2342}
\showDOI{\tempurl}


\bibitem[Ono et~al\mbox{.}(2021)]%
        {onoPipelineProfilerVisualAnalytics2021}
\bibfield{author}{\bibinfo{person}{Jorge~Piazentin Ono}, \bibinfo{person}{Sonia
  Castelo}, \bibinfo{person}{Roque Lopez}, \bibinfo{person}{Enrico Bertini},
  \bibinfo{person}{Juliana Freire}, {and} \bibinfo{person}{Claudio Silva}.}
  \bibinfo{year}{2021}\natexlab{}.
\newblock \showarticletitle{{{PipelineProfiler}}: {{A Visual Analytics Tool}}
  for the {{Exploration}} of {{AutoML Pipelines}}}.
\newblock \bibinfo{journal}{\emph{IEEE Transactions on Visualization and
  Computer Graphics}}  \bibinfo{volume}{27} (\bibinfo{date}{Feb.}
  \bibinfo{year}{2021}).
\newblock


\bibitem[Papineni et~al\mbox{.}(2002)]%
        {papineniBleuMethodAutomatic2002}
\bibfield{author}{\bibinfo{person}{Kishore Papineni}, \bibinfo{person}{Salim
  Roukos}, \bibinfo{person}{Todd Ward}, {and} \bibinfo{person}{Wei-Jing Zhu}.}
  \bibinfo{year}{2002}\natexlab{}.
\newblock \showarticletitle{Bleu: A Method for Automatic Evaluation of Machine
  Translation}. In \bibinfo{booktitle}{\emph{Proceedings of the 40th Annual
  Meeting of the {{Association}} for {{Computational Linguistics}}}}.
\newblock


\bibitem[Park et~al\mbox{.}(2022)]%
        {parkVANTVisualAnalytics2022}
\bibfield{author}{\bibinfo{person}{Sebeom Park}, \bibinfo{person}{Soohyun Lee},
  \bibinfo{person}{Youngtaek Kim}, \bibinfo{person}{Hyeon Jeon},
  \bibinfo{person}{Seokweon Jung}, \bibinfo{person}{Jinwook Bok}, {and}
  \bibinfo{person}{Jinwook Seo}.} \bibinfo{year}{2022}\natexlab{}.
\newblock \showarticletitle{{{VANT}}: {{A Visual Analytics System}} for
  {{Refining Parallel Corpora}} in {{Neural Machine Translation}}}. In
  \bibinfo{booktitle}{\emph{2022 {{IEEE}} 15th {{Pacific Visualization
  Symposium}} ({{PacificVis}})}}.
\newblock
\urldef\tempurl%
\url{https://doi.org/10.1109/pacificvis53943.2022.00029}
\showDOI{\tempurl}


\bibitem[Patel et~al\mbox{.}(2008)]%
        {patelInvestigatingStatisticalMachine2008}
\bibfield{author}{\bibinfo{person}{Kayur Patel}, \bibinfo{person}{James
  Fogarty}, \bibinfo{person}{James~A. Landay}, {and} \bibinfo{person}{Beverly
  Harrison}.} \bibinfo{year}{2008}\natexlab{}.
\newblock \showarticletitle{Investigating Statistical Machine Learning as a
  Tool for Software Development}. In \bibinfo{booktitle}{\emph{Proceeding of
  the Twenty-Sixth Annual {{CHI}} Conference on {{Human}} Factors in Computing
  Systems - {{CHI}} '08}}.
\newblock
\urldef\tempurl%
\url{https://doi.org/10.1145/1357054.1357160}
\showDOI{\tempurl}


\bibitem[Pearson(1901)]%
        {pearsonLinesPlanesClosest1901}
\bibfield{author}{\bibinfo{person}{Karl Pearson}.}
  \bibinfo{year}{1901}\natexlab{}.
\newblock \showarticletitle{On Lines and Planes of Closest Fit to Systems of
  Points in Space}.
\newblock \bibinfo{journal}{\emph{The London, Edinburgh, and Dublin
  Philosophical Magazine and Journal of Science}}  \bibinfo{volume}{2}
  (\bibinfo{year}{1901}).
\newblock
\urldef\tempurl%
\url{https://doi.org/10.1080/14786440109462720}
\showDOI{\tempurl}


\bibitem[Pedregosa et~al\mbox{.}(2011)]%
        {pedregosaScikitlearnMachineLearning2011}
\bibfield{author}{\bibinfo{person}{Fabian Pedregosa}, \bibinfo{person}{Ga{\"e}l
  Varoquaux}, \bibinfo{person}{Alexandre Gramfort}, \bibinfo{person}{Vincent
  Michel}, \bibinfo{person}{Bertrand Thirion}, \bibinfo{person}{Olivier
  Grisel}, \bibinfo{person}{Mathieu Blondel}, \bibinfo{person}{Peter
  Prettenhofer}, \bibinfo{person}{Ron Weiss}, \bibinfo{person}{Vincent
  Dubourg}, {et~al\mbox{.}}} \bibinfo{year}{2011}\natexlab{}.
\newblock \showarticletitle{Scikit-Learn: {{Machine}} Learning in Python}.
\newblock \bibinfo{journal}{\emph{the Journal of machine Learning research}}
  \bibinfo{volume}{12} (\bibinfo{year}{2011}).
\newblock


\bibitem[Polyzotis et~al\mbox{.}(2019)]%
        {polyzotisDataValidationMachine2019}
\bibfield{author}{\bibinfo{person}{Neoklis Polyzotis}, \bibinfo{person}{Martin
  Zinkevich}, \bibinfo{person}{Sudip Roy}, \bibinfo{person}{Eric Breck}, {and}
  \bibinfo{person}{Steven Whang}.} \bibinfo{year}{2019}\natexlab{}.
\newblock \showarticletitle{Data Validation for Machine Learning}. In
  \bibinfo{booktitle}{\emph{Proceedings of Machine Learning and Systems}},
  Vol.~\bibinfo{volume}{1}.
\newblock


\bibitem[Popovi{\'c}(2015)]%
        {popovicChrFCharacterNgram2015}
\bibfield{author}{\bibinfo{person}{Maja Popovi{\'c}}.}
  \bibinfo{year}{2015}\natexlab{}.
\newblock \showarticletitle{{{chrF}}: Character n-Gram {{F-score}} for
  Automatic {{MT}} Evaluation}. In \bibinfo{booktitle}{\emph{Proceedings of the
  Tenth Workshop on Statistical Machine Translation}}.
\newblock
\urldef\tempurl%
\url{https://doi.org/10.18653/v1/w15-3049}
\showDOI{\tempurl}


\bibitem[Popovi{\'c}(2021)]%
        {popovicAgreeDisagreeAnalysis2021}
\bibfield{author}{\bibinfo{person}{Maja Popovi{\'c}}.}
  \bibinfo{year}{2021}\natexlab{}.
\newblock \showarticletitle{Agree to Disagree: {{Analysis}} of Inter-Annotator
  Disagreements in Human Evaluation of Machine Translation Output}. In
  \bibinfo{booktitle}{\emph{Proceedings of the 25th Conference on Computational
  Natural Language Learning}}.
\newblock
\urldef\tempurl%
\url{https://doi.org/10.18653/v1/2021.conll-1.18}
\showDOI{\tempurl}


\bibitem[Popovi{\'c} and Castilho(2019)]%
        {popovicChallengeTestSets2019}
\bibfield{author}{\bibinfo{person}{Maja Popovi{\'c}} {and}
  \bibinfo{person}{Sheila Castilho}.} \bibinfo{year}{2019}\natexlab{}.
\newblock \showarticletitle{Challenge Test Sets for {{MT}} Evaluation}. In
  \bibinfo{booktitle}{\emph{Proceedings of Machine Translation Summit {{XVII}}:
  {{Tutorial}} Abstracts}}.
\newblock


\bibitem[Post(2018)]%
        {postCallClarityReporting2018}
\bibfield{author}{\bibinfo{person}{Matt Post}.}
  \bibinfo{year}{2018}\natexlab{}.
\newblock \showarticletitle{A Call for Clarity in Reporting {{BLEU}} Scores}.
  In \bibinfo{booktitle}{\emph{Proceedings of the Third Conference on Machine
  Translation: {{Research}} Papers}}.
\newblock
\urldef\tempurl%
\url{https://doi.org/10.18653/v1/w18-6319}
\showDOI{\tempurl}


\bibitem[Prabhu et~al\mbox{.}(2021)]%
        {prabhuDidTheyDirect2021}
\bibfield{author}{\bibinfo{person}{Vinay Prabhu}, \bibinfo{person}{Ryan
  Teehan}, \bibinfo{person}{Eniko Srivastava}, {and} \bibinfo{person}{Abdul
  Nimeri}.} \bibinfo{year}{2021}\natexlab{}.
\newblock \showarticletitle{Did They Direct the Violence or Admonish It? {{A}}
  Cautionary Tale on Contronomy, Androcentrism and Back-Translation Foibles}.
  In \bibinfo{booktitle}{\emph{{{AfricaNLP}} Workshop at {{EACL}}}}.
\newblock


\bibitem[Radford et~al\mbox{.}(2021)]%
        {radfordLearningTransferableVisual2021}
\bibfield{author}{\bibinfo{person}{Alec Radford}, \bibinfo{person}{Jong~Wook
  Kim}, \bibinfo{person}{Chris Hallacy}, \bibinfo{person}{Aditya Ramesh},
  \bibinfo{person}{Gabriel Goh}, \bibinfo{person}{Sandhini Agarwal},
  \bibinfo{person}{Girish Sastry}, \bibinfo{person}{Amanda Askell},
  \bibinfo{person}{Pamela Mishkin}, \bibinfo{person}{Jack Clark},
  \bibinfo{person}{Gretchen Krueger}, {and} \bibinfo{person}{Ilya Sutskever}.}
  \bibinfo{year}{2021}\natexlab{}.
\newblock \showarticletitle{Learning Transferable Visual Models from Natural
  Language Supervision}. In \bibinfo{booktitle}{\emph{Proceedings of the 38th
  International Conference on Machine Learning}}
  \emph{(\bibinfo{series}{Proceedings of Machine Learning Research},
  Vol.~\bibinfo{volume}{139})}.
\newblock
\urldef\tempurl%
\url{https://proceedings.mlr.press/v139/radford21a.html}
\showURL{%
\tempurl}


\bibitem[Raganato et~al\mbox{.}(2019)]%
        {raganatoMuCoWTestSuite2019}
\bibfield{author}{\bibinfo{person}{Alessandro Raganato}, \bibinfo{person}{Yves
  Scherrer}, {and} \bibinfo{person}{J{\"o}rg Tiedemann}.}
  \bibinfo{year}{2019}\natexlab{}.
\newblock \showarticletitle{The {{MuCoW}} Test Suite at {{WMT}} 2019:
  {{Automatically}} Harvested Multilingual Contrastive Word Sense
  Disambiguation Test Sets for Machine Translation}. In
  \bibinfo{booktitle}{\emph{Proceedings of the Fourth Conference on Machine
  Translation (Volume 2: {{Shared}} Task Papers, Day 1)}}.
\newblock
\urldef\tempurl%
\url{https://doi.org/10.18653/v1/w19-5354}
\showDOI{\tempurl}


\bibitem[Raji et~al\mbox{.}(2020)]%
        {rajiSavingFaceInvestigating2020}
\bibfield{author}{\bibinfo{person}{Inioluwa~Deborah Raji},
  \bibinfo{person}{Timnit Gebru}, \bibinfo{person}{Margaret Mitchell},
  \bibinfo{person}{Joy Buolamwini}, \bibinfo{person}{Joonseok Lee}, {and}
  \bibinfo{person}{Emily Denton}.} \bibinfo{year}{2020}\natexlab{}.
\newblock \showarticletitle{Saving Face: {{Investigating}} the Ethical Concerns
  of Facial Recognition Auditing}. In \bibinfo{booktitle}{\emph{Proceedings of
  the {{AAAI}}/{{ACM}} Conference on {{AI}}, Ethics, and Society}}
  \emph{(\bibinfo{series}{{{AIES}} '20})}.
\newblock
\urldef\tempurl%
\url{https://doi.org/10.1145/3375627.3375820}
\showDOI{\tempurl}


\bibitem[Rajpurkar et~al\mbox{.}(2018)]%
        {rajpurkarKnowWhatYou2018}
\bibfield{author}{\bibinfo{person}{Pranav Rajpurkar}, \bibinfo{person}{Robin
  Jia}, {and} \bibinfo{person}{Percy Liang}.} \bibinfo{year}{2018}\natexlab{}.
\newblock \showarticletitle{Know What You Don't Know: {{Unanswerable}}
  Questions for {{SQuAD}}}. In \bibinfo{booktitle}{\emph{Proceedings of the
  56th Annual Meeting of the Association for Computational Linguistics (Volume
  2: {{Short}} Papers)}}.
\newblock
\urldef\tempurl%
\url{https://doi.org/10.18653/v1/p18-2124}
\showDOI{\tempurl}


\bibitem[Rebuffi et~al\mbox{.}(2021)]%
        {rebuffiDataAugmentationCan2021}
\bibfield{author}{\bibinfo{person}{Sylvestre-Alvise Rebuffi},
  \bibinfo{person}{Sven Gowal}, \bibinfo{person}{Dan~Andrei Calian},
  \bibinfo{person}{Florian Stimberg}, \bibinfo{person}{Olivia Wiles}, {and}
  \bibinfo{person}{Timothy~A Mann}.} \bibinfo{year}{2021}\natexlab{}.
\newblock \showarticletitle{Data Augmentation Can Improve Robustness}. In
  \bibinfo{booktitle}{\emph{Advances in Neural Information Processing
  Systems}}, Vol.~\bibinfo{volume}{34}.
\newblock


\bibitem[Recht et~al\mbox{.}(2019)]%
        {rechtImageNetClassifiersGeneralize2019}
\bibfield{author}{\bibinfo{person}{Benjamin Recht}, \bibinfo{person}{Rebecca
  Roelofs}, \bibinfo{person}{Ludwig Schmidt}, {and} \bibinfo{person}{Vaishaal
  Shankar}.} \bibinfo{year}{2019}\natexlab{}.
\newblock \showarticletitle{Do {{ImageNet}} Classifiers Generalize to
  {{ImageNet}}?}. In \bibinfo{booktitle}{\emph{Proceedings of the 36th
  International Conference on Machine Learning}}
  \emph{(\bibinfo{series}{Proceedings of Machine Learning Research},
  Vol.~\bibinfo{volume}{97})}.
\newblock


\bibitem[Reimers and Gurevych(2019)]%
        {reimersSentenceBERTSentenceEmbeddings2019}
\bibfield{author}{\bibinfo{person}{Nils Reimers} {and} \bibinfo{person}{Iryna
  Gurevych}.} \bibinfo{year}{2019}\natexlab{}.
\newblock \showarticletitle{Sentence-{{BERT}}: {{Sentence Embeddings}} Using
  {{Siamese BERT-Networks}}}. In \bibinfo{booktitle}{\emph{EMNLP}}.
\newblock
\urldef\tempurl%
\url{https://doi.org/10.18653/v1/d19-1410}
\showDOI{\tempurl}


\bibitem[Reiter(2018)]%
        {reiterStructuredReviewValidity2018}
\bibfield{author}{\bibinfo{person}{Ehud Reiter}.}
  \bibinfo{year}{2018}\natexlab{}.
\newblock \showarticletitle{A Structured Review of the Validity of {{BLEU}}}.
\newblock \bibinfo{journal}{\emph{Computational Linguistics}}
  \bibinfo{volume}{44} (\bibinfo{date}{Sept.} \bibinfo{year}{2018}).
\newblock
\urldef\tempurl%
\url{https://doi.org/10.1162/coli_a_00322}
\showDOI{\tempurl}


\bibitem[Ribeiro and Lundberg(2022)]%
        {ribeiroAdaptiveTestingDebugging2022}
\bibfield{author}{\bibinfo{person}{Marco~Tulio Ribeiro} {and}
  \bibinfo{person}{Scott Lundberg}.} \bibinfo{year}{2022}\natexlab{}.
\newblock \showarticletitle{Adaptive Testing and Debugging of {{NLP}} Models}.
  In \bibinfo{booktitle}{\emph{Proceedings of the 60th Annual Meeting of the
  Association for Computational Linguistics (Volume 1: {{Long}} Papers)}}.
\newblock
\urldef\tempurl%
\url{https://doi.org/10.18653/v1/2022.acl-long.230}
\showDOI{\tempurl}


\bibitem[Ribeiro et~al\mbox{.}(2016)]%
        {ribeiroWhyShouldTrust2016}
\bibfield{author}{\bibinfo{person}{Marco~Tulio Ribeiro},
  \bibinfo{person}{Sameer Singh}, {and} \bibinfo{person}{Carlos Guestrin}.}
  \bibinfo{year}{2016}\natexlab{}.
\newblock \showarticletitle{"{{Why Should I Trust You}}?": {{Explaining}} the
  {{Predictions}} of {{Any Classifier}}}. In
  \bibinfo{booktitle}{\emph{Proceedings of the 22nd {{ACM SIGKDD International
  Conference}} on {{Knowledge Discovery}} and {{Data Mining}}}}.
\newblock
\urldef\tempurl%
\url{https://doi.org/10.1145/2939672.2939778}
\showDOI{\tempurl}


\bibitem[Ribeiro et~al\mbox{.}(2020)]%
        {ribeiroAccuracyBehavioralTesting2020}
\bibfield{author}{\bibinfo{person}{Marco~Tulio Ribeiro},
  \bibinfo{person}{Tongshuang Wu}, \bibinfo{person}{Carlos Guestrin}, {and}
  \bibinfo{person}{Sameer Singh}.} \bibinfo{year}{2020}\natexlab{}.
\newblock \showarticletitle{Beyond {{Accuracy}}: {{Behavioral Testing}} of
  {{NLP Models}} with {{CheckList}}}. In \bibinfo{booktitle}{\emph{Proceedings
  of the 58th {{Annual Meeting}} of the {{Association}} for {{Computational
  Linguistics}}}}.
\newblock
\urldef\tempurl%
\url{https://doi.org/10.18653/v1/2020.acl-main.442}
\showDOI{\tempurl}


\bibitem[Rikters et~al\mbox{.}(2017)]%
        {riktersVisualizingNeuralMachine2017}
\bibfield{author}{\bibinfo{person}{Mat{\=i}ss Rikters}, \bibinfo{person}{Mark
  Fishel}, {and} \bibinfo{person}{Ond{\v r}ej Bojar}.}
  \bibinfo{year}{2017}\natexlab{}.
\newblock \showarticletitle{Visualizing {{Neural Machine Translation
  Attention}} and {{Confidence}}}.
\newblock \bibinfo{journal}{\emph{The Prague Bulletin of Mathematical
  Linguistics}}  \bibinfo{volume}{109} (\bibinfo{date}{Oct.}
  \bibinfo{year}{2017}).
\newblock
\urldef\tempurl%
\url{https://doi.org/10.1515/pralin-2017-0037}
\showDOI{\tempurl}


\bibitem[R{\"o}ttger et~al\mbox{.}(2021)]%
        {rottgerHateCheckFunctionalTests2021}
\bibfield{author}{\bibinfo{person}{Paul R{\"o}ttger}, \bibinfo{person}{Bertie
  Vidgen}, \bibinfo{person}{Dong Nguyen}, \bibinfo{person}{Zeerak Waseem},
  \bibinfo{person}{Helen Margetts}, {and} \bibinfo{person}{Janet
  Pierrehumbert}.} \bibinfo{year}{2021}\natexlab{}.
\newblock \showarticletitle{{{HateCheck}}: {{Functional}} Tests for Hate Speech
  Detection Models}. In \bibinfo{booktitle}{\emph{Proceedings of the 59th
  Annual Meeting of the Association for Computational Linguistics and the 11th
  International Joint Conference on Natural Language Processing (Volume 1:
  {{Long}} Papers)}}.
\newblock
\urldef\tempurl%
\url{https://doi.org/10.18653/v1/2021.acl-long.4}
\showDOI{\tempurl}


\bibitem[Rudinger et~al\mbox{.}(2018)]%
        {rudingerGenderBiasCoreference2018}
\bibfield{author}{\bibinfo{person}{Rachel Rudinger}, \bibinfo{person}{Jason
  Naradowsky}, \bibinfo{person}{Brian Leonard}, {and} \bibinfo{person}{Benjamin
  Van~Durme}.} \bibinfo{year}{2018}\natexlab{}.
\newblock \showarticletitle{Gender Bias in Coreference Resolution}. In
  \bibinfo{booktitle}{\emph{Proceedings of the 2018 Conference of the North
  {{American}} Chapter of the Association for Computational Linguistics:
  {{Human}} Language Technologies, Volume 2 (Short Papers)}}.
\newblock
\urldef\tempurl%
\url{https://doi.org/10.18653/v1/n18-2002}
\showDOI{\tempurl}


\bibitem[Sai et~al\mbox{.}(2022)]%
        {saiSurveyEvaluationMetrics2022}
\bibfield{author}{\bibinfo{person}{Ananya~B. Sai}, \bibinfo{person}{Akash~Kumar
  Mohankumar}, {and} \bibinfo{person}{Mitesh~M. Khapra}.}
  \bibinfo{year}{2022}\natexlab{}.
\newblock \showarticletitle{A Survey of Evaluation Metrics Used for {{NLG}}
  Systems}.
\newblock \bibinfo{journal}{\emph{Acm Computing Surveys}}  \bibinfo{volume}{55}
  (\bibinfo{date}{Jan.} \bibinfo{year}{2022}).
\newblock
\urldef\tempurl%
\url{https://doi.org/10.1145/3485766}
\showDOI{\tempurl}


\bibitem[Sald{\'i}as~Fuentes et~al\mbox{.}(2022)]%
        {saldiasfuentesMoreEffectiveHuman2022}
\bibfield{author}{\bibinfo{person}{Bel{\'e}n Sald{\'i}as~Fuentes},
  \bibinfo{person}{George Foster}, \bibinfo{person}{Markus Freitag}, {and}
  \bibinfo{person}{Qijun Tan}.} \bibinfo{year}{2022}\natexlab{}.
\newblock \showarticletitle{Toward More Effective Human Evaluation for Machine
  Translation}. In \bibinfo{booktitle}{\emph{Proceedings of the 2nd Workshop on
  Human Evaluation of {{NLP}} Systems ({{HumEval}})}}.
\newblock
\urldef\tempurl%
\url{https://doi.org/10.18653/v1/2022.humeval-1.7}
\showDOI{\tempurl}


\bibitem[Schelter et~al\mbox{.}(2019)]%
        {schelterUnitTestingData2019}
\bibfield{author}{\bibinfo{person}{Sebastian Schelter}, \bibinfo{person}{Felix
  Biessmann}, \bibinfo{person}{Dustin Lange}, \bibinfo{person}{Tammo Rukat},
  \bibinfo{person}{Phillipp Schmidt}, \bibinfo{person}{Stephan Seufert},
  \bibinfo{person}{Pierre Brunelle}, {and} \bibinfo{person}{Andrey Taptunov}.}
  \bibinfo{year}{2019}\natexlab{}.
\newblock \showarticletitle{Unit Testing Data with Deequ}. In
  \bibinfo{booktitle}{\emph{Proceedings of the 2019 International Conference on
  Management of Data}} \emph{(\bibinfo{series}{{{SIGMOD}} '19})}.
\newblock
\urldef\tempurl%
\url{https://doi.org/10.1145/3299869.3320210}
\showDOI{\tempurl}


\bibitem[Scott(2015)]%
        {scottMultivariateDensityEstimation2015}
\bibfield{author}{\bibinfo{person}{David~W. Scott}.}
  \bibinfo{year}{2015}\natexlab{}.
\newblock \bibinfo{booktitle}{\emph{Multivariate Density Estimation:
  {{Theory}}, Practice, and Visualization}}.
\newblock
\urldef\tempurl%
\url{https://doi.org/10.1002/9781118575574}
\showDOI{\tempurl}


\bibitem[Sculley(2022)]%
        {sculleyDataCentricViewTechnical2022}
\bibfield{author}{\bibinfo{person}{D. Sculley}.}
  \bibinfo{year}{2022}\natexlab{}.
\newblock \bibinfo{title}{A {{Data-Centric View}} of {{Technical Debt}} in
  {{AI}}}.
\newblock
\newblock
\urldef\tempurl%
\url{https://datacentricai.org/data-in-deployment/}
\showURL{%
\tempurl}


\bibitem[Sculley et~al\mbox{.}(2015)]%
        {sculleyHiddenTechnicalDebt2015}
\bibfield{author}{\bibinfo{person}{D. Sculley}, \bibinfo{person}{Gary Holt},
  \bibinfo{person}{Daniel Golovin}, \bibinfo{person}{Eugene Davydov},
  \bibinfo{person}{Todd Phillips}, \bibinfo{person}{Dietmar Ebner},
  \bibinfo{person}{Vinay Chaudhary}, \bibinfo{person}{Michael Young},
  \bibinfo{person}{Jean-Fran{\c c}ois Crespo}, {and} \bibinfo{person}{Dan
  Dennison}.} \bibinfo{year}{2015}\natexlab{}.
\newblock \showarticletitle{Hidden {{Technical Debt}} in {{Machine Learning
  Systems}}}. In \bibinfo{booktitle}{\emph{Advances in {{Neural Information
  Processing Systems}}}}, Vol.~\bibinfo{volume}{28}.
\newblock


\bibitem[Shen et~al\mbox{.}(2021)]%
        {shen2021everyday}
\bibfield{author}{\bibinfo{person}{Hong Shen}, \bibinfo{person}{Alicia DeVos},
  \bibinfo{person}{Motahhare Eslami}, {and} \bibinfo{person}{Kenneth
  Holstein}.} \bibinfo{year}{2021}\natexlab{}.
\newblock \showarticletitle{Everyday Algorithm Auditing: {{Understanding}} the
  Power of Everyday Users in Surfacing Harmful Algorithmic Behaviors}.
\newblock \bibinfo{journal}{\emph{Proc. ACM Hum.-Comput. Interact.}}
  \bibinfo{volume}{5}, Article \bibinfo{articleno}{433} (\bibinfo{date}{Oct.}
  \bibinfo{year}{2021}).
\newblock
\urldef\tempurl%
\url{https://doi.org/10.1145/3479577}
\showDOI{\tempurl}


\bibitem[Silverman(2018)]%
        {silvermanDensityEstimationStatistics2018}
\bibfield{author}{\bibinfo{person}{Bernard~W Silverman}.}
  \bibinfo{year}{2018}\natexlab{}.
\newblock \bibinfo{booktitle}{\emph{Density Estimation for Statistics and Data
  Analysis}}.
\newblock
\urldef\tempurl%
\url{https://doi.org/10.1201/9781315140919}
\showDOI{\tempurl}


\bibitem[Soares et~al\mbox{.}(2018)]%
        {data-scientific}
\bibfield{author}{\bibinfo{person}{Felipe Soares}, \bibinfo{person}{Viviane
  Moreira}, {and} \bibinfo{person}{Karin Becker}.}
  \bibinfo{year}{2018}\natexlab{}.
\newblock \showarticletitle{A Large Parallel Corpus of Full-Text Scientific
  Articles}. In \bibinfo{booktitle}{\emph{Proceedings of the Eleventh
  International Conference on Language Resources and Evaluation (LREC-2018)}}
  (Miyazaki, Japan). \bibinfo{publisher}{European Language Resource
  Association}.
\newblock
\urldef\tempurl%
\url{http://aclweb.org/anthology/L18-1546}
\showURL{%
\tempurl}


\bibitem[Spinner et~al\mbox{.}(2019)]%
        {spinnerExplAInerVisualAnalytics2019}
\bibfield{author}{\bibinfo{person}{Thilo Spinner}, \bibinfo{person}{Udo
  Schlegel}, \bibinfo{person}{Hanna Schafer}, {and}
  \bibinfo{person}{Mennatallah {El-Assady}}.} \bibinfo{year}{2019}\natexlab{}.
\newblock \showarticletitle{{{explAIner}}: {{A Visual Analytics Framework}} for
  {{Interactive}} and {{Explainable Machine Learning}}}.
\newblock \bibinfo{journal}{\emph{IEEE Transactions on Visualization and
  Computer Graphics}} (\bibinfo{year}{2019}).
\newblock
\urldef\tempurl%
\url{https://doi.org/10.1109/tvcg.2019.2934629}
\showDOI{\tempurl}


\bibitem[Stanovsky et~al\mbox{.}(2019)]%
        {stanovskyEvaluatingGenderBias2019}
\bibfield{author}{\bibinfo{person}{Gabriel Stanovsky}, \bibinfo{person}{Noah~A.
  Smith}, {and} \bibinfo{person}{Luke Zettlemoyer}.}
  \bibinfo{year}{2019}\natexlab{}.
\newblock \showarticletitle{Evaluating {{Gender Bias}} in {{Machine
  Translation}}}. In \bibinfo{booktitle}{\emph{ACL}}.
\newblock
\urldef\tempurl%
\url{https://doi.org/10.18653/v1/p19-1164}
\showDOI{\tempurl}


\bibitem[Steele and Specia(2018)]%
        {steeleVisevalMetricViewer2018}
\bibfield{author}{\bibinfo{person}{David Steele} {and} \bibinfo{person}{Lucia
  Specia}.} \bibinfo{year}{2018}\natexlab{}.
\newblock \showarticletitle{Vis-Eval Metric Viewer: {{A}} Visualisation Tool
  for Inspecting and Evaluating Metric Scores of Machine Translation Output}.
  In \bibinfo{booktitle}{\emph{Proceedings of the 2018 Conference of the North
  {{American}} Chapter of the Association for Computational Linguistics:
  {{Demonstrations}}}}.
\newblock


\bibitem[Stent et~al\mbox{.}(2005)]%
        {stentEvaluatingEvaluationMethods2005}
\bibfield{author}{\bibinfo{person}{Amanda Stent}, \bibinfo{person}{Matthew
  Marge}, {and} \bibinfo{person}{Mohit Singhai}.}
  \bibinfo{year}{2005}\natexlab{}.
\newblock \showarticletitle{Evaluating Evaluation Methods for Generation in the
  Presence of Variation}. In \bibinfo{booktitle}{\emph{Proceedings of the 6th
  International Conference on Computational Linguistics and Intelligent Text
  Processing}} \emph{(\bibinfo{series}{{{CICLing}}'05})}.
\newblock
\urldef\tempurl%
\url{https://doi.org/10.1007/978-3-540-30586-6_38}
\showDOI{\tempurl}


\bibitem[Strobelt et~al\mbox{.}(2018)]%
        {strobeltEq2sEqv2018}
\bibfield{author}{\bibinfo{person}{Hendrik Strobelt},
  \bibinfo{person}{Sebastian Gehrmann}, \bibinfo{person}{Michael Behrisch},
  \bibinfo{person}{Adam Perer}, \bibinfo{person}{Hanspeter Pfister}, {and}
  \bibinfo{person}{Alexander~M Rush}.} \bibinfo{year}{2018}\natexlab{}.
\newblock \showarticletitle{S Eq 2s Eq-v Is: {{A}} Visual Debugging Tool for
  Sequence-to-Sequence Models}.
\newblock \bibinfo{journal}{\emph{IEEE transactions on visualization and
  computer graphics}}  \bibinfo{volume}{25} (\bibinfo{year}{2018}).
\newblock
\urldef\tempurl%
\url{https://doi.org/10.1109/tvcg.2018.2865044}
\showDOI{\tempurl}


\bibitem[Taira et~al\mbox{.}(2021)]%
        {tairaPragmaticAssessmentGoogle2021}
\bibfield{author}{\bibinfo{person}{Breena~R Taira}, \bibinfo{person}{Vanessa
  Kreger}, \bibinfo{person}{Aristides Orue}, {and} \bibinfo{person}{Lisa~C
  Diamond}.} \bibinfo{year}{2021}\natexlab{}.
\newblock \showarticletitle{A Pragmatic Assessment of Google Translate for
  Emergency Department Instructions}.
\newblock \bibinfo{journal}{\emph{Journal of Geneneral Internal Medicine}}
  \bibinfo{volume}{36} (\bibinfo{date}{Nov.} \bibinfo{year}{2021}).
\newblock
\urldef\tempurl%
\url{https://doi.org/10.1007/s11606-021-06666-z}
\showDOI{\tempurl}


\bibitem[Tiedemann(2020)]%
        {tiedemann-2020-tatoeba}
\bibfield{author}{\bibinfo{person}{J{\"o}rg Tiedemann}.}
  \bibinfo{year}{2020}\natexlab{}.
\newblock \showarticletitle{The {T}atoeba {T}ranslation {C}hallenge {--}
  {R}ealistic Data Sets for Low Resource and Multilingual {MT}}. In
  \bibinfo{booktitle}{\emph{Proceedings of the Fifth Conference on Machine
  Translation}}. \bibinfo{publisher}{Association for Computational
  Linguistics}, \bibinfo{address}{Online}, \bibinfo{pages}{1174--1182}.
\newblock
\urldef\tempurl%
\url{https://www.aclweb.org/anthology/2020.wmt-1.139}
\showURL{%
\tempurl}


\bibitem[Tiedemann and Thottingal(2020)]%
        {tiedemannOPUSMTBuildingOpen2020}
\bibfield{author}{\bibinfo{person}{J{\"o}rg Tiedemann} {and}
  \bibinfo{person}{Santhosh Thottingal}.} \bibinfo{year}{2020}\natexlab{}.
\newblock \showarticletitle{{{OPUS-MT}} \textemdash{} {{Building}} Open
  Translation Services for the {{World}}}. In
  \bibinfo{booktitle}{\emph{Proceedings of the 22nd Annual Conferenec of the
  European Association for Machine Translation ({{EAMT}})}}.
\newblock


\bibitem[Troles and Schmid(2021)]%
        {trolesExtendingChallengeSets2021}
\bibfield{author}{\bibinfo{person}{Jonas-Dario Troles} {and}
  \bibinfo{person}{Ute Schmid}.} \bibinfo{year}{2021}\natexlab{}.
\newblock \showarticletitle{Extending Challenge Sets to Uncover Gender Bias in
  Machine Translation: {{Impact}} of Stereotypical Verbs and Adjectives}. In
  \bibinfo{booktitle}{\emph{Proceedings of the Sixth Conference on Machine
  Translation}}.
\newblock


\bibitem[Tufte(2013)]%
        {tufteVisualDisplayQuantitative2013}
\bibfield{author}{\bibinfo{person}{Edward~R. Tufte}.}
  \bibinfo{year}{2013}\natexlab{}.
\newblock \bibinfo{booktitle}{\emph{The Visual Display of Quantitative
  Information} (\bibinfo{edition}{2nd ed., 8th print} ed.)}.
\newblock


\bibitem[{van der Maaten} and Hinton(2008)]%
        {vandermaatenVisualizingDataUsing2008}
\bibfield{author}{\bibinfo{person}{Laurens {van der Maaten}} {and}
  \bibinfo{person}{Geoffrey Hinton}.} \bibinfo{year}{2008}\natexlab{}.
\newblock \showarticletitle{Visualizing Data Using T-{{SNE}}}.
\newblock \bibinfo{journal}{\emph{Journal of Machine Learning Research}}
  \bibinfo{volume}{9} (\bibinfo{year}{2008}).
\newblock
\urldef\tempurl%
\url{http://jmlr.org/papers/v9/vandermaaten08a.html}
\showURL{%
\tempurl}


\bibitem[Vanmassenhove and Monti(2021)]%
        {vanmassenhoveGENderITAnnotatedEnglishItalian2021}
\bibfield{author}{\bibinfo{person}{Eva Vanmassenhove} {and}
  \bibinfo{person}{Johanna Monti}.} \bibinfo{year}{2021}\natexlab{}.
\newblock \showarticletitle{{{gENder-IT}}: {{An}} Annotated {{English-Italian}}
  Parallel Challenge Set for Cross-Linguistic Natural Gender Phenomena}. In
  \bibinfo{booktitle}{\emph{Proceedings of the 3rd Workshop on Gender Bias in
  Natural Language Processing}}.
\newblock
\urldef\tempurl%
\url{https://doi.org/10.18653/v1/2021.gebnlp-1.1}
\showDOI{\tempurl}


\bibitem[Vermeulen et~al\mbox{.}(2021)]%
        {vermeulenApplicationUniformManifold2021}
\bibfield{author}{\bibinfo{person}{Marc Vermeulen}, \bibinfo{person}{Kate
  Smith}, \bibinfo{person}{Katherine Eremin}, \bibinfo{person}{Georgina
  Rayner}, {and} \bibinfo{person}{Marc Walton}.}
  \bibinfo{year}{2021}\natexlab{}.
\newblock \showarticletitle{Application of {{Uniform Manifold Approximation}}
  and {{Projection}} ({{UMAP}}) in Spectral Imaging of Artworks}.
\newblock \bibinfo{journal}{\emph{Spectrochimica Acta Part A: Molecular and
  Biomolecular Spectroscopy}}  \bibinfo{volume}{252} (\bibinfo{year}{2021}).
\newblock
\urldef\tempurl%
\url{https://doi.org/10.1016/j.saa.2021.119547}
\showDOI{\tempurl}


\bibitem[Vilar et~al\mbox{.}(2006)]%
        {vilarErrorAnalysisStatistical2006}
\bibfield{author}{\bibinfo{person}{David Vilar}, \bibinfo{person}{Jia Xu},
  \bibinfo{person}{Luis~Fernando D'Haro}, {and} \bibinfo{person}{Hermann Ney}.}
  \bibinfo{year}{2006}\natexlab{}.
\newblock \showarticletitle{Error Analysis of Statistical Machine Translation
  Output}. In \bibinfo{booktitle}{\emph{Proceedings of the Fifth International
  Conference on Language Resources and Evaluation ({{LREC}}'06)}}.
\newblock


\bibitem[Wang et~al\mbox{.}(2019)]%
        {wangVizSeqVisualAnalysis2019}
\bibfield{author}{\bibinfo{person}{Changhan Wang}, \bibinfo{person}{Anirudh
  Jain}, \bibinfo{person}{Danlu Chen}, {and} \bibinfo{person}{Jiatao Gu}.}
  \bibinfo{year}{2019}\natexlab{}.
\newblock \showarticletitle{{{VizSeq}}: A Visual Analysis Toolkit for Text
  Generation Tasks}. In \bibinfo{booktitle}{\emph{Proceedings of the 2019
  {{Conference}} on {{Empirical Methods}} in {{Natural Language Processing}}
  and the 9th {{International Joint Conference}} on {{Natural Language
  Processing}} ({{EMNLP-IJCNLP}}): {{System Demonstrations}}}}.
\newblock
\urldef\tempurl%
\url{https://doi.org/10.18653/v1/d19-3043}
\showDOI{\tempurl}


\bibitem[Wexler et~al\mbox{.}(2019)]%
        {wexlerWhatIfToolInteractive2019}
\bibfield{author}{\bibinfo{person}{James Wexler}, \bibinfo{person}{Mahima
  Pushkarna}, \bibinfo{person}{Tolga Bolukbasi}, \bibinfo{person}{Martin
  Wattenberg}, \bibinfo{person}{Fernanda Viegas}, {and} \bibinfo{person}{Jimbo
  Wilson}.} \bibinfo{year}{2019}\natexlab{}.
\newblock \showarticletitle{The {{What-If Tool}}: {{Interactive Probing}} of
  {{Machine Learning Models}}}.
\newblock \bibinfo{journal}{\emph{IEEE TVCG}}  \bibinfo{volume}{26}
  (\bibinfo{year}{2019}).
\newblock
\urldef\tempurl%
\url{https://doi.org/10.1109/tvcg.2019.2934619}
\showDOI{\tempurl}


\bibitem[Wu et~al\mbox{.}(2019)]%
        {wuErruditeScalableReproducible2019}
\bibfield{author}{\bibinfo{person}{Tongshuang Wu}, \bibinfo{person}{Marco~Tulio
  Ribeiro}, \bibinfo{person}{Jeffrey Heer}, {and} \bibinfo{person}{Daniel
  Weld}.} \bibinfo{year}{2019}\natexlab{}.
\newblock \showarticletitle{Errudite: {{Scalable}}, {{Reproducible}}, and
  {{Testable Error Analysis}}}. In \bibinfo{booktitle}{\emph{Proceedings of the
  57th {{Annual Meeting}} of the {{Association}} for {{Computational
  Linguistics}}}}.
\newblock
\urldef\tempurl%
\url{https://doi.org/10.18653/v1/p19-1073}
\showDOI{\tempurl}


\bibitem[Wu et~al\mbox{.}(2021)]%
        {wuPolyjuiceGeneratingCounterfactuals2021}
\bibfield{author}{\bibinfo{person}{Tongshuang Wu}, \bibinfo{person}{Marco~Tulio
  Ribeiro}, \bibinfo{person}{Jeffrey Heer}, {and} \bibinfo{person}{Daniel
  Weld}.} \bibinfo{year}{2021}\natexlab{}.
\newblock \showarticletitle{Polyjuice: {{Generating}} Counterfactuals for
  Explaining, Evaluating, and Improving Models}. In
  \bibinfo{booktitle}{\emph{Proceedings of the 59th Annual Meeting of the
  Association for Computational Linguistics and the 11th International Joint
  Conference on Natural Language Processing (Volume 1: {{Long}} Papers)}}.
\newblock
\urldef\tempurl%
\url{https://doi.org/10.18653/v1/2021.acl-long.523}
\showDOI{\tempurl}


\bibitem[Yates(2006)]%
        {yatesScalingTowerBabel2006}
\bibfield{author}{\bibinfo{person}{Sarah Yates}.}
  \bibinfo{year}{2006}\natexlab{}.
\newblock \showarticletitle{Scaling the Tower of {{Babel Fish}}: {{An}}
  Analysis of the Machine Translation of Legal Information}.
\newblock \bibinfo{journal}{\emph{Law Library Journal}}  \bibinfo{volume}{98}
  (\bibinfo{year}{2006}).
\newblock


\bibitem[Zhang et~al\mbox{.}(2022)]%
        {zhang2022sliceteller}
\bibfield{author}{\bibinfo{person}{Xiaoyu Zhang},
  \bibinfo{person}{Jorge~Piazentin Ono}, \bibinfo{person}{Huan Song},
  \bibinfo{person}{Liang Gou}, \bibinfo{person}{Kwan-Liu Ma}, {and}
  \bibinfo{person}{Liu Ren}.} \bibinfo{year}{2022}\natexlab{}.
\newblock \showarticletitle{SliceTeller : A Data Slice-Driven Approach for
  Machine Learning Model Validation}.
\newblock \bibinfo{journal}{\emph{IEEE Transactions on Visualization and
  Computer Graphics}} (\bibinfo{year}{2022}), \bibinfo{pages}{1--11}.
\newblock
\urldef\tempurl%
\url{https://doi.org/10.1109/TVCG.2022.3209465}
\showDOI{\tempurl}


\bibitem[Zhao et~al\mbox{.}(2018)]%
        {zhaoGenderBiasCoreference2018}
\bibfield{author}{\bibinfo{person}{Jieyu Zhao}, \bibinfo{person}{Tianlu Wang},
  \bibinfo{person}{Mark Yatskar}, \bibinfo{person}{Vicente Ordonez}, {and}
  \bibinfo{person}{Kai-Wei Chang}.} \bibinfo{year}{2018}\natexlab{}.
\newblock \showarticletitle{Gender Bias in Coreference Resolution:
  {{Evaluation}} and Debiasing Methods}. In
  \bibinfo{booktitle}{\emph{Proceedings of the 2018 Conference of the North
  {{American}} Chapter of the Association for Computational Linguistics:
  {{Human}} Language Technologies, Volume 2 (Short Papers)}}.
\newblock
\urldef\tempurl%
\url{https://doi.org/10.18653/v1/n18-2003}
\showDOI{\tempurl}


\end{thebibliography}

\end{CJK*}
\end{document}